\documentclass[%
 reprint,
superscriptaddress,
 amsmath,amssymb,
 aps,
]{revtex4-2}

\usepackage[colorlinks=true, allcolors=blue]{hyperref}

\usepackage[utf8]{inputenc}
\usepackage[T1]{fontenc}
\usepackage[english]{babel}

\usepackage{graphicx}
\usepackage{multirow}
\usepackage{bm}
\usepackage{url}
\usepackage{mathtools}
\usepackage{float}
\usepackage{subcaption}

\usepackage{xcolor}

\makeatletter

\newcommand{\ket}[1]{\ensuremath{|{#1}\rangle}}
\newcommand{\bra}[1]{\ensuremath{\langle{#1}|}}

\newcommand{\avr}[1]{\ensuremath{\langle{#1}\rangle}}

\newcommand{\cnj}[1]{{#1}^{\ast}}
\newcommand{\hcnj}[1]{{#1}^{\dagger}}

\newcommand{\Tr}{\mathop{\rm Tr}\nolimits}

 \newcommand{\bs}[1]{\boldsymbol{#1}}
 \newcommand{\vc}[1]{\mathbf{#1}}

 \newcommand{\ind}[1]{\mathrm{#1}}

 \newcommand{\e}{\mathrm{e}}
 \newcommand{\ee}{\mathrm{e}}

\makeatother

\begin{document}

%%%%%%%%%%%%%%%%%%%%%
\title{Quantum repeaters and teleportation via entangled phase-modulated multimode coherent states}
%%%%%%%%%%%%%%%%%%%%%

 \author{R. Goncharov}
\email[Email address: ]{rkgoncharov@itmo.ru}
\affiliation{Quantum Information Laboratory, ITMO University, Kadetskaya Line, 3, Saint Petersburg, 199034, Russia}
\affiliation{Leading Research Center "National Center for Quantum Internet", ITMO University, Birzhevaya Line,
  16, Saint Petersburg, 199034, Russia}
\affiliation{SMARTS-Quanttelecom LLC, 6th Vasilyevskogo Ostrova Line, 59, Saint Petersburg, 199178, Russia}

\author{Alexei D. Kiselev}
\email[Email address: ]{alexei.d.kiselev@gmail.com}
\affiliation{Laboratory of Quantum Processes and Measurements, ITMO
  University,199034 Kadetskaya Line 3b, Saint Petersburg, Russia} 
\affiliation{Leading Research Center "National Center for Quantum Internet", ITMO University, Birzhevaya Line,
  16, Saint Petersburg, 199034, Russia}
\affiliation{Quantum Information Laboratory, ITMO University, Kadetskaya Line, 3, Saint Petersburg, 199034, Russia}

\author{E. S. Moiseev}
%  \email[Email address: ]{}
\affiliation{Kazan Quantum Center, Kazan National Research Technical University, 18a Chetaeva str.,  Kazan, 420111, Russia}

\author{E. Samsonov}
%  \email[Email address: ]{}
\affiliation{Quantum Information Laboratory, ITMO University, Kadetskaya Line, 3, Saint Petersburg,
  199034, Russia}
\affiliation{SMARTS-Quanttelecom LLC, 6th Vasilyevskogo Ostrova Line, 59, Saint Petersburg, 199178, Russia}
\affiliation{Leading Research Center "National Center for Quantum Internet", ITMO University, Birzhevaya Line,
  16, Saint Petersburg, 199034, Russia}

\author{S. A. Moiseev}
%  \email[Email address: ]{}
\affiliation{Kazan Quantum Center, Kazan National Research Technical University, 18a Chetaeva str.,  Kazan, 420111, Russia}

\author{F. Kiselev}
%  \email[Email address: ]{}
\affiliation{Quantum Information Laboratory, ITMO University, Kadetskaya Line, 3, Saint Petersburg,
  199034, Russia}
\affiliation{SMARTS-Quanttelecom LLC, 6th Vasilyevskogo Ostrova Line, 59, Saint Petersburg, 199178, Russia}
\affiliation{Leading Research Center "National Center for Quantum Internet", ITMO University, Birzhevaya Line,
  16, Saint Petersburg, 199034, Russia}

\author{V. Egorov}
%  \email[Email address: ]{}
\affiliation{Quantum Information Laboratory, ITMO University, Kadetskaya Line, 3, Saint Petersburg,
  199034, Russia}
\affiliation{SMARTS-Quanttelecom LLC, 6th Vasilyevskogo Ostrova Line, 59, Saint Petersburg, 199178, Russia}
\affiliation{Leading Research Center "National Center for Quantum Internet", ITMO University, Birzhevaya Line,
  16, Saint Petersburg, 199034, Russia}

\date{\today}

\begin{abstract}
 We present a scheme of quantum repeater that uses entangled  multimode coherent states which are
  obtained by electro-optic modulation of symmetric and antisymmetric Schr\"odinger cat states.  
  Part of generated entangled frequency modes are sent to a symmetric beam splitter at the central
  node, while the remaining modes are stored locally in quantum memories.      
  The entangled coherent states between remote quantum memories are conditionally prepared by photon
  counting measurements at the output channels of the beam splitter.  
  We study how the effects of decoherence in the quantum channel affect statistics of photocounts
  and, for the heralding outcomes determined by the parity of photocounts,
  evaluate the probability of success and
  the fidelity of the prepared entanglement depending on the symmetry of input cat states.
  \textcolor{black}{It is demonstrated that the generated 
  entanglement can be employed for teleportation of the phase information
  from the modulated states which are utilized
  in quantum key distributions with subcarrier wave encoding.}
\end{abstract}

\maketitle

%%%%%%%%%%%%%%%%
\section{Introduction}
\label{sec:introduction}
%%%%%%%%%%%%%%%%%

Transferring a quantum state between remote parties
is the primary purpose of quantum communication~\cite{Gisin:nphot:2007,Krenn2016}.
It lies at the heart of a variety of applications
that include secure transfer of classical messages using quantum key
distribution (QKD)~\cite{Gisin:rmp:2002,Xu:rmp:2020}, quantum
metrology~\cite{Giovannetti:nphot:2011,Toth:jpa:2014,Khabiboulline2019} and 
distributed computations~\cite{Meter2016,Yimsiriwattana2004}.  

One of the key problems of quantum communication is
the generation of high-fidelity quantum states entangled between distant sites~\cite{Gisin:nphot:2007}.
In the majority of approaches to this problem,
photons represent information carriers that can function as flying qubits and
the fundamental difficulty
is that the photons are subject to
losses (optical absorption) and other noise-induced perturbations
present in photonic channels such as optical fibers and turbulent atmosphere.
This noise has a detrimental effect on the quality of entanglement generated
between two remote parties leading to exponential decay of
the entanglement degree with the channel length.
Owing to the exponential losses, for the photons propagating in optical fibers,
the achievable distances are limited to about $200$~km
and the transmission of entanglement over global distances
(thousands of kilometers) becomes a challenging task.
The concept of quantum repeaters (QRs)
was put forward in~\cite{Briegel:prl:1998}
as the method to overcome this limitation.

The mode of operation of QRs
assumes that the transmission channel is divided
into several segments (elementary links).
The first step is to
prepare entanglement between the two nodes (at the ends) of each link.
Then, at the next step,
entanglement swapping between neighboring links is used to transfer
entanglement over significant distances to the target points of a quantum network.

There is a number of reviews focusing on different aspects of
QRs~\cite{Sangouard:rmp:2011,Munro2015,Muralidharan:scirep:2016,Azuma:arxiv:2022}.
For instance,
the review~\cite{Sangouard:rmp:2011}
focuses on the so-called
DLCZ protocol (for Duan, Lukin, Cirac, and Zoller)
developed in~\cite{Duan:nature:2001}
and its improvements.
In this protocol,
linear optics is combined with photon counting
to perform needed operations and
atomic ensembles are used as quantum memories (see~\cite{Lvovsky:nature:2009} for a review).
The primitives
and fundamental components needed for QRs
along with
classification of QR protocols into three relevant generations
are reviewed in~\cite{Munro2015}.
Advantages and challenges of each generation of QRs
determined by the methods utilized to suppress loss and operation errors
are analyzed in~\cite{Muralidharan:scirep:2016}.
A more recent review~\cite{Azuma:arxiv:2022}
additionally discusses newly emerging
classes of repeaters such as memoryless, error-corrected,
and all-photonic repeaters~\cite{Azuma:nature:2015,Zwerger:prl:2018,Su:pra:2018}
and put particular emphasis
on increasingly important role of QRs for
development of
long-distance quantum networks (quantum internet)~\cite{Wehner2018,Azuma:avs:2021,Wei:lpor:2022}.
For such entanglement-assisted networks,
the primary challenge  is  to go beyond the limit of
point-to-point quantum communication,
achieving high-rate
secure communication without using trusted relay nodes.

Among a variety of photonic quantum states used in QR protocols
the coherent states
have been attracted considerable attention~\cite{Look:pra:2008,Sangouard:josab:2010,Ghasemi:laserph:2019}
as the states that are relatively easy to produce and control.
The QR protocol analyzed in~\cite{Look:pra:2008}
uses hybrid entanglement where the coherent states are entangled with atomic (spin) qubits.
On the other hand, 
the entanglement generation and swapping protocols
studied in~\cite{Sangouard:josab:2010,Ghasemi:laserph:2019}
are based on the entangled coherent states.

These states were originally introduced in~\cite{Sanders:pra:1992}
(see Ref.~\cite{Sanders:jpa:2012} for a review) and
there is a number of quantum information processing
tasks that can be performed using
the entangled coherent
states~\cite{Munro:pra:2000,Jeong:pra:2001,Enk:pra:2001,Liu:chphys:2016,Sisodia:qinf:2017}.
Multimode coherent states also provide a potentially promising source of multipartite
entanglement~\cite{Miry:tmph:2019,Ra:nature:2020} required by the quantum networks.
In this paper, the approach to QRs based on such states will be our primary concern.

%\cite{Dong:jitp:2014,Huang:pra:2015}

More specifically,
we present and theoretically study
the scheme of a quantum repeater that uses
multimode coherent states generated by electro-optic modulation of
Schr\"odinger cat states.
This is the method that
produces phase-coded multimode signals to 
perform the subcarrier wave (SCW) encoding
which is proved to be useful in point-to-point~\cite{Merolla:prl:1999,Gleim:16,Miroshnichenko:optexp:2018,Gaidash:22},
plug-and-play (P\&P)
\cite{Bannik:21}, continuous-variable (CV) quantum key distribution
(QKD)~\cite{Melnik2018,Samsonov:scirep:2020,Samsonov:josab:2021} and twin-field QKD
\cite{Chistiakov:19}.
The scope of this method is not limited to the above protocols
and, owing to its robustness to
environmental distortion of fiber line, interferometer-free scheme and multiplexing capacity,
the SCW encoding is one of promising approaches for quantum communication.  

For such
electro-optic phase modulation-based method, we examine both
the heralded entanglement generation in
elementary links and the entanglement swapping procedures  
utilized for the creation and distribution
of entanglement between broadband quantum
memories (QMs)
(reports on recent experimental developments in the rapidly developing field of the QMs can be found in, e.g.,
Refs.~\cite{Saglamyurek:nature:2011,Sinclair:prl:2014,Moiseev2016,Kaczmarek:pra:2018,Ikuta:nature:2018,Davidson:pra:2020,Moiseev_2021,Lago:nature:2021,Askarani:prl:2021,Wang:josab:2021,Bustard:prl:2022,Businger:nature:2022}
).
As compared to QR schemes based on
controlled beam splitters (see e.g.~\cite{Sangouard:josab:2010}),
the electro-optic phase modulator provides a more flexible tool.
In practical implementation, using such tool
opens up new possibilities such as ability to control entanglement
swapping by choosing different sidebands and
quick adjustment of the mutual phase between Alice and Bob.
An important feature of our QR model is that it can be
associated with a real-life experimental detection scheme and
its applicability goes beyond the scope of QR protocols.
In order to demonstrate the latter,  
we additionally discuss the task of teleportation of the coherent state phase. 

The paper is organized as follows.
In Section~\ref{sec:elementary-link} we present
our modulator-based scheme of the elementary link
and analytical model for the heralded entanglement generation procedure
employed to
produce coherent cat states entangled between remote nodes of the link.
In Section~\ref{sec:fidel-prob-phot}
we shall analyze performance of the proposed scheme and quality of the generated
entanglement by evaluating
the fidelity and probabilities of photocounts for heralding events. 
In this analysis, we
consider nonideal photodetectors and take into account effects of decoherence
using the decoherence model which is typical for
CV quantum channels with energy transfer to environmental modes.
In Section~\ref{sec:entangl-swapping}
we discuss the entanglement swapping technique used to create
long-distance entanglement and its performance. 
Section~\ref{sec:teleporation} presents a teleportation scheme
that employs created entangled coherent states
to transfer phase information
encoded into SCW states between remote parties 
Finally, we discuss and summarize our results in Section~\ref{sec:disc}.

%%%%%%%%%%%%%%%%%%
\section{Elementary link}
\label{sec:elementary-link}
%%%%%%%%%%%%%%%%%%

\begin{figure*}[!htb]
   \centering
     \includegraphics[width=0.7\textwidth]{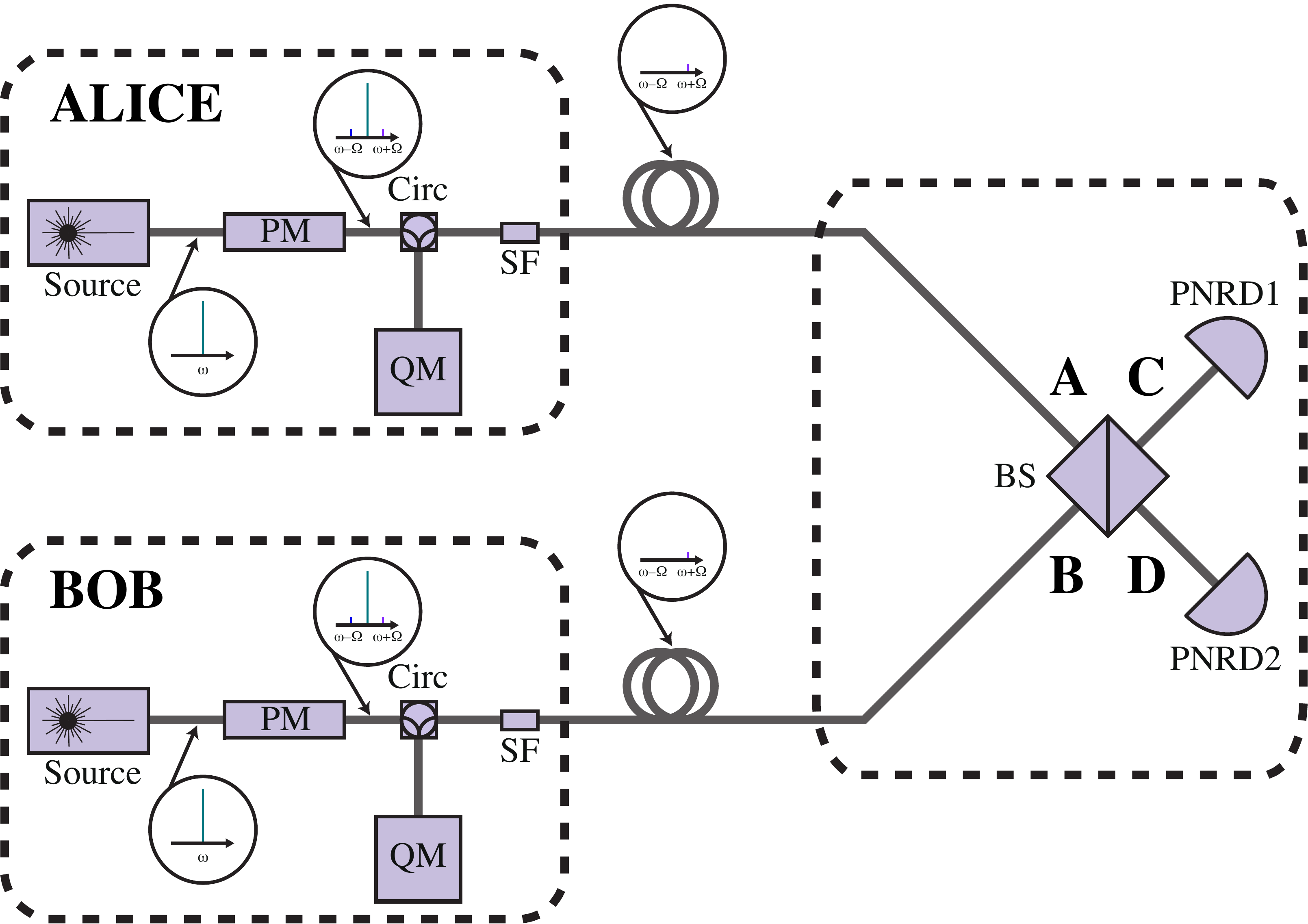}
\caption{Optical schematic of the subcarrier wave quantum repeater elementary link. Source is a
  source of Schr\"odinger cat states, PM is the electro-optic modulator, Circ is the circulator, SF is the
  spectral filter, QM is the quantum memory, BS is the beam splitter, PNRD is the photon-number-resolving
  detector. Diagrams in circles show the absolute value of signal spectrum taking into account only
  the first-order sidebands.} 
\label{scheme-0}
\end{figure*}

Figure~\ref{scheme-0} presents
the optical scheme of an elementary
link that creates entanglement between two remote nodes
(Alice and Bob) using the phase modulation method.
In this method,
Alice and Bob utilize electro-optic phase modulators
to produce local multimode entangled states
by modulating superpositions of one-mode coherent states
known as the Schr\"odinger cat states.
An important advantage of using fast phase
modulators is that Alice and Bob
can actively control the
output states by changing the modulation index and phase.

Referring to Fig.~\ref{scheme-0},
after generation of the local modulated states,
one of the frequency modes is sent to the central
relay via spectral filtering while keeping other modes stored in quantum memories.
In this section, 
we perform our analysis assuming that
the input (non-modulated) state of the (central) carrier wave mode
for Alice's and Bob's modulators,
$\ket{S_A}$ and $\ket{S_B}$,
are generally two different one-mode Schr\"odinger cat states of the following form:
\begin{align}
  &
  \label{eq:Schr_ini}
    \ket{S_A}=\ket{\Psi_{\nu'}(\alpha)},
    \:
    \ket{S_B}=\ket{\Psi_{\nu}(\beta)},
    \:
    \nu,\nu'\in\{+,-\},
\end{align}
where
$\ket{\Psi_{+}(\alpha)}$ and $\ket{\Psi_{-}(\alpha)}$
are the symmetric (even) and antisymmetric (odd) Schr\"odinger cats,
respectively. These states were originally introduced in~\cite{Dodonov:physica1974}
as even and odd coherent states given by
\begin{align}
  &
  \label{eq:Sch_cats}
    \ket{\Psi_{\pm}(\alpha)}=\frac{1}{\sqrt{M_{\pm}(\alpha)}}
    \ket{\alpha^{(\pm)}},\quad
    \ket{\alpha^{(\pm)}}\equiv\ket{\alpha}\pm\ket{-\alpha}
\end{align}
where
\begin{align}
  \label{eq:M_pm}
  M_{\pm}(\alpha)=\avr{\alpha^{(\pm)}|\alpha^{(\pm)}}=
  2(1\pm\exp(-2 |\alpha|^2)).
\end{align}

There is a variety of experimental techniques used to generate optical
Schr\"odinger cats~\cite{Polzik:prl:2006,Ourjoumtsev:nature:2009,Lund:pra:2013,Serikawa:prl:2018,Takase:pra:2021,Ourjoumtsev:nature:2007,Puri2017,Moiseev_2020,Grimm2020,Sychev:aipconf:2018,Zhiling:sciadv:2022}. 
It includes the method based on
photon subtraction from the squeezed vacuum
state~\cite{Polzik:prl:2006,Ourjoumtsev:nature:2009,Lund:pra:2013,Serikawa:prl:2018,Takase:pra:2021}, 
the protocol that uses
homodyne detection and photon number states~\cite{Ourjoumtsev:nature:2007},
reservoir engineering \cite{Puri2017,Moiseev_2020,Grimm2020}
and the methods that
involve making quadrature measurements of one of the modes of a biphoton NOON
state~\cite{Sychev:aipconf:2018}
and reflecting coherent-state photons
from a microwave cavity containing a superconducting qubit~\cite{Zhiling:sciadv:2022}. 
%\cite{Lee:pra:2012}

According to the model of electro-optic modulator~\cite{Kiselev:josab:2017},
for the input states $\ket{\pm\alpha}_A$ and $\ket{\pm\beta}_B$,
the modulated states
also known as the SCW states
can be described as the multimode coherent states
given by
\begin{align}
\ket{\pm\alpha}_A\to\ket{\pm\bs{\alpha}}_A=\otimes_{\mu=-S}^{S}\ket{\pm\alpha_\mu}_A, \notag \\
     \label{eq:cs_alp-bet}\ket{\pm\beta}_B\to\ket{\pm\bs{\beta}}_B=\otimes_{\mu=-S}^{S}\ket{\pm\beta_\mu}_B,
  \\
    \label{eq:U-mu}
    \alpha_{\mu}=\cnj{U_{\mu 0}^{(A)}} \alpha,\quad
    \beta_{\mu}=\cnj{U_{\mu 0}^{(B)}} \beta,
\end{align}
where $U_{\mu 0}^{(A,B)}$ are
the elements of the evolution matrix.
In the large $S$ limit,
these elements can be approximated as follows~\cite{Kiselev:josab:2017}
\begin{align}
  \label{eq:U-mu-approx}
  U_{\mu 0}^{(A,B)}\approx\ee^{-i\mu\phi_{A,B}} J_{\mu}(m_{A,B}),
\end{align}
where $\phi_{A,B}$ and $m_{A,B}$
are the phases and the modulation indices, respectively.

From Eq.~\eqref{eq:cs_alp-bet} it can be readily seen that
the electro-optic modulator acts like a multiport beam splitter
that transforms the one-mode cat states~\eqref{eq:Schr_ini}
into the multimode Schr\"odinger cat states of the following form:
\begin{align}
  &
  \label{eq:Schr_mod}
    \ket{S_A}\to\ket{\Psi_A}=\ket{\Psi_{\nu'}^{(A)}(\bs{\alpha})},
    \:
    \ket{S_B}\to\ket{\Psi_B}=\ket{\Psi_{\nu}^{(B)}(\bs{\beta})},
\end{align}
where
\begin{align}
&
                \ket{\Psi_{\pm}^{(A)}(\bs{\alpha})}=\frac{1}{\sqrt{M_{\pm}(\bs{\alpha})}}\ket{\bs{\alpha}^{(\pm)}}_A,
                \notag
  \\
&
    \label{eq:Schr_alp-bet1}
                            \ket{\Psi_{\pm}^{(B)}(\bs{\beta})}=\frac{1}{\sqrt{M_{\pm}(\bs{\beta})}}
\{\ket{\bs{\beta}}_B\pm\ket{-\bs{\beta}}_B\}
    \\
    &
    \label{eq:Schr_alp-bet2}
    \ket{\bs{\alpha}^{(\pm)}}_A=\ket{\bs{\alpha}}_A\pm\ket{-\bs{\alpha}}_A,
    \\
    &
      \label{eq:M_pm-mult}
    M_{\pm}(\bs{\alpha})=2(1\pm\exp(-2 |\bs{\alpha}|^2)),
    \:
    |\bs{\alpha}|^2=\sum_{\mu=-S}^{S}|\alpha_\mu|^2.
\end{align}

Thus the modulated state shared by Alice and Bob, $\ket{\Psi_{AB}}$,
is the tensor product of two multimode
Schr\"odinger cat states after the Alice's and Bob's modulators:
$\ket{\Psi}_{AB}=\ket{\Psi_{\nu'}^{(A)}(\bs{\alpha})}\otimes\ket{\Psi_{\nu}^{(B)}(\bs{\beta})}$.
This state can be written in the form
%\begin{widetext}
\begin{align}
  &
  \label{eq:Psi-AB}
  \ket{\Psi}_{AB}=
  \frac{1}{N_{AB}}
  \Bigl\{
  \ket{{\bs{\alpha}},{\bs{\beta}}}_{AB}
  + \nu'\nu \ket{-{\bs{\alpha}},-{\bs{\beta}}}_{AB}
    \notag
  \\
&
    +\nu \ket{{\bs{\alpha}},-{\bs{\beta}}}_{AB}
  +\nu'
 \ket{-{\bs{\alpha}},{\bs{\beta}}}_{AB}
    \Bigr\}
    =\sqrt{\frac{M_{\nu'\nu}(\bs{\alpha},\bs{\beta})}{M_{\nu'}(\bs{\alpha})M_{\nu}(\bs{\beta})}}
        \notag
  \\
  &
    \times\bigl\{
    \ket{\Psi_{\nu'\nu}^{(AB)}(\bs{\alpha},\bs{\beta})}+\nu\ket{\Psi_{\nu'\nu}^{(AB)}(\bs{\alpha},-\bs{\beta})}
    \bigr\},
  \\
  &
      M_{\pm}({\bs{\alpha}},{\bs{\beta}})=
    2
    \left\{
    1\pm\exp[-2(|\bs{\alpha}|^2+|\bs{\beta}|^2)]
    \right\},
\end{align}
%\end{widetext}
where
$N_{AB}=\sqrt{M_{\nu'}(\alpha)M_{\nu}(\beta)}$
and
$\ket{{\bs{\alpha}},{\bs{\beta}}}_{AB}$
represents $\ket{\bs{\alpha}}_{A}\otimes\ket{\bs{\beta}}_{B}$
rearranged into the tensor product of modes.

From now on we shall restrict our analysis to the important special case where
$\bs{\alpha}=\bs{\beta}$ and,
following the general approach~\cite{Sangouard:rmp:2011,Sangouard:josab:2010}
to preparation of entangled states shared by Alice and Bob,
we shall assume that the modes $\bs{\alpha}$ that enter
the modulated states,
$\ket{\Psi_{\nu}(\bs{\alpha})}\equiv\ket{\Psi_{\nu}(\bs{\alpha}_{\ind{qm}},\bs{\alpha}_{\ind{bs}})}$,
are divided into two groups:
the modes $\bs{\alpha}_{\ind{qm}}$
stored in the quantum memory
 and
 the modes $\bs{\alpha}_{\ind{bs}}$
put to interfere onto a symmetric $50:50$ beam splitter
with the output channels $C$ and $D$.
The above modes
brought into interference onto the beam splitter
appear to be transformed as follows
\begin{align}
  &
  \label{eq:BS-out}
  \hat{T}_{\bs{\alpha}_{\ind{bs}}\bs{\alpha}_{\ind{bs}}\to
  CD}\ket{\pm \bs{\alpha}_{\ind{bs}}}_A\otimes\ket{\pm\bs{\alpha}_{\ind{bs}}}_B=
  \ket{\pm\bs{\gamma}_{\ind{bs}}}_C\otimes\ket{\vc{0}}_D,
  \notag
  \\
  &
\hat{T}_{\bs{\alpha}_{\ind{bs}}\bs{\alpha}_{\ind{bs}}\to
  CD}\ket{\pm \bs{\alpha}_{\ind{bs}}}_A\otimes\ket{\mp\bs{\alpha}_{\ind{bs}}}_B=
    \ket{\vc{0}}_C\otimes\ket{\pm\bs{\gamma}_{\ind{bs}}}_D,
\end{align}
where $\bs{\gamma}_{\ind{bs}}=\sqrt{2}\bs{\alpha}_{\ind{bs}}$.

We can now apply
the transformation~\eqref{eq:BS-out}
to
the state shared by Alice and Bob (see~\eqref{eq:Psi-AB}).
The result reads
\begin{widetext}
\begin{align}
  &
  \label{eq:T_Psi_gen}
    \hat{T}_{\bs{\alpha}_{\ind{bs}}\bs{\alpha}_{\ind{bs}}\to CD}\ket{\Psi_{\nu'}^{(A)}(\bs{\alpha}_{\ind{qm}},\bs{\alpha}_{\ind{bs}})}\otimes
  \ket{\Psi_{\nu}^{(B)}(\bs{\alpha}_{\ind{qm}},\bs{\alpha}_{\ind{bs}})}
 =
    \frac{1}{2{\sqrt{M_{\nu'}(\bs{\alpha}_{\ind{qm}},\bs{\alpha}_{\ind{bs}})M_{\nu}(\bs{\alpha}_{\ind{qm}},\bs{\alpha}_{\ind{bs}})}}}
    \notag
  \\
  &
    \times\sum_{\mu=\pm}
    \sqrt{M_{\mu'}(\bs{\gamma}_{\ind{bs}})M_{\mu}(\bs{\alpha}_{\ind{qm}},\bs{\alpha}_{\ind{qm}})}
    \Bigl\{
    \ket{\Psi_{\mu'}^{(C)}(\bs{\gamma}_{\ind{bs}})}\otimes\ket{\vc{0}}_D\otimes\ket{\Psi_{\mu}^{(AB)}(\bs{\alpha}_{\ind{qm}},\bs{\alpha}_{\ind{qm}})}
  \notag
  \\
  &
+\nu
\ket{\vc{0}}_C\otimes\ket{\Psi_{\mu'}^{(D)}(\bs{\gamma}_{\ind{bs}})}\otimes
    \ket{\Psi_{\mu}^{(AB)}(\bs{\alpha}_{\ind{qm}},-\bs{\alpha}_{\ind{qm}})}
    \Bigr\},\quad
    \mu'=\nu\nu'\mu,
\end{align}
\end{widetext}
where $\bs{\gamma}_{\ind{bs}}\equiv\sqrt{2}\bs{\alpha}_{\ind{bs}}$.

If, for instance, we now perform a measurement on
the output mode $C$ to distinguish the states
$\ket{\Psi_{\nu'\nu}^{(C)}(\bs{\gamma}_{\ind{bs}})}$ and $\ket{\Psi_{-\nu'\nu}^{(C)}(\bs{\gamma}_{\ind{bs}})}$,
the multimode state will collapse onto either
$  \ket{\Psi_{+}^{(AB)}({\bs{\alpha}}_{\ind{qm}},{\bs{\alpha}}_{\ind{qm}})}$
or
$  \ket{\Psi_{-}^{(AB)}({\bs{\alpha}}_{\ind{qm}},{\bs{\alpha}}_{\ind{qm}})}$,
respectively.
Thus preparation of  the entangled coherent cat states 
is heralded by
the parity of clicks of a photon-number-resolving detector placed at the output channel $C$.

\begin{figure*}[!htp]
\centering
\begin{subfigure}{.5\textwidth}
  \centering
  \includegraphics[width=.8\linewidth]{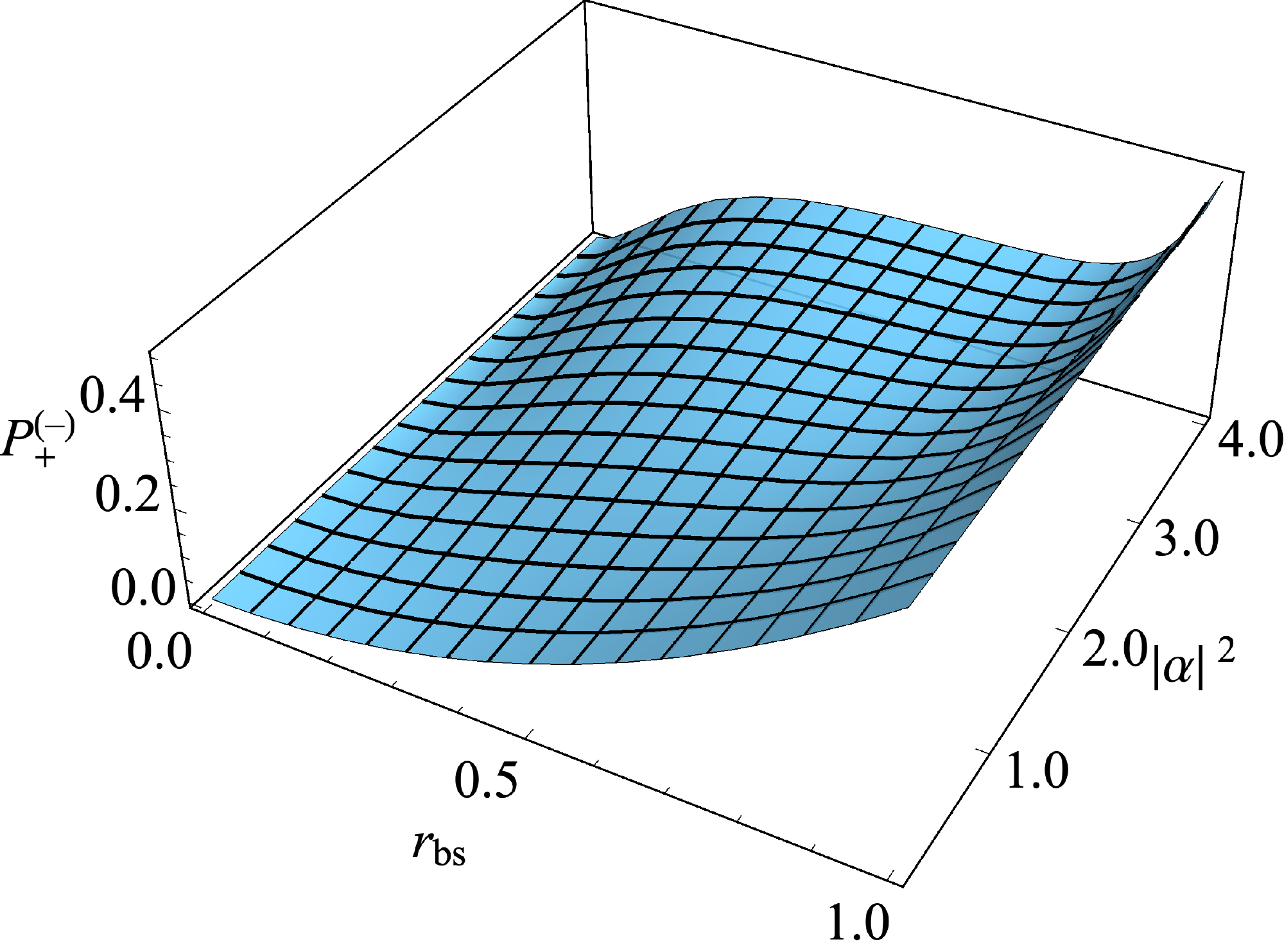}
  \label{subfig:p-p-minus}
   \caption{}
 \end{subfigure}%
\begin{subfigure}{.5\textwidth}
  \centering
  \includegraphics[width=.8\linewidth]{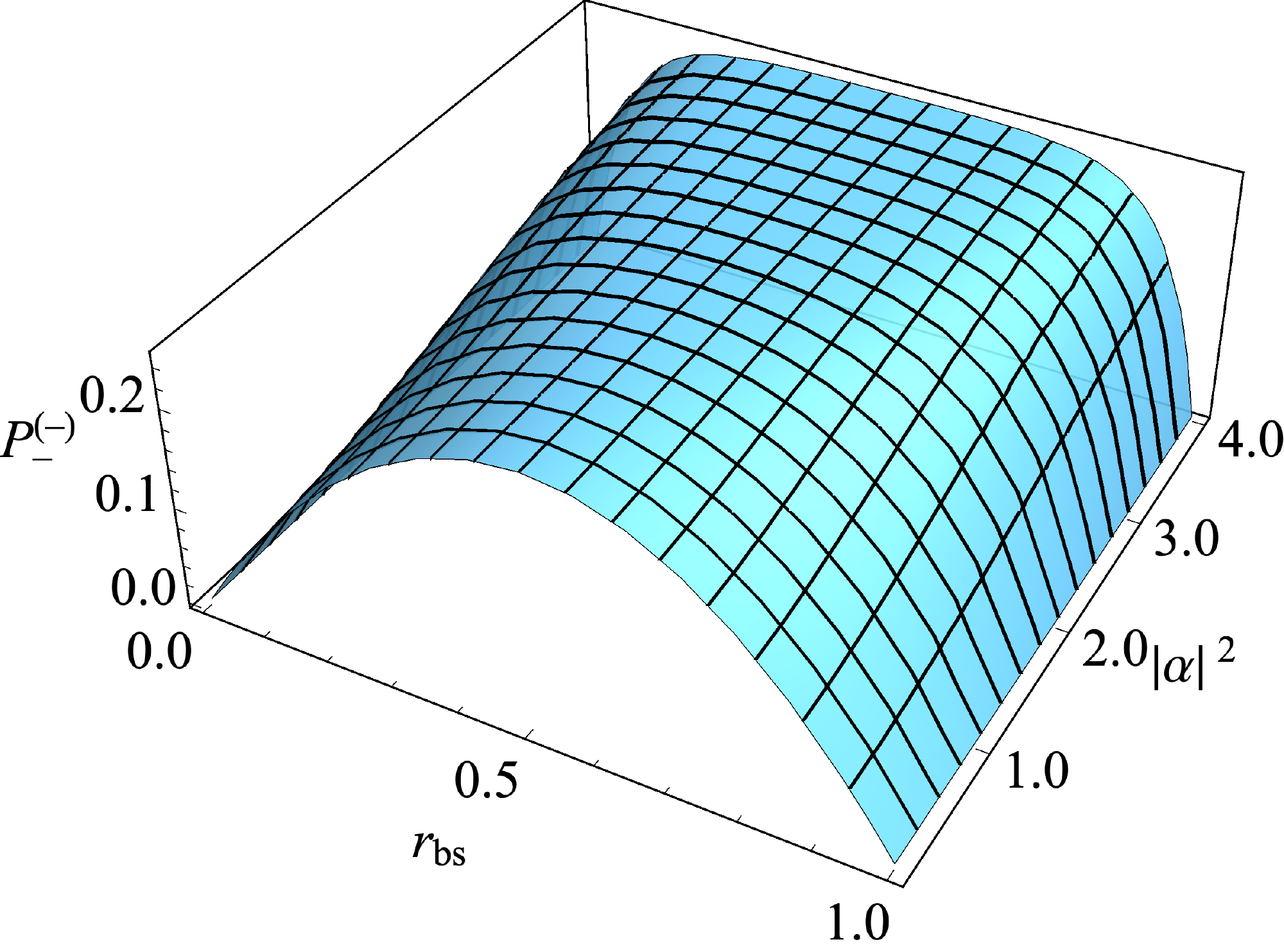}
  \label{subfig:p-m-minus}
   \caption{}
 \end{subfigure}
 \\
\begin{subfigure}{.5\textwidth}
  \centering
  \includegraphics[width=.8\linewidth]{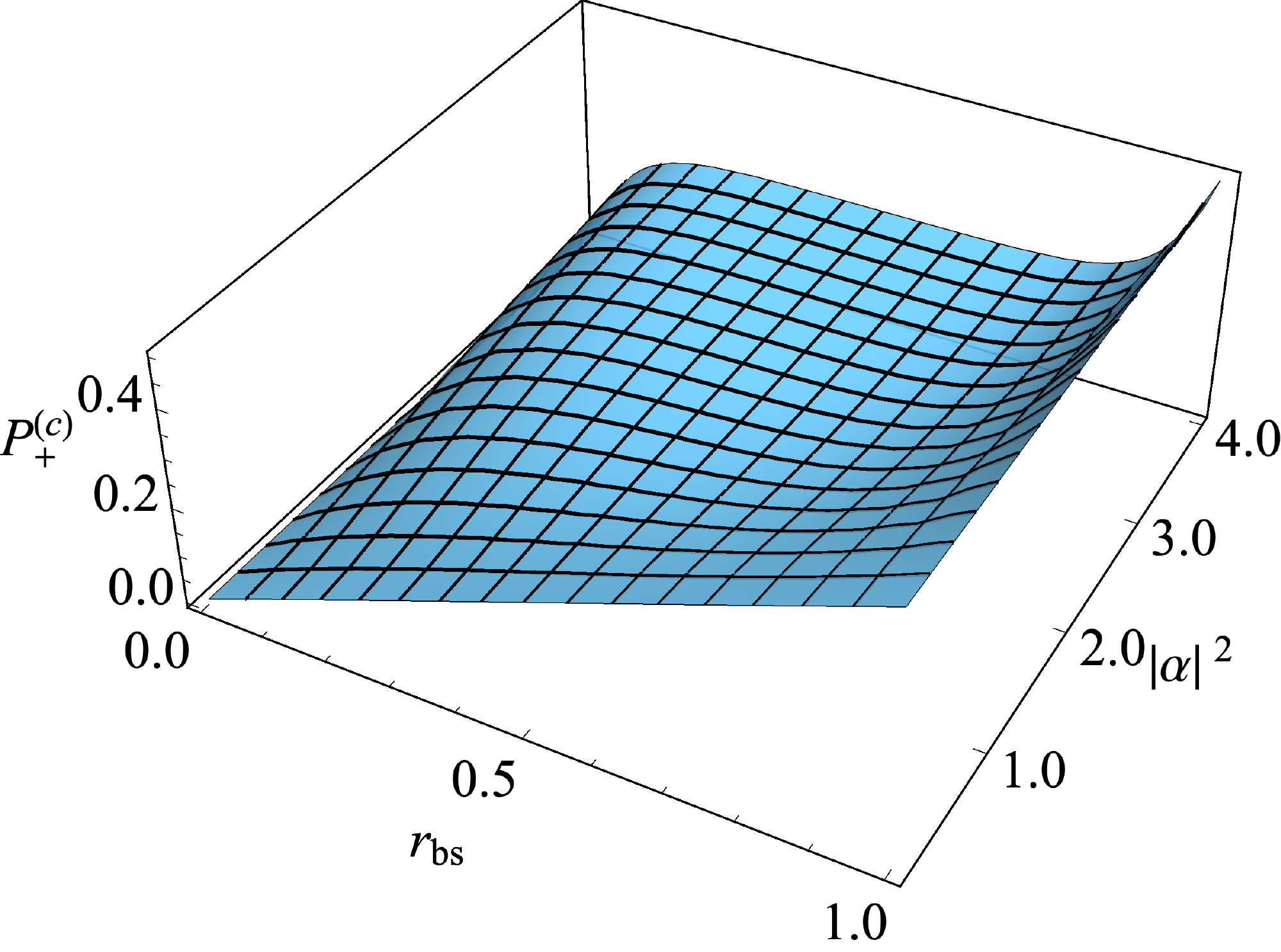}
  \label{subfig:p-plus-c}
   \caption{}
\end{subfigure}%
\begin{subfigure}{.5\textwidth}
  \centering
  \includegraphics[width=.8\linewidth]{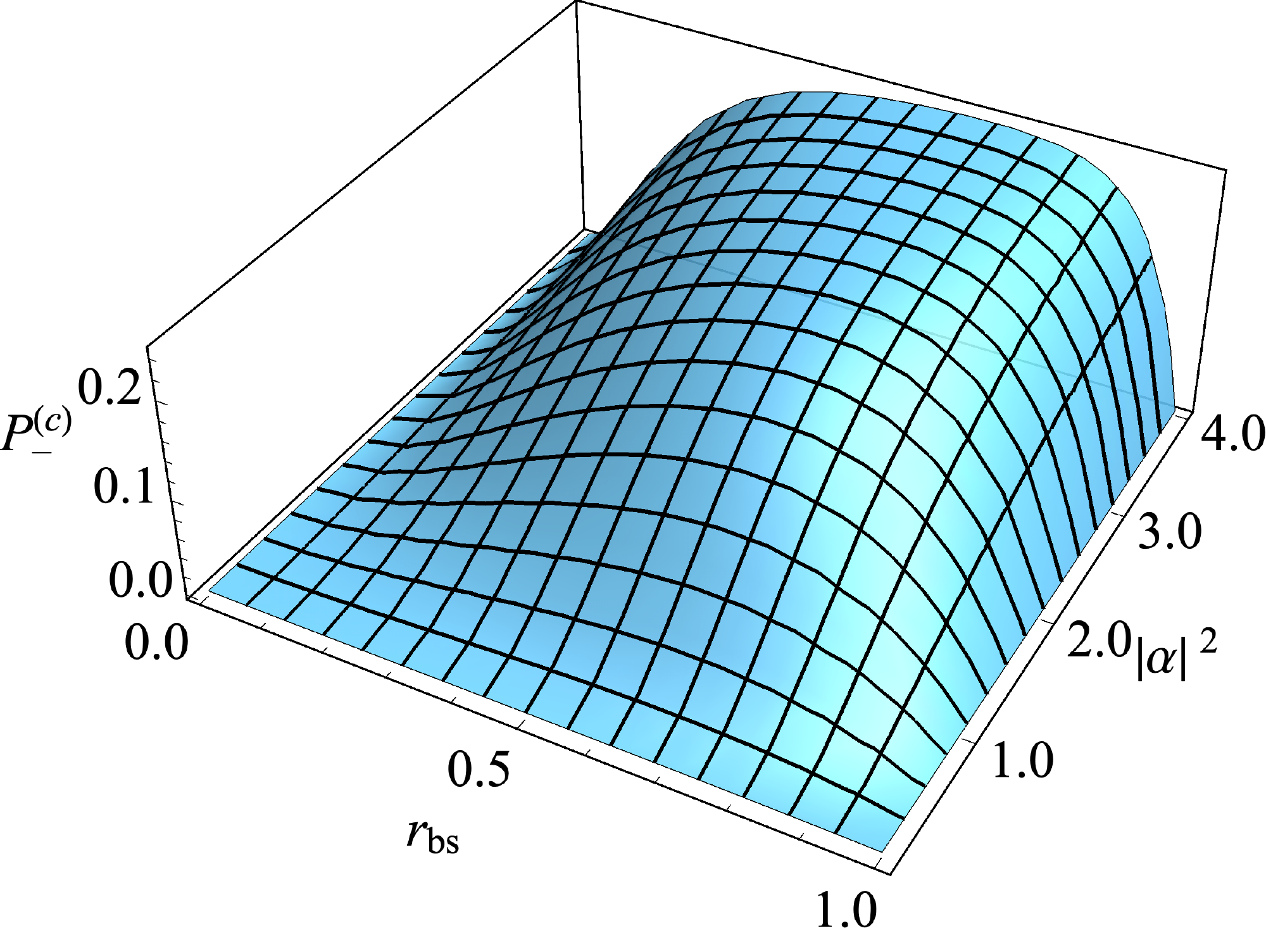}
  \label{subfig:p-minus-c}
  \caption{}
\end{subfigure}
\caption{
  Probabilities $P_{\pm}^{(-)}$ and $P_{\mp}^{(c)}$ associated with
  the cat states $\ket{\Phi_{\pm}(\bs{\gamma}_{\ind{bs}})}$
  at an output port of the beam splitter
  computed
from Eqs.~\eqref{eq:Pmum} and~\eqref{eq:Pmuc}
  as a function of $|\alpha|^{2}$ and
  $r_{\mathrm{bs}}=|\bs{\alpha}_{\ind{bs}}|^{2}/|\bs{\alpha}|^{2}$.
}  
\label{p-v}
\end{figure*}

Now we introduce the modified symmetric cat state
\begin{align}
  \ket{\tilde{\Psi}_{+}(\bs{\alpha})}&=
\frac{1}{\sqrt{\tilde{M}_{+}(\bs{\alpha})}}\Bigl[
  \ket{\tilde{\bs{\alpha}}}+\ket{-\tilde{\bs{\alpha}}}
  \Bigr],\notag \\
     \label{eq:tPsip}
     \ket{\pm\tilde{\bs{\alpha}}}&\equiv\ket{\pm\bs{\alpha}}-\e^{-|\bs{\alpha}|^2/2}\ket{\vc{0}},
\end{align}
where $\tilde{M}_{+}(\bs{\alpha})=M_{+}(\bs{\alpha})-4\exp(-|\bs{\alpha}|^2)$,
defined in terms of the coherent states renormalized by subtracting the vacuum contribution.
Then we have
a set of three orthonormal states:
$\ket{\Psi_{-}(\sqrt{2}\bs{\alpha}_{\ind{bs}})}\equiv\ket{\Phi_{-}(\bs{\gamma}_{\ind{bs}})}$,
$\ket{\tilde{\Psi}_{+}(\sqrt{2}\bs{\alpha}_{\ind{bs}})}\equiv\ket{\Phi_{+}(\bs{\gamma}_{\ind{bs}})}$,
 and
$\ket{\vc{0}}$
that can be conveniently used to render
the state~\eqref{eq:T_Psi_gen} into the form
%\begin{widetext}
\begin{align}
&
                \label{eq:Psi_BS}
\hat{T}_{\bs{\alpha}_{\ind{bs}}\bs{\alpha}_{\ind{bs}}\to CD}\ket{\Psi_{\nu'}^{(A)}(\bs{\alpha}_{\ind{qm}},\bs{\alpha}_{\ind{bs}})}\otimes
  \ket{\Psi_{\nu}^{(B)}(\bs{\alpha}_{\ind{qm}},\bs{\alpha}_{\ind{bs}})}
   \notag
 \\
  &
                =\sqrt{P_0^{(\nu'\nu)}}\ket{\vc{0},\vc{0}}_{CD}
    \otimes\ket{\Psi_{\nu'}^{(A)}(\bs{\alpha}_{\ind{qm}})}\otimes\ket{\Psi_{\nu}^{(B)}(\bs{\alpha}_{\ind{qm}})}
   \notag
 \\
  &
    + \sum_{\mu=\pm} 
    \sqrt{P_{\mu}^{(\nu'\nu)}}
    \Bigl\{
    \ket{{\Phi}_{\mu'}^{(C)}(\bs{\gamma}_{\ind{bs}})}\otimes\ket{\vc{0}}_{D}\otimes\ket{\Psi_{\mu}^{(AB)}(\bs{\alpha}_{\ind{qm}},\bs{\alpha}_{\ind{qm}})}
    \notag
   \\
  &
   +\nu
\ket{\vc{0}}_C\otimes\ket{\Phi_{\mu'}^{(D)}(\bs{\gamma}_{\ind{bs}})}\otimes
    \ket{\Psi_{\mu}^{(AB)}(\bs{\alpha}_{\ind{qm}},-\bs{\alpha}_{\ind{qm}})}
    \Bigr\},
\end{align}
%\end{widetext}
where
$P_{\mu}^{(\nu'\nu)}$ is the probability
for the states  $\ket{\Phi_{\mu'}(\bs{\gamma}_{\ind{bs}})}_C\otimes\ket{\vc{0}}_{D}$
and $\ket{\vc{0}}_C\otimes\ket{\Phi_{\mu'}(\bs{\gamma}_{\ind{bs}})}_{D}$
to be detected at the output ports of the beam splitter, whereas
$P_0^{(\nu'\nu)}$ is the probability to detect the vacuum state
$\ket{\vc{0}}_C\otimes\ket{\vc{0}}_{D}$.
So, we have
\begin{align}
  \label{eq:Pmp-sum}
        2 P_{+}^{(\nu'\nu)}+ 2 P_{-}^{(\nu'\nu)}+P_{0}^{(\nu'\nu)}=1
\end{align}
and the expressions for $P_{\mu}^{(\nu'\nu)}$ are given by 
\begin{subequations}
  \label{eq:Pmp0-gen}
\begin{align}
    &P_{\mu}^{(\nu'\nu)}=\ind{Prob}(\mu|\nu'\nu)\notag \\
    &
      \label{eq:Pmu-gen}
    =\frac{1}{4}
    \frac{\tilde{M}_{\mu'}(\bs{\gamma}_{\ind{bs}})M_{\mu}(\bs{\alpha}_{\ind{qm}},\bs{\alpha}_{\ind{qm}})}{%
    M_{\nu'}(\bs{\alpha}_{\ind{qm}},\bs{\alpha}_{\ind{bs}})M_{\nu}(\bs{\alpha}_{\ind{qm}},\bs{\alpha}_{\ind{bs}})},\quad
    \mu\in\{+,-\},
  \\
      &
    \notag
    P_{0}^{(\nu'\nu)}=\ind{Prob}(0|\nu'\nu)\\
    &
    \label{eq:P0-gen}=
    \frac{M_{\nu'}(\bs{\alpha}_{\ind{qm}})M_{\nu}(\bs{\alpha}_{\ind{qm}})}{%
    M_{\nu'}(\bs{\alpha}_{\ind{qm}},\bs{\alpha}_{\ind{bs}})M_{\nu}(\bs{\alpha}_{\ind{qm}},\bs{\alpha}_{\ind{bs}})}
    \exp(-|\bs{\gamma}_{\ind{bs}}|^2).
\end{align}
\end{subequations}
In the case, where the modulated cat states are identical
with $\nu=\nu'$,
formulas~\eqref{eq:Pmp0-gen} can be simplified giving
the probabilities $P_{\mu}^{(\nu\nu)}\equiv P_{\mu}^{(\nu)}$
in the form
\begin{subequations}
  \label{eq:Pmum}
  \begin{align}
    &P_{\pm}^{(+)}=\tanh^2{|{\alpha}|^2}P_{\pm}^{(-)}, \\
    \label{eq:Pmup-vs-Pmum}&|\alpha|^2=|\bs{\alpha}|^2=|\bs{\alpha}_{\ind{qm}}|^2+|\bs{\alpha}_{\ind{bs}}|^2,
    \quad
    |\bs{\alpha}_{\ind{bs}}|^2= r_{\ind{bs}} |\bs{\alpha}|^2
    \\
  &
  \label{eq:Pmm}
    P_{-}^{(-)}(r_{\ind{bs}},|{\alpha}|^2)
    =\frac{\sinh(2(1-r_{\ind{bs}})|{\alpha}|^2)\sinh(2 r_{\ind{bs}}|{\alpha}|^2)}{
    4\sinh^2(|{\alpha}|^2)}, 
  \\
  &
    \label{eq:Ppm}
    P_{+}^{(-)}(r_{\ind{bs}},|{\alpha}|^2)
    =\frac{\cosh(2(1-r_{\ind{bs}})|{\alpha}|^2)\sinh^2(r_{\ind{bs}}|{\alpha}|^2)}{
    2\sinh^2(|{\alpha}|^2)},
  \\
 &P_0^{(-)}(r_{\ind{bs}},|{\alpha}|^2)=\frac{\sinh^2((1-r_{\ind{bs}})|{\alpha}|^2)}{\sinh^2(|{\alpha}|^2)}, \notag
    \\
    \label{eq:P0m}
    &P_0^{(+)}(r_{\ind{bs}},|{\alpha}|^2)=\frac{\cosh^2((1-r_{\ind{bs}})|{\alpha}|^2)}{\cosh^2(|{\alpha}|^2)},
  \end{align}
\end{subequations}
where the parameter $r_{\ind{bs}}$
is the ratio of the average number of photons for the modes sent to the beam splitter,
$|\bs{\alpha}_{\ind{bs}}|^2/|\bs{\alpha}|^2$,
and the input mean photon number, $|\bs{\alpha}|^2=|\alpha|^2$.
In the opposite case with  modulated cat states that differ in symmetry,
the probabilities
$P_{\mu}^{(+-)}=P_{\mu}^{(-+)}\equiv P_{\mu}^{(c)}$
are given by
\begin{subequations}
  \label{eq:Pmuc}
\begin{align}
  &
  \label{eq:Pmc}
      P_{-}^{(c)}(r_{\ind{bs}},|{\alpha}|^2)
      =
\frac{\sinh(2(1-r_{\ind{bs}})|{\alpha}|^2)\sinh^2(r_{\ind{bs}}|{\alpha}|^2)}{\sinh(2|{\alpha}|^2)}, 
  \\
  &
    \label{eq:Ppc}
    P_{+}^{(c)}(r_{\ind{bs}},|{\alpha}|^2)
    =\frac{\cosh(2(1-r_{\ind{bs}})|{\alpha}|^2)\sinh(2r_{\ind{bs}}|{\alpha}|^2)}{2\sinh(2|{\alpha}|^2)},
  \\
 &
  \label{eq:P0c}
      P_0^{(c)}(r_{\ind{bs}},|{\alpha}|^2)
      =\frac{\sinh(2(1-r_{\ind{bs}})|{\alpha}|^2)}{\sinh(2|{\alpha}|^2)}.
\end{align}
\end{subequations}

From Eq.~\eqref{eq:Pmup-vs-Pmum},
the probabilities for two antisymmetric cat states $P_{\pm}^{(-)}$
are higher than those describing
the case of two symmetric cats, $P_{\pm}^{(+)}$.
In particular, at small values of $|\alpha|$,
the latter is proportional to $|\alpha|^2$,
whereas, for  $P_{\pm}^{(-)}$, we have
\begin{align}
  \label{eq:Ppmm-small}
  P^{(-)}_{-}\approx r_{\ind{bs}}(1-r_{\ind{bs}}),\quad
P^{(-)}_{+}\approx r_{\ind{bs}}^2/2.  
\end{align}
Note that the relations
\begin{align}
  \label{eq:Ppc-small}
P^{(c)}_{-}\approx r_{\ind{bs}}(1-r_{\ind{bs}})^2|\alpha|^4, \quad
P^{(c)}_{+}\approx r_{\ind{bs}}/2
\end{align}
describe behavior of $P_{\pm}^{(c)}$ in the region
of small amplitudes $|\alpha|$.

Now we briefly discuss how the photon number ratio
$r_{\ind{bs}}$ affects $|{\alpha}|^2$-dependence of
the probabilities.
Figure~\ref{p-v} shows
the surfaces representing 
the probabilities $P_{\pm}^{(-)}$ and $P_{\pm}^{(c)}$
plotted in the $r_{\ind{bs}}$-$|\alpha|^2$ plane.

At $0<r_{\ind{bs}}<1$,
all the probabilities
$P_{\pm}^{(+)}$, $P_{\pm}^{(-)}$ and $P_{\pm}^{(c)}$
approach the value $1/4$ in the limit of large amplitudes with $|\alpha|\to\infty$.
They also vanish provided the ratio $r_{\ind{bs}}$ is zero.
Obviously, in this case, we cannot produce entangled states.

From Eqs.~\eqref{eq:Pmum}
and~\eqref{eq:Pmuc}, at $r_{\ind{bs}}=1$,
the probabilities $P_{-,0}^{(-)}$ and $P_{-,0}^{(c)}$ equal zero,
whereas $P_{+}^{(-)}=P_{+}^{(c)}=1/2$.
Another case where the probabilities
$P_{-}^{(-)}$ and $P_{+}^{(c)}$ (see Eqs.~\eqref{eq:Pmm} and \eqref{eq:Ppc})
are independent of $|\alpha|$, $P_{-}^{(-)}=P_{+}^{(c)}=1/4$ ,
occurs at the ratio equal to one $1/2$, $r_{\ind{bs}}=1/2$.

%%%%%%%%%%%%%%%%%%%%%%%%%%%%%%%
\section{Photodetection and decoherence}
\label{sec:fidel-prob-phot}
%%%%%%%%%%%%%%%%%%%%%%%%%%

In the previous section,
we have found that
generation of the entangled cat states
is heralded by outcomes of
measurements  performed at the relay node
in the orthonormal basis $\{\ket{\Phi_{+}},\ket{\Phi_{-}},\ket{\vc{0}}\}$.
Such measurements would perfectly discriminate between
the states at the output ports of the beam splitter
producing the entangled states, $\ket{\Psi^{(AB)}_{\pm}}$,
shared by the remote nodes of the link
with the probabilities $P_{\pm}^{(\nu\nu')}$
conditioned on the symmetry of
the input local states: $\ket{\Psi^{(A)}_{\nu'}}\otimes \ket{\Psi^{(B)}_{\nu}}$.
In this section, we
extend our analysis to the cases where
outcomes of the measurements are determined by
the statistics
of photocounts registered by the photodetector
and decoherence-induced effects are taken into consideration. 

%%%%%%%%%%%%%%%%%
\subsection{Probability of photocounts}
\label{subsec:prob-photon}
%%%%%%%%%%%%%%%%%%

We begin with the decoherence effects for the states transmitted through the fiber
channels from Alice and Bob to the beam splitter (central) node. The decoherence processes will
affect the output states of the beam splitter, 
$\ket{\Phi_{\pm}^{(S)}(\bs{\gamma}_{\ind{bs}})}=\ket{\Phi_{\pm}^{(S)}(\gamma)}$,
where $S$ stands for the output channel of the beam splitter
($S$ is either $C$ or $D$),
that enter the right hand side of
Eq.~\eqref{eq:Psi_BS}.
For simplicity, we
restrict ourselves to the one-mode case
with $\ket{\Phi_{\pm}^{(S)}(\bs{\gamma}_{\ind{bs}})}=\ket{\Phi_{\pm}^{(S)}(\gamma)}$
and assume that
the noisy channels are identical and
\textcolor{black}{represented by the well-known pure-loss bosonic channel}
that can be described by introducing additional
environmental (ancillary) mode (see, e.g., Refs.~\cite{Enk:pra:2001,Ghasemi:laserph:2019}). 
An isometry representing the Stinespring dilation of the fiber channel
where the signal mode is supplemented with an extra system $E$
(the channel's environment) is given by the following mapping
\begin{align}
  &
  \label{eq:channel}
    \ket{\gamma}_S\to\ket{\gamma_s}_S\otimes\ket{\gamma_e}_E=
    \ket{\gamma_s,\gamma_e}\equiv\ket{\bs{\gamma}},
    \notag
  \\
  &
    |\gamma|^2=|\gamma_s|^2+|\gamma_e|^2,
\end{align}
where $|\gamma_s|^2=\eta |\gamma|^2$ and
$|\gamma_e|^2=(1-\eta) |\gamma|^2$;
\textcolor{black}{
$\eta=\exp(-L/L_{\ind{att}})$ is the fiber transmittance,
$L$ is the fiber length and $L_{\ind{att}}$ is
the attenuation length.
Note that,
the pure-loss channel provides a simplified model of
light propagation in a fiber, where a number of
polarization and dispersion dependent effects
are not taken in account.
}

One of the methods to discriminate between
the symmetric and antisymmetric cat states
is to perform photocounting measurements
that may distinguish the parity of the photon number registered by a photodetector.
For a photon-number resolving photodetector,
the probability to detect 
$k$ photons in the signal mode conditioned on the event that
the initial state
$\ket{\Psi_{\nu'}^{(A)}}\otimes\ket{\Psi_{\nu}^{(B)}}$
is projected onto
the entangled state $\ket{\Psi_{\mu}^{(AB)}}$  
\begin{align}
  \label{eq:Prob_k}
  \ind{Prob}(k|\mu,\nu'\nu)=P(k|\mu'),\quad
  \mu'=\nu'\nu\mu
\end{align}
that expressed in terms of
the probability of clicks
$P(k|\mu)$
determined by the statistics of photocounts for
the cat states $\ket{\Phi_\mu(\bs{\gamma})}$.
According to the well-known Kelley-Kleiner formula~\cite{Kelley:pr:1964,Vogel:bk:2006},
this probability reads
\begin{align}
  &
    \label{eq:P_k_mu}
    P(k|\mu)=\bra{\Phi_\mu(\bs{\gamma})}\hat{\Pi}_k\ket{\Phi_\mu(\bs{\gamma})},
    \quad P(k|0)=\delta_{k0},
\end{align}
where $\hat\Pi_k$ is the positive-operator-valued measure
for a number-resolving detector given by
\begin{align}
  &
  \label{eq:Pi_k}
    \hat{\Pi}_k=:\frac{(\xi\hat{n}_s)^k}{k!}\e^{-\xi\hat{n}_s}:,
    \quad
    \hat{n}_s=\hcnj{\hat{a}}_s \hat{a}_s,
\end{align}
where $\xi$ is the efficiency of the detector.
We can now apply the algebraic identity
\begin{align}
&\Tr_{E}\ket{\bs{\gamma}^{(\pm)}}\bra{\bs{\gamma}^{(\pm)}}=
  \frac{1}{4}
  \left\{
M_{\pm}(\gamma_e)\ket{\gamma_s^{(+)}}\bra{\gamma_s^{(+)}}\right. \notag \\  \label{eq:gamma-gamma_pm}&\left.+M_{\mp}(\gamma_e)\ket{\gamma_s^{(-)}}\bra{\gamma_s^{(-)}}
  \right\}
\end{align}
to deduce the following expressions of the probabilities~\eqref{eq:P_k_mu}
%\begin{widetext}
\begin{align}
&
  \label{eq:P-k-plus}
  P(k|+)=\frac{1}{4\tilde{M}_{+}(\bs{\gamma})}
  \bigl\{
                M_{+}(\gamma_e) C_{+}(k,\gamma_s)
                \notag
  \\
  &
                +
  M_{-}(\gamma_e) C_{-}(k,\gamma_s)
  \bigr\},\quad
  k>0,
  \\
  &
  \label{eq:P-k-minus}
  P(k|-)=\frac{1}{4{M}_{-}(\bs{\gamma})}
  \bigl\{
    M_{-}(\gamma_e) C_{+}(k,\gamma_s)
    \notag
  \\
  &
    +
  M_{+}(\gamma_e) C_{-}(k,\gamma_s)
    \bigr\},\quad P(k|0)=0,
\end{align}
%\end{widetext}
where
\begin{align}
     &
  \label{eq:C_nu_k}
  C_{\mu}(k,\gamma_s)=\bra{\gamma_s^{(\mu)}}\hat{\Pi}_k\ket{\gamma_s^{(\mu)}}=
       2\frac{(\xi|\gamma_s|^2)^k}{k!}\e^{-|\gamma_s|^2}
       \notag
  \\
     &
       \times
  \bigl\{
  \e^{(1-\xi)|\gamma_s|^2}
+(-1)^k \mu
  \e^{-(1-\xi)|\gamma_s|^2}
    \bigr\}.
\end{align}

In what follows, we consider the heralding outcomes
described by the two different parity of clicks, $p_c\in\{\ind{even},\ind{odd}\}$,
and thus discriminate between the cases,
where the number of registered photons is either odd or even.
The sum of the corresponding probabilities
\begin{align}
  \notag
  &P(\ind{even}|\pm,0)=
  \sum_{n=1}^{\infty}P(2n|\pm,0),
  \\
  \label{eq:P-even-odd-mu}
  &P(\ind{odd}|\pm,0)=
  \sum_{n=0}^{\infty}P(2n+1|\pm,0),
\end{align}
and the no-click probability, $P(0|\mu)$,
gives unity and
we have the completeness identity
\begin{align}
  \label{eq:P_pc-sum}
  P(0|\pm,0)+
  P(\ind{even}|\pm,0)+
  P(\ind{odd}|\pm,0)=1.
\end{align}

By substituting
Eqs.~\eqref{eq:P-k-plus}--\eqref{eq:C_nu_k}
into relations~\eqref{eq:P-even-odd-mu}
we obtain the expressions for the probabilities of
the detection outcomes determined by the parity of clicks
%\begin{widetext}
\begin{subequations}
    \label{eq:P-pc}
\begin{align}
&
  \label{eq:P-pc-plus}
  P(p_c|+)=\frac{1}{4\tilde{M}_{+}(\bs{\gamma})}
  \bigl\{
                M_{+}(\gamma_e) C_{+}(p_c,\gamma_s)
                \notag
  \\
  &
                +
  M_{-}(\gamma_e) C_{-}(p_c,\gamma_s)
  \bigr\},\quad
  p_c\in\{\ind{even},\ind{odd}\},
  \\
  &
  \label{eq:P-pc-minus}
  P(p_c|-)=\frac{1}{4{M}_{-}(\bs{\gamma})}
  \bigl\{
    M_{-}(\gamma_e) C_{+}(p_c,\gamma_s)
    \notag
  \\
  &
    +
  M_{+}(\gamma_e) C_{-}(p_c,\gamma_s)
  \bigr\},\quad P(p_c|0)=0,
\end{align}
\end{subequations}
%\end{widetext}
where
\begin{subequations}
    \label{eq:C_nu_all}
\begin{align}
  &
    \label{eq:C_nu_even}
    C_{\pm}(\ind{even},\gamma_s)=
    2 \e^{-|\gamma_s|^2} (\cosh(\xi |\gamma_s|^2)-1)
    \notag
  \\
  &
    \times
    \bigl\{
  \e^{(1-\xi)|\gamma_s|^2}
\pm
  \e^{-(1-\xi)|\gamma_s|^2}
    \bigr\},
  \\
   &
    \label{eq:C_nu_odd}
    C_{\pm}(\ind{odd},\gamma_s)=
     2 \e^{-|\gamma_s|^2} \sinh(\xi |\gamma_s|^2)
     \notag
  \\
  &
    \times
    \bigl\{
  \e^{(1-\xi)|\gamma_s|^2}
\mp
  \e^{-(1-\xi)|\gamma_s|^2}
    \bigr\}.
\end{align}
\end{subequations}

\begin{figure*}[!htp]
\centering
\begin{subfigure}{.5\textwidth}
  \centering
  \includegraphics[width=.8\linewidth]{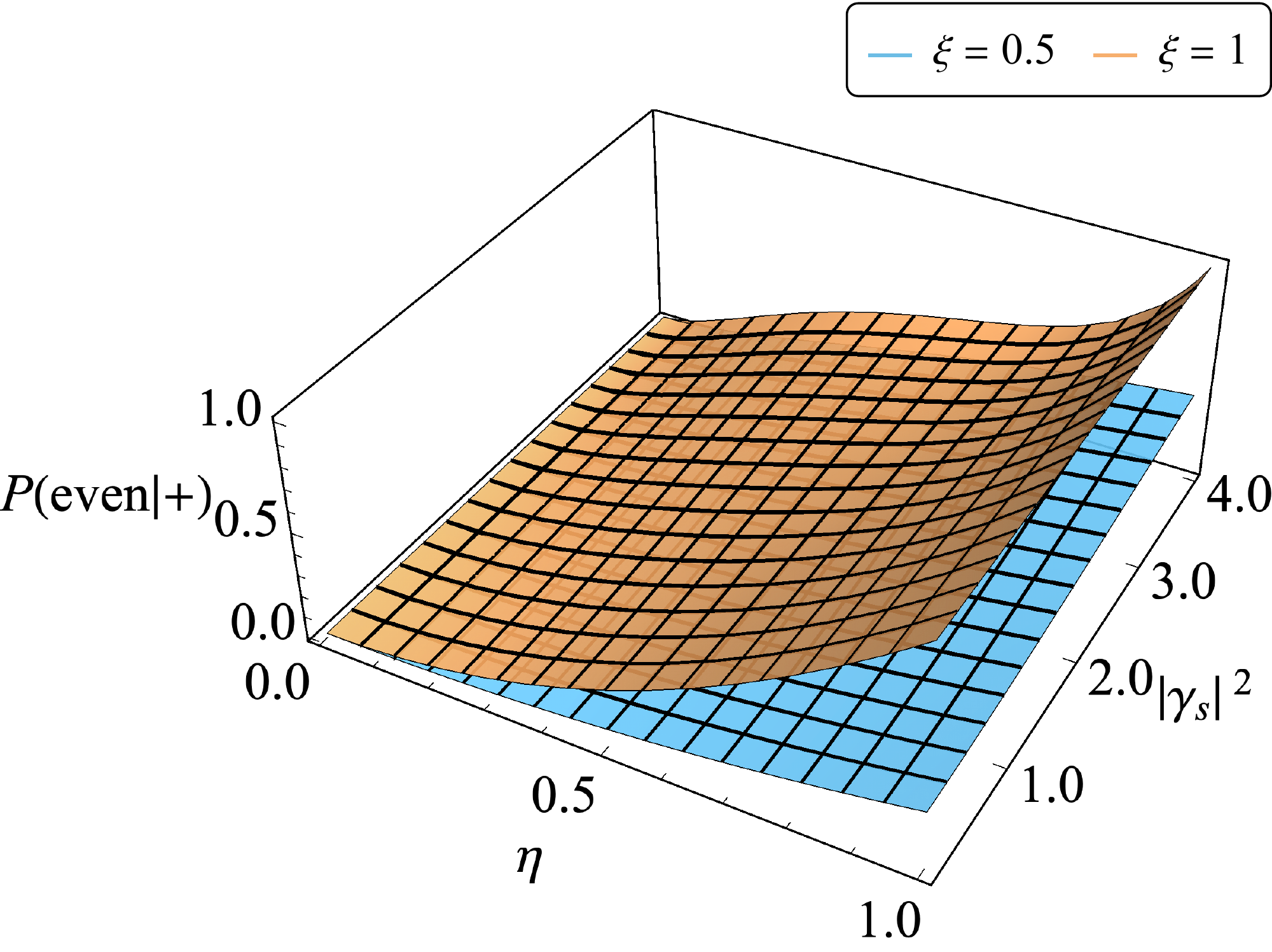}
  \caption{}
  \label{p-even-plus}
\end{subfigure}%
\begin{subfigure}{.5\textwidth}
  \centering
  \includegraphics[width=.8\linewidth]{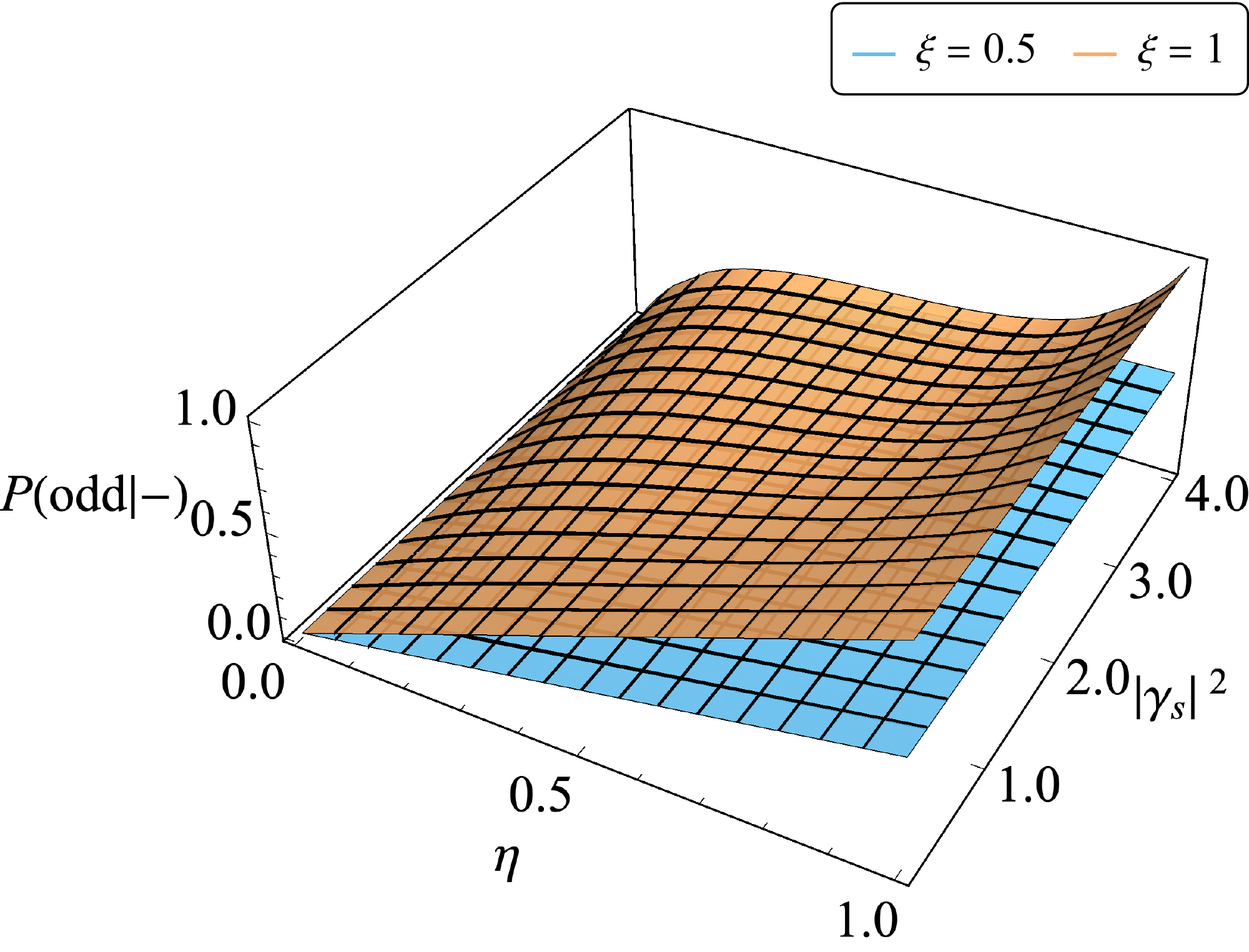}
  \caption{}
  \label{p-odd-minus}
\end{subfigure}
\caption{Conditional probabilities (a)~$P(\ind{even}|+)$
  and (b)~$P(\ind{odd}|-)$
  plotted against the mean photon
  number $|\gamma_s|^2$ and the channel transmittance $\eta$
  at different values of the detection efficiency $\xi$.  
} 
\label{p-cond}
\end{figure*}

It can be checked that
these relations meet the algebraic identity
\begin{align}
   &
    \label{eq:symmetry-fid-1}
      \frac{P(\ind{even}|-)P_{-}^{(\pm)}}{P(\ind{even}|+)P_{+}^{(\pm)}}=
    \frac{P(\ind{odd}|+)P_{-}^{(c)}}{P(\ind{odd}|-)P_{+}^{(c)}}.
\end{align}
They can also be readily generalized to the case of multimode signal,
where $\ket{\gamma_s}\to \ket{\bs{\gamma}_s}\equiv\ket{\gamma_1^{(s)},\ldots,\gamma_{N_s}^{(s)}}$
and $N_s$ is the number of signal modes registered by a broadband detector,
by replacing $|\gamma_s|^2$ and $\xi|\gamma_s|^2$ with
$|\bs{\gamma}_s|^2=\sum_{i=1}^{N_s}|\gamma_i^{(s)}|^2$
and $\sum_{i=1}^{N_s}\xi_i|\gamma_i^{(s)}|^2$.

Referring to Fig.~\ref{p-cond},
it is seen that,
for the limiting case of the ideal photodetector
and the lossless channel with the efficiency and the transmittance
both equal to unity, $\xi=\eta=1$, the probability of odd (even) number of clicks
vanishes for the symmetric (antisymmetric) states,
$P(\ind{odd}|+)=0$ ($P(\ind{even}|-)=0$),
and thus the cat states are perfectly distinguishable
with $P(\ind{even}|+)=1$ ($P(\ind{odd}|-)=1$).
It can also be noted that
low detector efficiency $\xi$
has a noticeable detrimental effect on
the conditional probabilities.

%%%%%%%%%%%%%%%%%
\subsection{Performance and entanglement}
\label{subsec:performance}
%%%%%%%%%%%%%%%%%%

We can now evaluate the probability of success
for the heralding outcomes labeled by the parity $p_c$.
Similar to all other probabilities,
this probability is conditioned on
the symmetry of the initial states
and can be written in the following general form:
\begin{align}
  &
  P_{s}^{(\nu'\nu)}(p_c)=
 \sum_{\mu=\pm}\ind{Prob}(p_c|\mu,\nu'\nu)\ind{Prob}(\mu|\nu'\nu)\notag \\ \label{eq:P-success}
 &=\sum_{\mu=\pm}P(p_c|\mu')P^{(\nu'\nu)}_{\mu}.
\end{align}
After substituting formulas
given by Eqs.~\eqref{eq:Pmuc}--\eqref{eq:Pmum} and~\eqref{eq:P-pc}
into Eq.~\eqref{eq:P-success} 
and performing cumbersome but rather straightforward algebra,
it can be shown that
$P_{\pm}^{(\nu'\nu)}$ with the photon number ratio
$r_{\ind{bs}}$ replaced by the product
$\zeta=\xi \eta r_{\mathrm{bs}}$
give explicit expressions for the probabilities of success $P_{s}^{(\nu'\nu)}$
as follows
\begin{subequations}
  \label{eq:Ps-pc}
\begin{align}
  &
    \label{eq:Ps-p-vs-m}
  P_{s}^{(+)}(p_{c})=\tanh ^{2}(|\alpha|^2) P_{s}^{(-)}(p_{c}),
  \\
  &
  \label{eq:Ps-odd}
    P_{s}^{(-)}(\mathrm {odd })=P_{-}^{(-)}(\zeta,|\alpha|^2),
    \: P_{s}^{(c)}(\mathrm { odd })=P_{+}^{(c)}(\zeta,|\alpha|^2),
  \\
  &
    \label{eq:Ps-even}
  P_{s}^{(-)}(\mathrm {even })=P_{+}^{(-)}(\zeta,|\alpha|^2),\:
    P_{s}^{(c)}(\mathrm { even })=P_{-}^{(c)}(\zeta,|\alpha|^2),
\end{align}
\end{subequations}
where $P_{\mp}^{(-)}$ and $P_{\mp}^{(c)}$
are given by Eqs.~\eqref{eq:Pmm}--\eqref{eq:Ppm}
and Eqs.~\eqref{eq:Pmc}--\eqref{eq:Ppc},
respectively.

\begin{figure*}[!htp]
\centering
\begin{subfigure}{.5\textwidth}
  \centering
  \includegraphics[width=.8\linewidth]{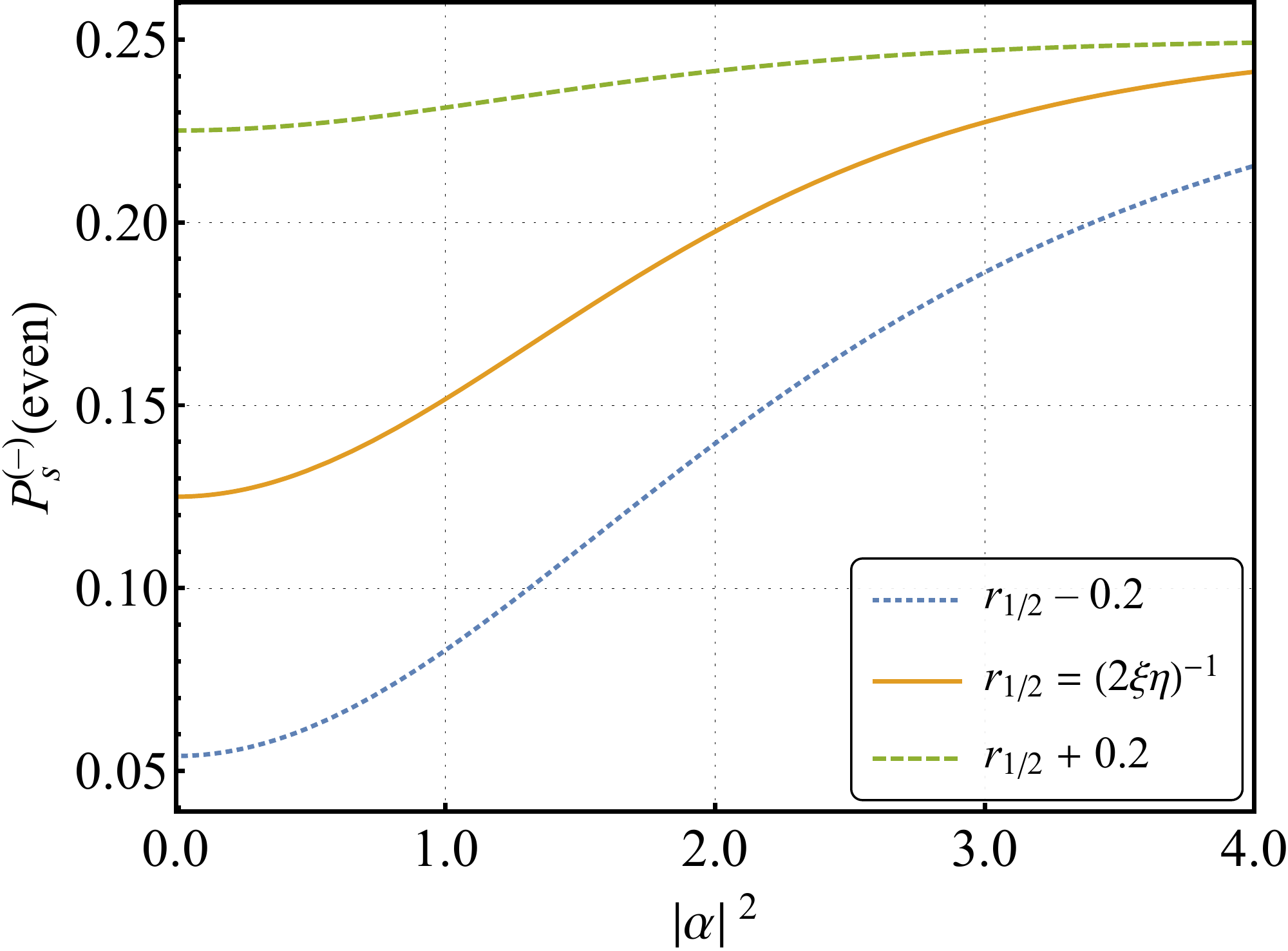}
  \label{subfig:ps-even}
  \caption{}
\end{subfigure}%
\begin{subfigure}{.5\textwidth}
  \centering
  \includegraphics[width=.8\linewidth]{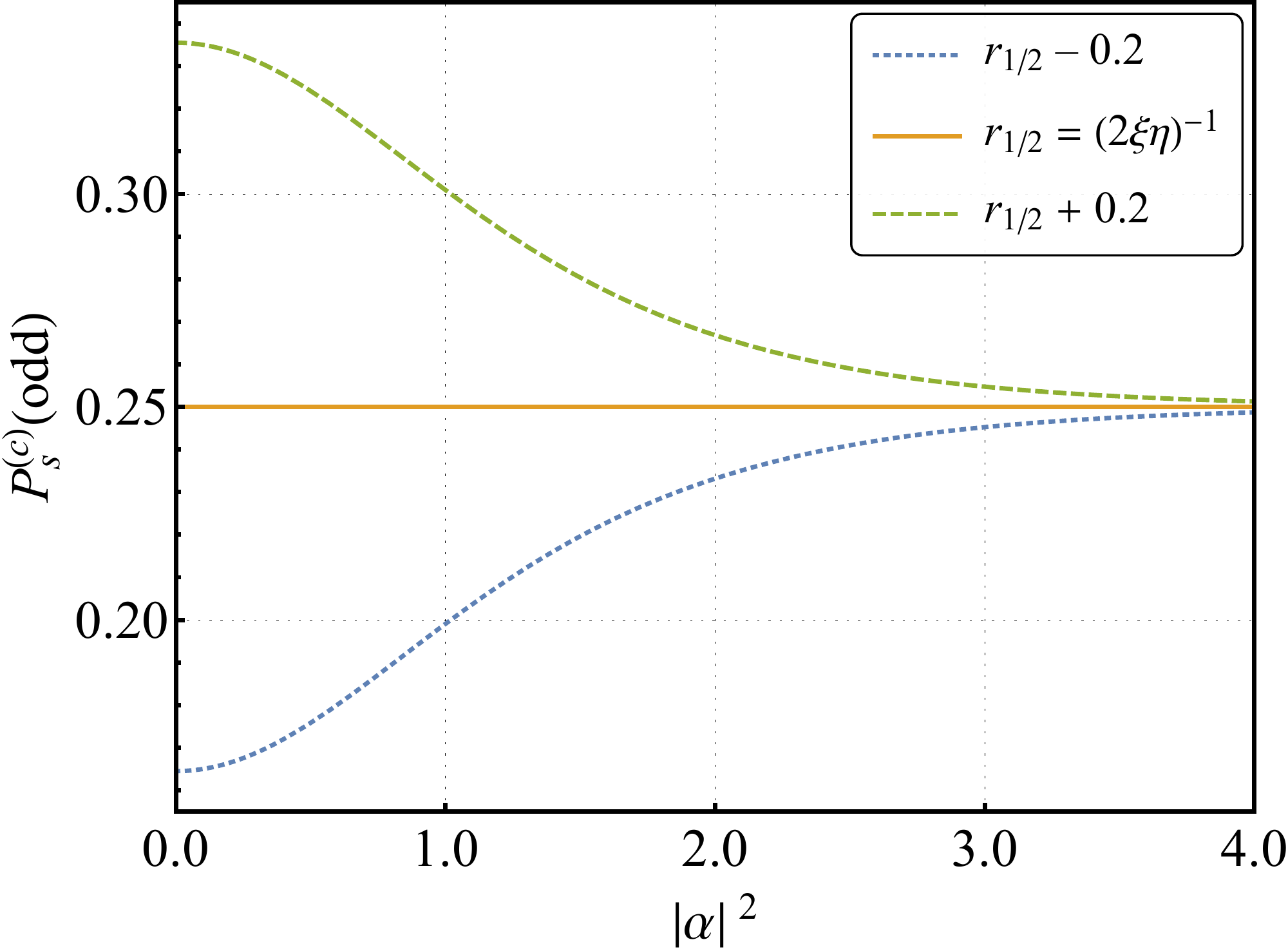}
  \label{subfig:ps-c-odd}
  \caption{}
\end{subfigure}
\\
\begin{subfigure}{.5\textwidth}
  \centering
  \includegraphics[width=.8\linewidth]{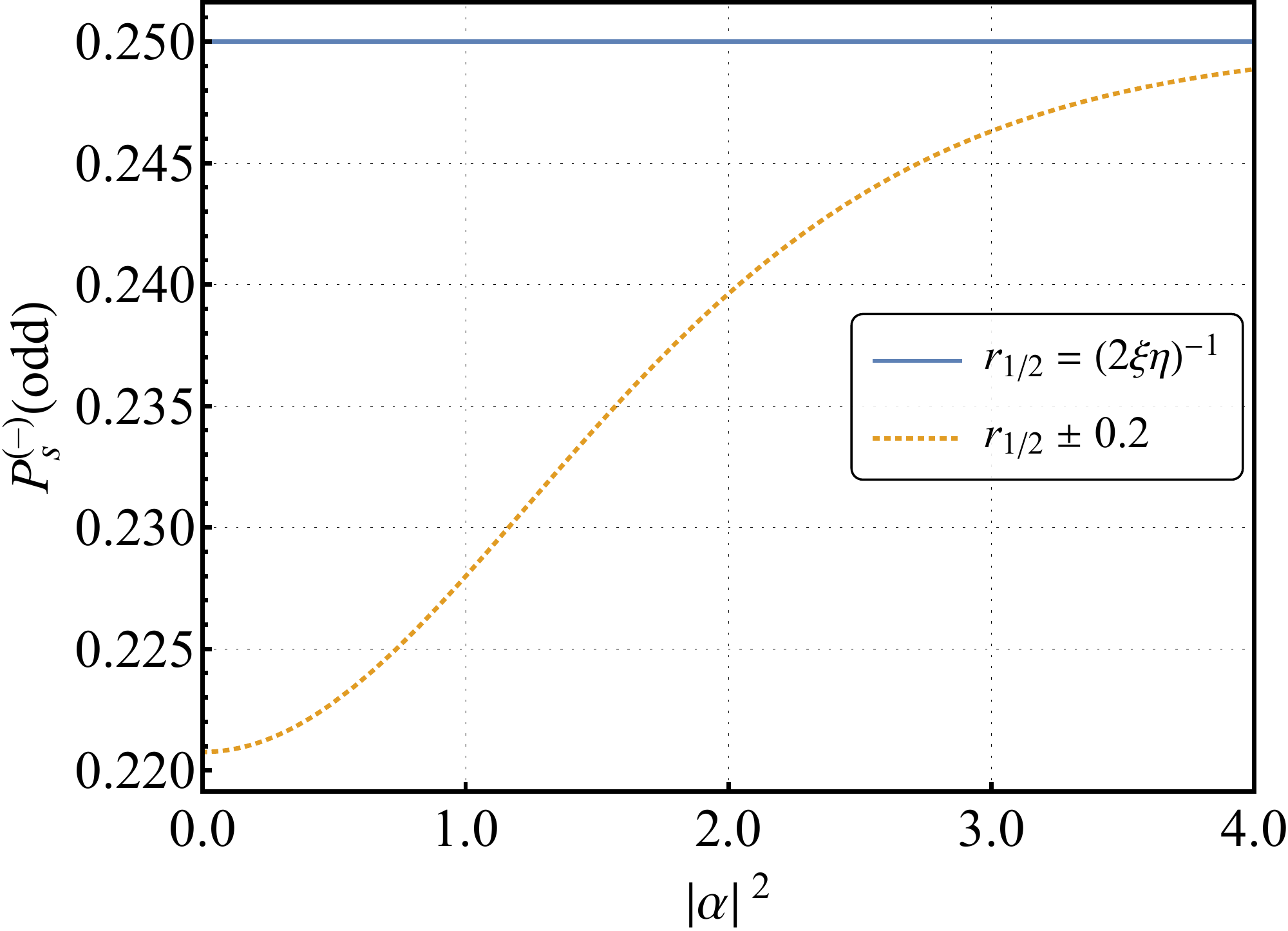}
  \label{subfig:ps-odd}
  \caption{}
\end{subfigure}%
\begin{subfigure}{.5\textwidth}
  \centering
  \includegraphics[width=.8\linewidth]{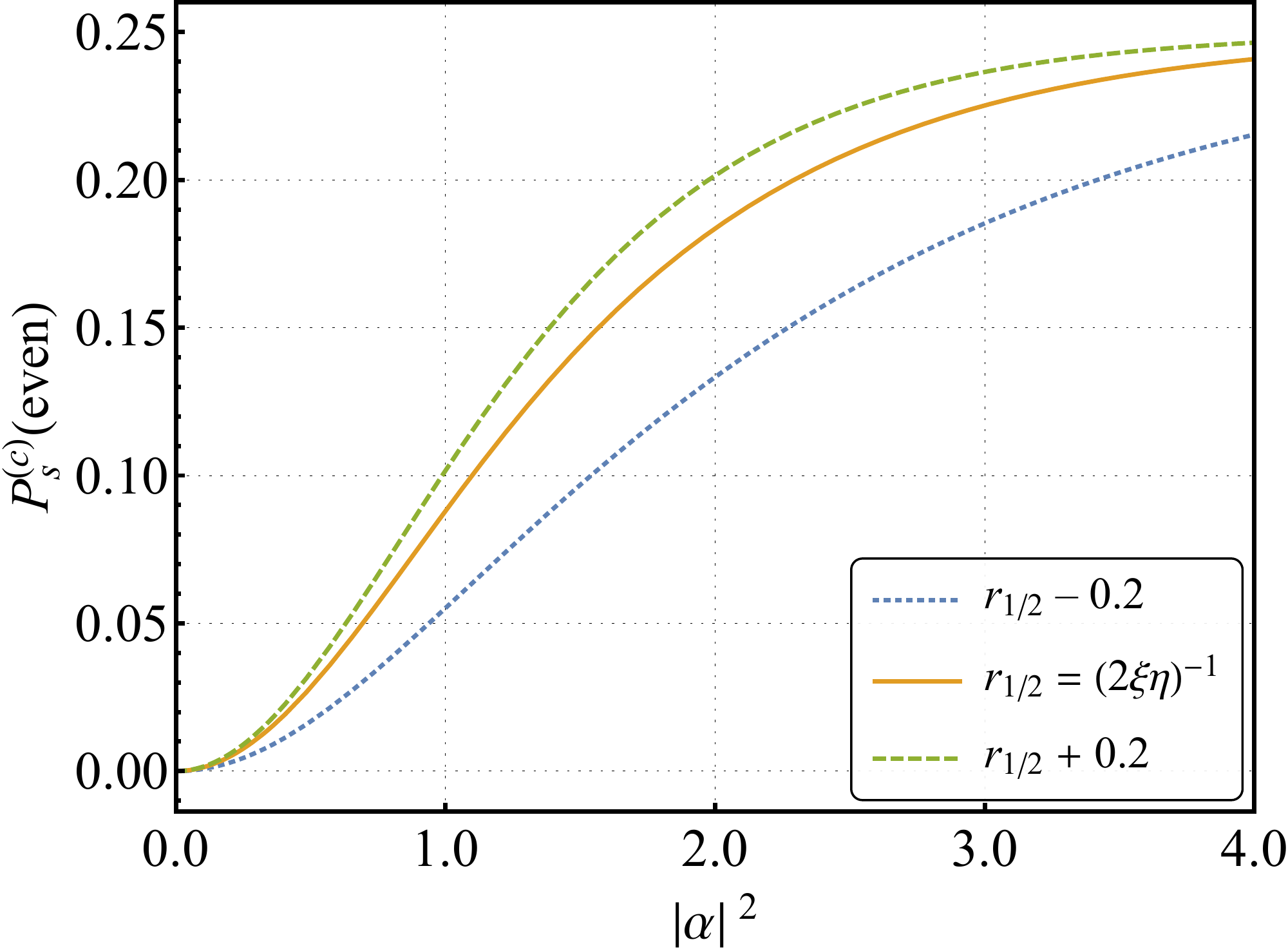}
  \label{subfig:ps-c-even}
  \caption{}
\end{subfigure}
\caption{The success probabilities, $P_{s}^{(-)}(p_{c})$
  and $P_{s}^{(c)}(p_{c})$,
(the parity of photocounts is labeled by $p_c\in\{\ind{even,odd}\}$),
  of outcomes that herald generation of predominantly (a)-(b)~symmetric and (c)-(d)~antisymmetric
   cat states with $F_{+}^{(\nu'\nu)}>1/2$ and $F_{-}^{(\nu'\nu)}>1/2$, respectively.
   The curves are computed at 
$r_{\mathrm{bs}}\in\{r_{1/2},r_{1/2}\pm 0.2\}$ 
with $\eta=0.95$ and $\xi=0.9$,
where $r_{1/2}\equiv(2\xi \eta)^{-1}$ ($\zeta=1/2$).
} 
\label{ps-prob}
\end{figure*}

\begin{figure*}[!htb]
  \centering
\begin{subfigure}{.5\textwidth}
  \centering
  \includegraphics[width=.8\linewidth]{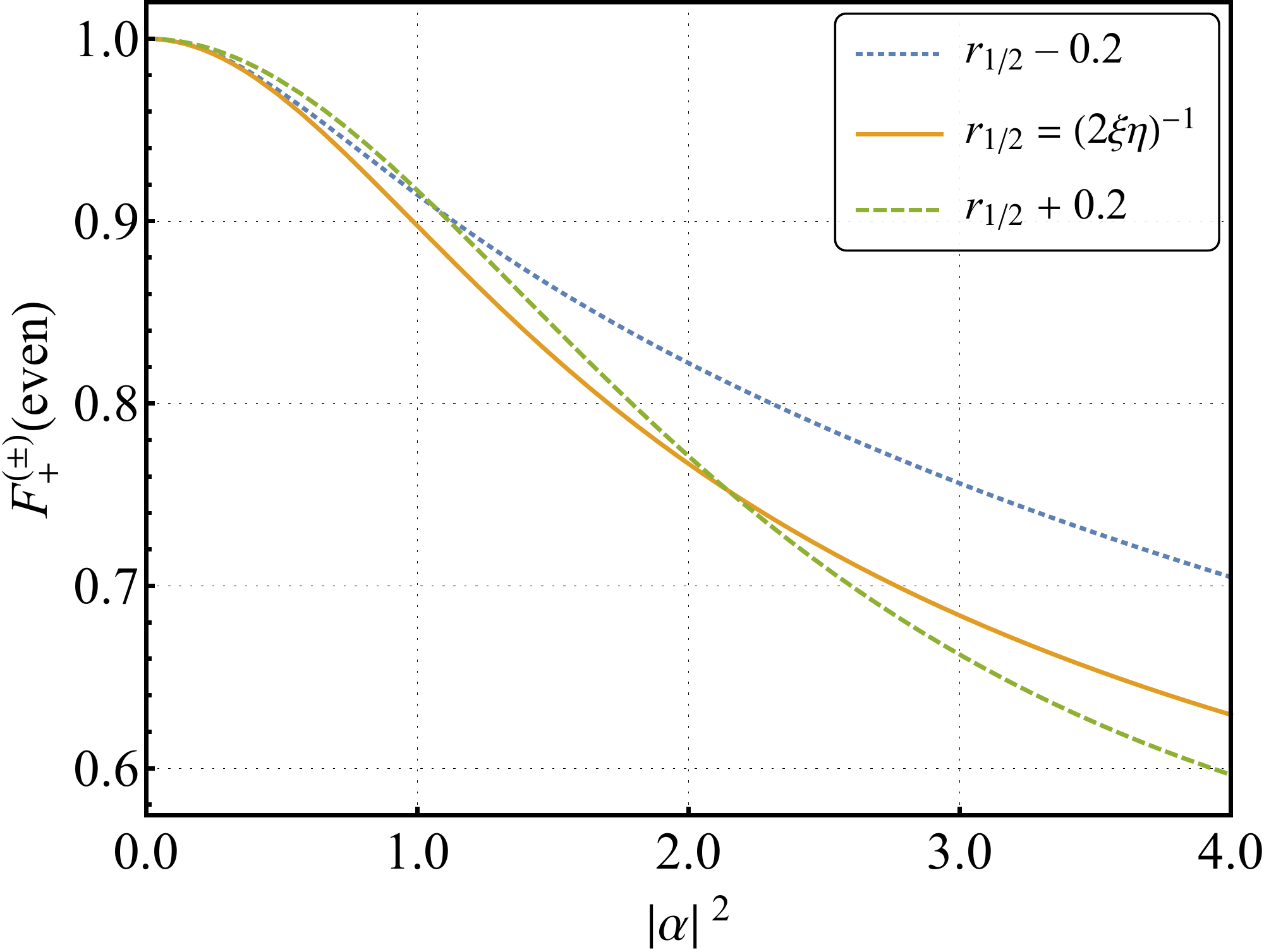}
  \label{subfig:fid-even}
  \caption{}
\end{subfigure}%
\begin{subfigure}{.5\textwidth}
  \centering
  \includegraphics[width=.8\linewidth]{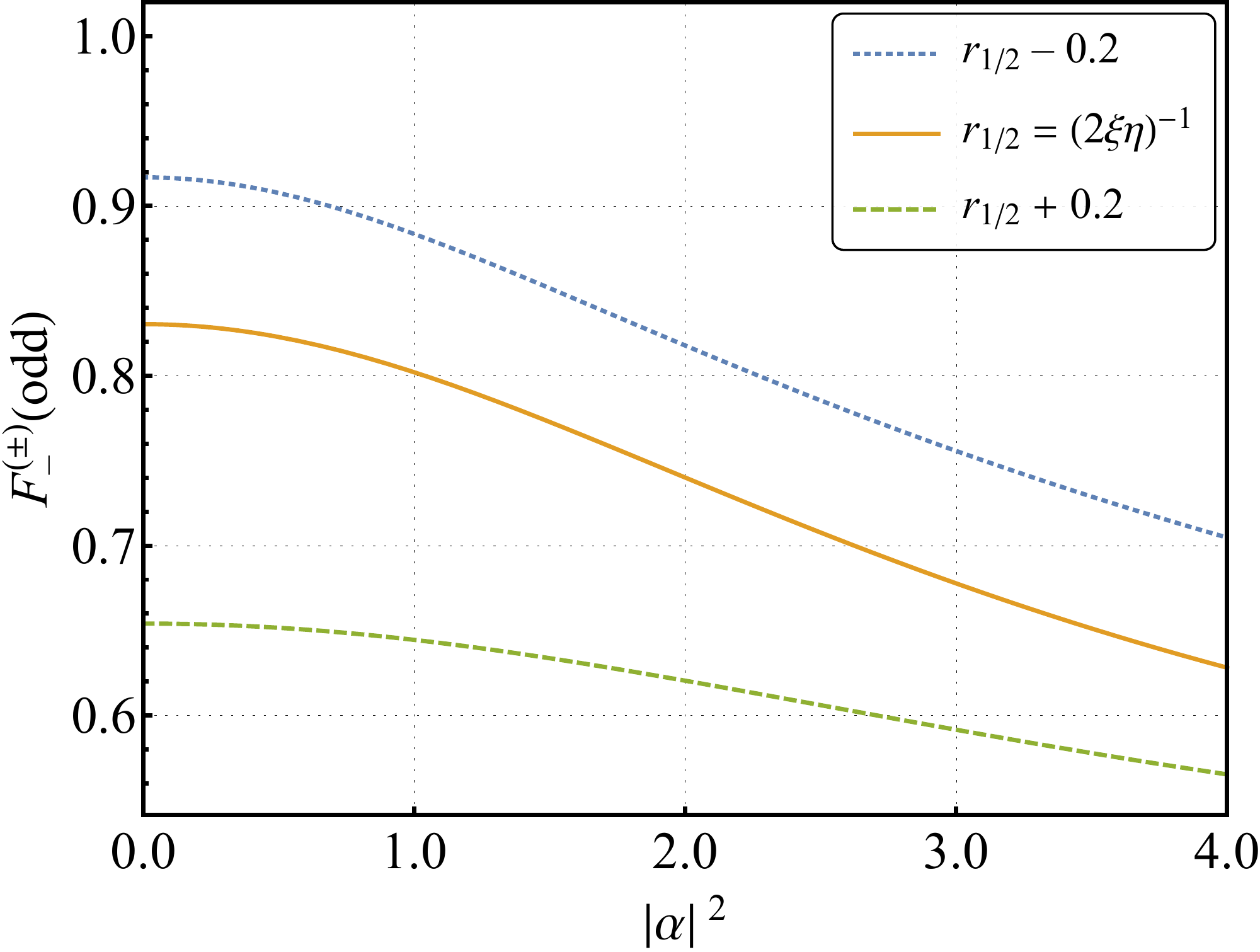}
  \label{subfig:fid-odd}
  \caption{}
\end{subfigure}
\caption{Fidelities of heralded symmetric and antisymmetric cat states:
  (a)~$F_{+}^{(\pm)}(\ind{even})$ and (b)~$F_{-}^{(\pm)}(\ind{odd})$
  as a function of $|\alpha|^2$ computed at
  different values of $r_{\ind{bs}}$
  (the parameters are listed in the caption of Fig.~\ref{ps-prob}).
} 
\label{fid-prob}
\end{figure*}

These relations imply that
the results presented at the end of Sec.~\ref{sec:fidel-prob-phot}
is directly applicable to the success probabilities.
As is shown in Fig.~\ref{ps-prob},
in the limit of vanishing $\alpha$,
the largest probabilities of success
$P_s^{(-)}(\ind{odd})$ and $P_s^{(c)}(\ind{odd})$
take the values $\zeta(1-\zeta)$ and $\zeta/2$,
respectively.
As the amplitude $|\alpha|$ increases,
the probabilities vary approaching the limiting value equal to $1/4$.

Given the heralding event $p_c$,
the heralded state is represented by
the density matrix 
% \begin{widetext}
\begin{align}
  &
    \label{eq:rho-pc}
    \hat{\rho}_{AB}^{(\nu'\nu)}(p_c)
    =\sum_{\mu=\pm} F_{\mu}^{(\nu'\nu)}(p_c)
    \notag
  \\
  &
    \times
    \ket{\Psi_{\mu}^{(AB)}(\bs{\alpha}_{\ind{qm}},\bs{\alpha}_{\ind{qm}})}\bra{\Psi_{\mu}^{(AB)}(\bs{\alpha}_{\ind{qm}},\bs{\alpha}_{\ind{qm}})},
\end{align}
% \end{widetext}
where
$F_{\mu}^{(\nu'\nu)}=\bra{\Psi_{\mu}^{(AB)}}\hat{\rho}_{AB}^{(\nu'\nu)} \ket{\Psi_{\mu}^{(AB)}}$
are
the fidelities that
can be computed using the well-known Bayesian formula
%\begin{widetext}
\begin{align}
  \label{eq:F-mu}
  &
    F_{\mu}^{(\nu'\nu)}(p_c)= \ind{Prob}(\mu|p_c,\nu'\nu)
  %   \notag
  % \\
  % &
    =
    \frac{\ind{Prob}(p_c|\mu,\nu'\nu)\ind{Prob}(\mu|\nu'\nu)}{P_{s}^{(\nu'\nu)}(p_c)}
    \notag
  \\
  &
    =
 \frac{P(p_c|\mu')P_{\mu}^{(\nu'\nu)}}{P_{s}^{(\nu'\nu)}(p_c)}
\end{align}
%\end{widetext}
and meet the symmetry condition
\begin{align}
  &
    \label{eq:symmetry-fid-2}
  F_{\mu}^{(\pm)}(\ind{even}/\ind{odd})=F_{\mu}^{(c)}(\ind{odd}/\ind{even})
\end{align}
which is a direct consequence of the identity~\eqref{eq:symmetry-fid-1}.
Similar to Eqs.~\eqref{eq:Ps-p-vs-m}--~\eqref{eq:Ps-even},
after tedious algebra, we derive the following expressions
for the fidelities:
\begin{subequations}
  \label{eq:F-mp}
\begin{align}
  &
    F_{-}^{(\pm)}(\ind{odd}) =F_{-}^{(c)}(\ind{even})\notag \\\label{eq:F-m-odd}&=
    \frac{\tanh(2(1-r_{\mathrm{bs}})|\alpha|^2)}{
    \tanh(2(1-r_{\mathrm{bs}})|\alpha|^2)+\tanh(2(r_{\mathrm{bs}}-\zeta)|\alpha|^2)},
      \\
  &
    F_{+}^{(\pm)}(\ind{even})=F_{+}^{(c)}(\ind{odd})\notag \\ \label{eq:F-p-even} &=
    \frac{1}{1+\tanh(2(1-r_{\mathrm{bs}})|\alpha|^2)\tanh(2(r_{\mathrm{bs}}-\zeta)|\alpha|^2)}.
\end{align}
\end{subequations}
Note that, owing to the completeness condition $F_{+}^{(\nu'\nu)}(p_c)+F_{-}^{(\nu'\nu)}(p_c)=1$
and the symmetry relations~\eqref{eq:symmetry-fid-2},
we need to specify only two fidelities.
These fidelities $F_{-}^{(\pm)}(\ind{odd})$ and $F_{+}^{(\pm)}(\ind{even})$
are both above 1/2 and
represent the predominantly antisymmetric and symmetric cat states, respectively.
From $|\alpha|^2$-dependencies of $F_{-}^{(\pm)}(\ind{odd})$ and $F_{+}^{(\pm)}(\ind{even})$ 
depicted Fig.~\ref{fid-prob},
both the fidelities start from their zero-amplitude values
\begin{align}
  &
    \label{eq:F-zer-alp}
      F_{+}^{(\pm)}(\ind{even},\alpha=0)= 1,\quad
    F_{-}^{(\pm)}(\ind{odd},\alpha=0)= \frac{1-r_{\mathrm{bs}}}{1-\zeta}.
\end{align}
and decay approaching $1/2$ with the amplitude $|\alpha|$.

At $r_{\ind{bs}}=r_{1/2}\equiv(2\eta\xi)^{-1}$, 
we have $\zeta=1/2$, so that
$2P_s^{(c)}(\ind{odd})$ and $2P_s^{(-)}(\ind{odd})$
equal $1/2$
(see Fig.~\ref{ps-prob})
as the probabilities of the most probable outcomes
heralding generation of
the symmetric and antisymmetric cat states with
the fidelities
$F_{+}^{(\pm)}(\ind{even})$ and $F_{-}^{(\pm)}(\ind{odd})$, respectively.
In this case,
formula~\eqref{eq:F-zer-alp} gives unity and $1/(2-\eta\xi)$ as
the maximum values of  $F_{+}^{(\pm)}(\ind{even})$ and $F_{-}^{(\pm)}(\ind{odd})$.

Interestingly,
$F_{+}^{(\pm)}(\ind{even},\alpha=0)$
does not depend on the parameters,
whereas the maximum value of
$2P_s^{(c)}(\ind{odd})$ is $\zeta$
provided the photon number ratio $r_{\ind{bs}}$,
is above $(2\eta\xi)^{-1}$.
By contrast,
when $r_{\ind{bs}}\ne (2\eta\xi)^{-1}$,
the probability $2P_s^{(-)}(\ind{odd})$
(it equals $(1-(2\zeta-1)^2)/2$ at $\alpha=0$)
cannot exceed one half
and the fidelity $F_{-}^{(\pm)}(\ind{odd},\alpha=0)$
grows as the ratio $r_{\ind{bs}}$ becomes smaller than
$(2\eta\xi)^{-1}$.

%%%%%%%%%%%%%%%%%%%%%%%
\section{Entanglement swapping}
\label{sec:entangl-swapping}
\begin{figure*}[!htb]
  \centering
     \includegraphics[width=0.7\textwidth]{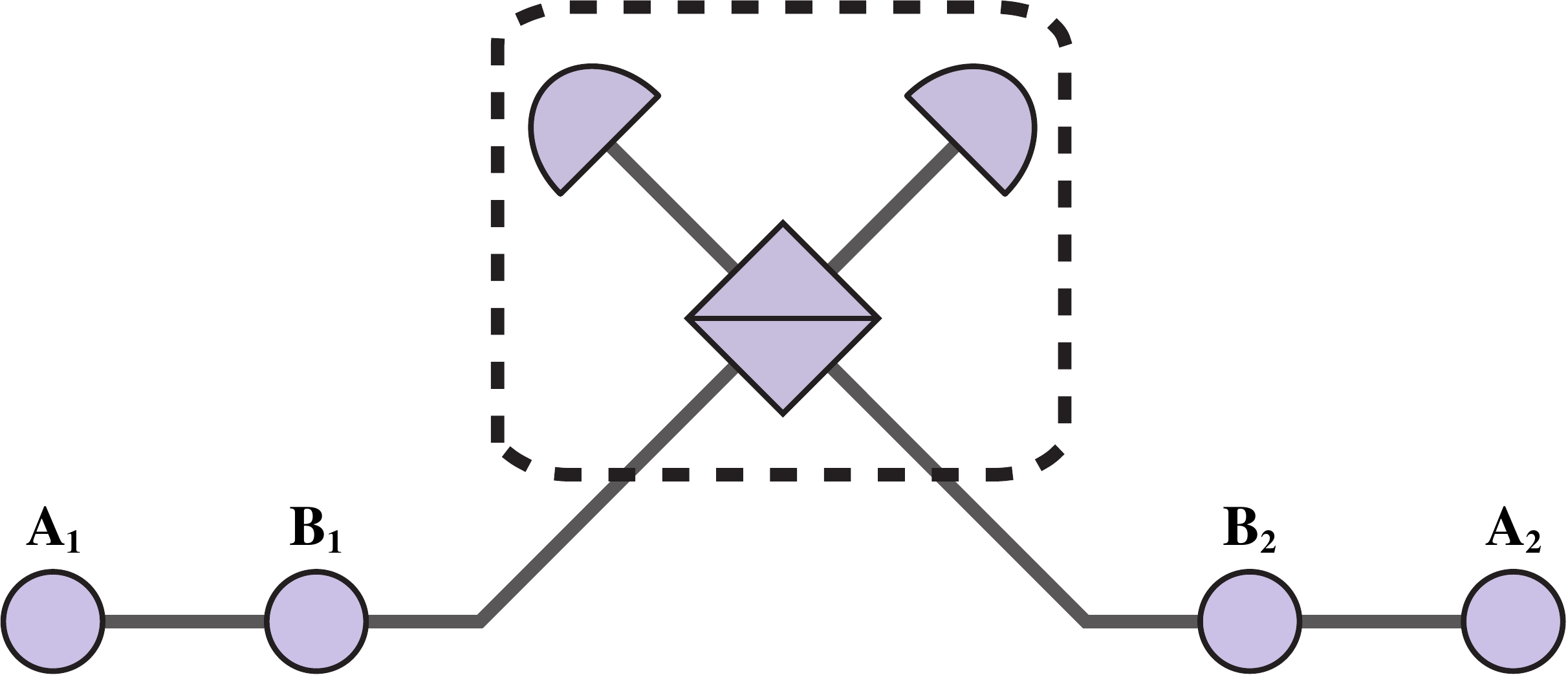}
\caption{Entanglement swapping scheme. $A_1-B_1$ and $B_2-A_2$ are adjacent elementary links.}
\label{scheme-1}
\end{figure*}

%%%%%%%%%%%%%%%%%%%%%%%%
In this section, we shall discuss the entanglement swapping protocol
that uses the heralded entanglement generation procedure
studied in the previous section
to create entanglement between remote
points of two elementary links, $A_1 - B_1$ and $B_2 - A_2$,
depicted in Fig.~\ref{scheme-1}.
\textcolor{black}{
In our subsequent analysis,
we consider the structure
where the neighboring end-nodes,
$B_1$ and $B_2$,
of the adjacent links,
$A_1-B_1$ and $B_2-A_2$,
are located at the repeater station
(the repeater node $B_1B_2$).
}

According to Eq.~\eqref{eq:rho-pc}),
the phase-modulated entangled states shared between the nodes of the links
are described by the density matrices
of the form:
%\begin{widetext}
\begin{subequations}
    \label{eq:rho1-rho2}
\begin{align}
  &
  \label{eq:rho1}
\hat{\rho}_{A_1B_1}^{(\nu'_1\nu_1)}(p_c)\equiv
  \hat{\rho}_1=
    \sum_{\mu_1=\pm}F_{\mu_1}^{(1)}
    \notag
  \\
  &
    \times
    \ket{\Psi_{\mu_1}^{(A_1B_1)}(\bs{\alpha}_{\ind{qm}},\bs{\alpha}_{\ind{qm}})}
    \bra{\Psi_{\mu_1}^{(A_1B_1)}(\bs{\alpha}_{\ind{qm}},\bs{\alpha}_{\ind{qm}})},
  \\
  &
      \label{eq:rho2}
    \hat{\rho}_{A_2B_2}^{(\nu'_2\nu_2)}(p_c)\equiv
  \hat{\rho}_2=
    \sum_{\mu_2=\pm}F_{\mu_2}^{(2)}
    \notag
  \\
  &
    \times
    \ket{\Psi_{\mu_2}^{(A_2B_2)}(\bs{\alpha}_{\ind{qm}},\bs{\alpha}_{\ind{qm}})}
    \bra{\Psi_{\mu_1}^{(A_2B_2)}(\bs{\alpha}_{\ind{qm}},\bs{\alpha}_{\ind{qm}})},
\end{align}
\end{subequations}
%\end{widetext}
where $F_{\mu_i}^{(i)}\equiv F_{\mu_i}^{(\nu'_i\nu_i)}$,
and
$  \hat{\rho}_1\otimes\hat{\rho}_2$ is the initial state of the two-link system.
So, the initial state
is the statistical ensemble
of the quantum states
\begin{align}
  &
    \label{eq:Psi_mu1mu2}
    \ket{\Psi_{\mu_1\mu_2}^{(A_1B_1,A_2B_2)}}
    \notag
  \\
    &
    \equiv
  \ket{\Psi_{\mu_1}^{(A_1B_1)}(\bs{\alpha}_{\ind{qm}},\bs{\alpha}_{\ind{qm}})}\otimes
  \ket{\Psi_{\mu_2}^{(A_2B_2)}(\bs{\alpha}_{\ind{qm}},\bs{\alpha}_{\ind{qm}})}
\end{align}
with the probabilities $F_{\mu_1}^{(1)}F_{\mu_2}^{(2)}$. 

When the modes stored in the quantum memories of the sites
$B_1$ and $B_2$ are released and sent to the beam splitter,
the transformed state for
each member of the ensemble is given
by general formula~\eqref{eq:Psi_BS}.
Similar to Sec.~\ref{subsec:prob-photon}, we assume that 
the heralding events of the photocounting measurements
are determined by the parity of photodetector clicks,
so that we may closely follow the line of reasoning presented in
Sec.~\ref{subsec:prob-photon} and~\ref{subsec:performance}
to deduce the heralded state for each
state $\ket{\Psi_{\mu_1\mu_2}^{(A_1B_1,A_2B_2)}}$
in the modified form of Eq.~\eqref{eq:rho-pc}
with $\ket{\Psi_{\mu}^{(AB)}}$ and $(\nu'\nu)$
replaced with $\ket{\Psi_{\mu}^{(A_1A_2)}}$ and
$(\mu_1\mu_2)$, respectively.

\textcolor{black}{
As for elementary links,
the fidelities,
$\tilde{F}^{(\mu_1\mu_2)}_{\mu}(p_c)$, and
and the probabilities,
$\tilde{P}^{(\mu_1\mu_2)}_{s}(p_c)$,
at  the repeater node, can be evaluated using
Eq.~\eqref{eq:F-mp}
and Eq.~\eqref{eq:Ps-pc}, respectively.
Since, at this node,
the modes are evenly partitioned
with $\bs{\alpha}_{\ind{bs}}=\bs{\alpha}_{\ind{qm}}$
and
$|\bs{\alpha}_{\ind{bs}}|^2=|\bs{\alpha}_{\ind{qm}}|^2=(1-r_{\ind{bs}})|\alpha|^2$,
the parameters $\{r_{\ind{bs}},|\alpha|^2\}$
should be accordingly changed as follows:
$\{r_{\ind{bs}},|\alpha|^2\}\to\{1/2,2(1-r_{\ind{bs}})|\alpha|^2\}$.
}

\textcolor{black}{
For the measurements performed locally
at the repeater node,
the fiber transmission losses are negligible and
the bulk part of losses can be attributed to
inefficiency of the quantum memories.
In order to take into account such losses,
we, following a widely used phenomenological
approach (see, e.g., Refs~\cite{Wu:pra:2020,Azuma:avs:2021,Semenenko:avs:2022}),
shall introduce the memory efficiency,
$\eta_{\ind{m}}$,
%which is given by the product of the storage and retrieval efficiencies.
as the parameter that replaces
the transmission coefficient $\eta$.
The latter implies that,
similar to the effects of the fiber transmission losses and inefficiency of the photodetectors,
decoherence effects in quantum memories
are modelled using a pure-loss bosonic channel
(see Eq.~\eqref{eq:channel} in Sec.~\ref{subsec:prob-photon}). 
As a result, at the repeater node,
the parameter $\zeta$ appears to be equal to the product
$\eta_{\ind{m}}\xi/2$:
$\zeta\to \zeta_{\ind{m}}=\eta_{\ind{m}}\xi/2$.
}

The final result for the entanglement swapping transformation
reads
\begin{align}
  &
\label{eq:rho_12}
    \hat{\rho}_1\otimes\hat{\rho}_2 \to
    \hat{\rho}_{12}(p_c)=\sum_{\mu=\pm}F_{\mu}^{(12)}(p_c)
    \notag
  \\
  &
    \times
    \ket{\Psi_{\mu}^{(A_1A_2)}(\bs{\alpha}_{\ind{qm}},\bs{\alpha}_{\ind{qm}})}
    \bra{\Psi_{\mu}^{(A_1A_2)}(\bs{\alpha}_{\ind{qm}},\bs{\alpha}_{\ind{qm}})},
\end{align}
where $F_{\mu}^{(12)}(p_c)$ is the fidelity given by
\begin{align}
  &
  \label{eq:F_12}
  F_{\mu}^{(12)}(p_c)=
  \frac{1}{P_{s}^{(12)}(p_c)}\sum_{\mu_1,\mu_2=\pm}P(p_c|\mu_1\mu_2\mu)P^{(\mu_1\mu_2)}_{\mu}
  F_{\mu_1}^{(1)}F_{\mu_2}^{(2)}
  \notag
  \\
  &
  =
    \frac{1}{P_{s}^{(12)}(p_c)}\sum_{\mu_1,\mu_2=\pm}
    \tilde{F}^{(\mu_1\mu_2)}_{\mu}(p_c)\tilde{P}^{(\mu_1\mu_2)}_{s}(p_c)
  F_{\mu_1}^{(1)}F_{\mu_2}^{(2)}.
\end{align}

It is rather straightforward to deduce
the expression for
the probabilities of success 
\begin{align}
  &
     \label{eq:Ps-12}
    P_{s}^{(12)}(p_c)
    \notag
  \\
   &=
  \sum_{\mu,\mu_1,\mu_2=\pm}\ind{Prob}(p_c|\mu,\mu_1\mu_2)\ind{Prob}(\mu|\mu_1\mu_2)\ind{Prob}(\mu_1,\mu_2)
  \notag
  \\
  &
    =\sum_{\mu,\mu_1,\mu_2=\pm}P(p_c|\mu_1\mu_2\mu)P^{(\mu_1\mu_2)}_{\mu}F_{\mu_1}^{(1)}F_{\mu_2}^{(2)}
    \notag
  \\
  &
    =\sum_{\mu_1,\mu_2=\pm}\tilde{P}^{(\mu_1\mu_2)}_{s}(p_c)F_{\mu_1}^{(1)}F_{\mu_2}^{(2)},
\end{align}
where $\tilde{F}^{(\mu_1\mu_2)}_{\mu}(p_c)$
($\tilde{P}^{(\mu_1\mu_2)}_{s}(p_c)$)
is the above-discussed fidelity (success probability)
given by Eq.~\eqref{eq:F-mp}
(Eq.~\eqref{eq:Ps-pc})
with the parameters
$\{r_{\ind{bs}},\zeta,|\alpha|^2\}$
changed to
$\{1/2,\eta_{\ind{m}}\xi/2,2(1-r_{\ind{bs}})|\alpha|^2\}$,
that determine performance of the protocol.

Our concluding remarks deal with
analysis of the performance of a repeater with two links. 
To this end, we shall follow
the method presented in~\cite{Wu:pra:2020}
and start with the probabilities of success for the links
\begin{align}
  \label{eq:P1-P2}
 \color{black} 2 P_{s}^{(\nu'_1\nu_1)}(p_c)|_{A_1B_1}\equiv p_1,
  \quad
    2P_{s}^{(\nu'_2\nu_2)}(p_c)|_{A_2B_2}\equiv p_2
\end{align}
that give the joint probability distribution for numbers of attempts is
\begin{align}
    P(n_1,n_2)=p_1q_1^{n_1-1}p_2q_2^{n_2-1},
\end{align}
where $q_i=1-p_i$.
An important point is that the entanglement swapping
can be performed
only after the entanglement is established in both links.
Similar to Ref.~\cite{Wu:pra:2020},
we take the assumption
that the time required for each attempt is
$T = L/c$,
\textcolor{black}{
where $L$
is  the distance between the end-nodes
and the central  node of an elementary link
giving the length of the fiber channel
and $c$ is the speed of light.
In our case,
the length of the elementary link is $L_{0}=2L$,
whereas
the total repeater length is $L_{\mathrm{tot}}=N_{L}L_{0}$,
where $N_{L}$ is the number of links.}

From Eq.~\eqref{eq:P1-P2}, it is not difficult to obtain
the probability distribution
for the magnitude of  the difference in number of attempts
\begin{align}
  &
  \label{eq:P_diff}
    \mathrm{Prob}(|n_1-n_2|=k)=\frac{p_1p_2(q_1^k+q_2^k)}{2(1-q_1q_2)}(2-\delta_{k0})
\end{align}
and evaluate the expectation values
%\begin{widetext}
\begin{align}
  &
    \label{eq:delta_n_avr}
    \avr{n_w}\equiv\avr{|n_1-n_2|}= \frac{p_2^2q_1+p_1^2q_2}{p_1p_2(1-q_1q_2)},
    \notag
  \\
  &
    \avr{n_t}\equiv\avr{n_1+n_2}= \frac{p_1+p_2}{p_1p_2},
  \\
  &
    \label{eq:n_max_min}
    \avr{n_{\ind{max}}}\equiv\avr{\max(n_1,n_2)}=\frac{\avr{n_t}+\avr{n_w}}{2},
    \notag
  \\
  &
    \avr{n_{\ind{min}}}\equiv\avr{\min(n_1,n_2)}=\frac{\avr{n_t}-\avr{n_w}}{2}
\end{align}
%\end{widetext}
that are related to the preparation and waiting times,
$T_{\ind{prep}}$ and $T_{\ind{w}}$,
as follows~\cite{Wu:pra:2020}
\begin{align}
  \label{eq:T_prep_wait}
  T_{\ind{prep}}=\avr{n_{\ind{max}}} T,\quad
  T_{\ind{w}}=\avr{n_{w}} T,\quad
  T=\frac{L}{c}.
\end{align}

\begin{figure}[!htb]
\centering
  \centering
  \includegraphics[width=\linewidth]{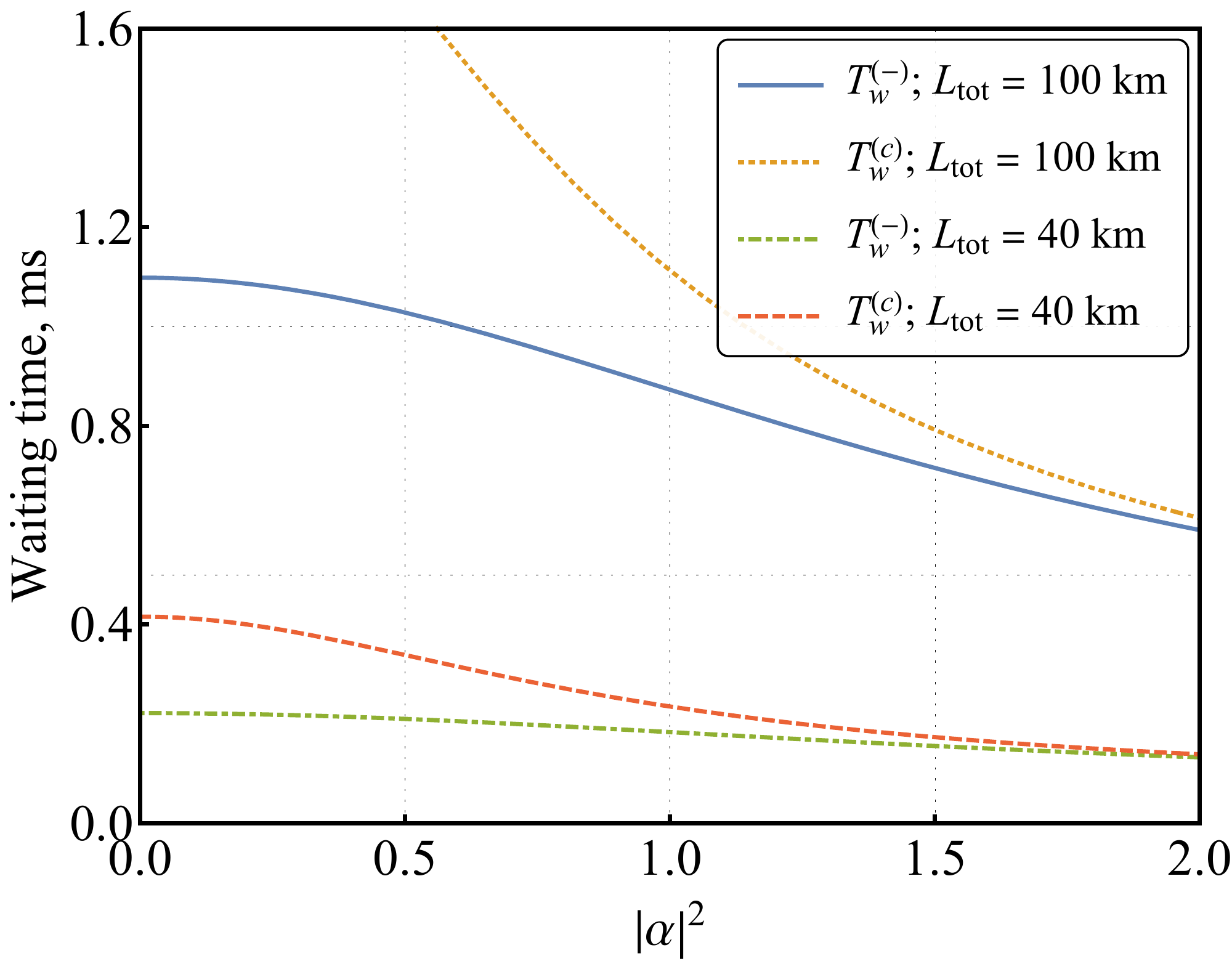}
  \caption{\textcolor{black}{Dependence of waiting times,
$T_{\ind{w}}^{(-)}$ and $T_{\ind{w}}^{(c)}$,
    on $|\alpha|^2$
  computed from Eq.~\eqref{eq:T_prep_wait}
  for swapping between identical links with
  $p_1=p_2=2P_s^{(-)}(\ind{odd})$ and $p_1=p_2=2P_s^{(c)}(\ind{odd})$,
  respectively. The quantum channel
  transmittance is taken to be $\eta(L)=10^{-\kappa L/10}$ with $\kappa=0.2$~dB/km, and
  the photon number ratio $r_{\ind{bs}}$ is $0.2$.}
}  
\label{t-wait}
\end{figure}

For successful operation of the repeater,
the waiting time is required to be shorter than
\textcolor{black}{
  the lifetime
  (alternatively, this time is called either the dephasing time or the coherence time) of the quantum memory:
$T_{\ind{w}}< T_{\ind{m}}$.
Under this condition,
degradation of the generated entanglement
has a negligible detrimental effect on
the swapping protocol.
}

Figure~\ref{t-wait} shows
the $|\alpha|^2$-dependencies of
the waiting times evaluated from
Eqs.~\eqref{eq:T_prep_wait} and~\eqref{eq:delta_n_avr}
assuming that
the success probabilities for the links
are both equal to either $P_s^{(-)}(\ind{odd})$
(the waiting time is $T_{\ind{w}}^{(-)}$)
or
$P_s^{(c)}(\ind{odd})$
(the waiting time is $T_{\ind{w}}^{(c)}$).
In our estimates,
in addition to $r_{\ind{bs}}=0.2$ and $\xi=0.9$,
we have used  the transmittance $\eta(L)=10^{-\kappa L/10}$ with
the attenuation coefficient $\kappa=0.2$~dB/km
describing losses in an optical fiber.
It can be seen that
the estimated waiting times
that fall within the range of miliseconds
may favorably compare with
the lifetimes reported in~\cite{Wu:pra:2020,Semenenko:avs:2022,Mol:qst:2023}.
%~\cite{Wu:pra:2020,Lago:nature:2021,Businger:nature:2022}
% by the order of magnitude.

%%%%%%%%%%%%%%%%%%%%%%%%%
\section{Teleportation}
\label{sec:teleporation}
%%%%%%%%%%%%%%%%%%%%%%%%%

It is well-known that shared entanglement is a resource
of vital importance in fundamental quantum communication
protocols such as quantum teleportation.
The above discussed
entanglement generation procedure
provides this resource through the phase-modulated
cat states entangled between the remote nodes.
In this section, we demonstrate
how such entanglement 
can be utilized to teleport phase information from
an SCW state resulting from
the output state of the electro-optic modulator
(see Eq.~\eqref{eq:cs_alp-bet}).
In order to develop some intuition on how
the teleportation procedure works,
we consider the simplest case where Alice and Bob
share a single-mode entangled coherent cat state
\begin{align}
  \label{eq:Psi-AB-tele}
  \ket{\Psi_{\nu}^{(AB)}(\alpha,\alpha)}=
  \frac{1}{\sqrt{M_{\nu}(\alpha,\alpha)}}
  \left\{
  \ket{\alpha,\alpha}_{AB} +\nu \ket{-\alpha,-\alpha}_{AB}
  \right\},
\end{align}
where
$M_{\pm}(\alpha,\alpha) = 2 ( 1 \pm \exp[-4|\alpha|^2])$ is the normalization constant.
Charlie at Alice's site possesses the SCW state
(see Eq.~\eqref{eq:cs_alp-bet}) with the amplitude $|\alpha|$ and
the phase $\phi_c$ of the SCW state
which is targeted for teleportation 
\begin{align}
  \label{eq:Psi_SCW-tele}
  \ket{\Psi_{\text{SCW}}}=\ket{\bs{\gamma}}_C,
  \quad
  \gamma_{\mu}=\cnj{U_{\mu 0}^{(C)}}\alpha
  \approx J_\mu(m_C) e^{i\mu \phi_c} \alpha,
\end{align}
where we assume that the phase coherence between Alice
and Charlie is maintained.
Thus the joint state of the tripartite system Alice, Bob, and Charlie
is
\begin{align}
  \label{eq:entagled-cat}
  \ket{S_{CAB}^{(\nu)}}=\ket{\Psi_{\text{SCW}}}\otimes\ket{\Psi_{\nu}^{(AB)}(\alpha,\alpha)},\:
  \ket{\Psi_{\text{SCW}}}=\ket{\bs{\gamma}}_C
  %\otimes\ket{\Psi_{\nu}^{(AB)}(\alpha,\alpha)}.
\end{align}

\begin{figure*}[!htb]
\centering
\begin{subfigure}{\textwidth}
\centering\includegraphics[width=.7\textwidth]{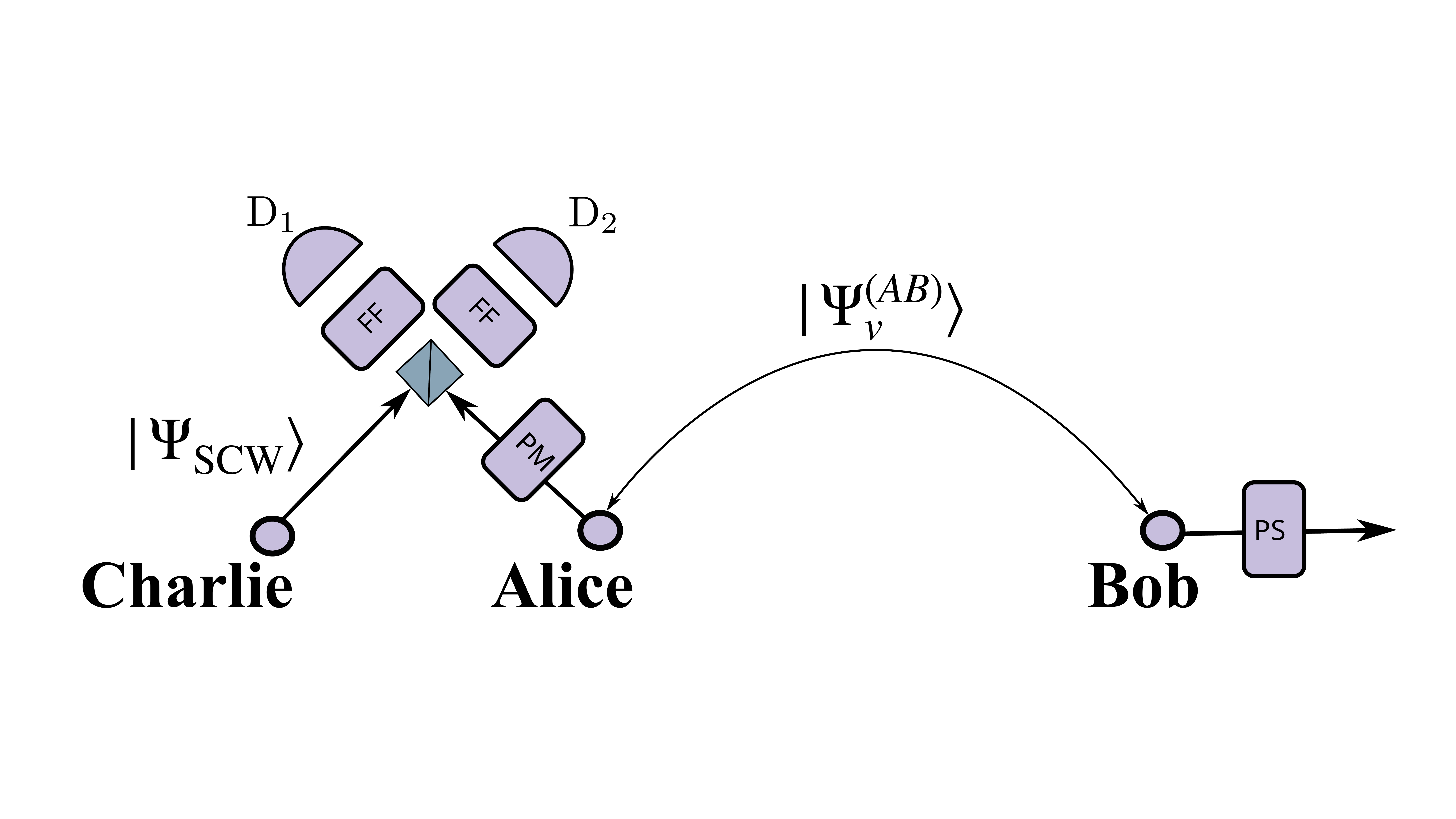} 
  \label{teleportation-scheme}
  \caption{}
\end{subfigure}\\
\begin{subfigure}{.5\textwidth}
  \centering
  \includegraphics[width=.8\textwidth]{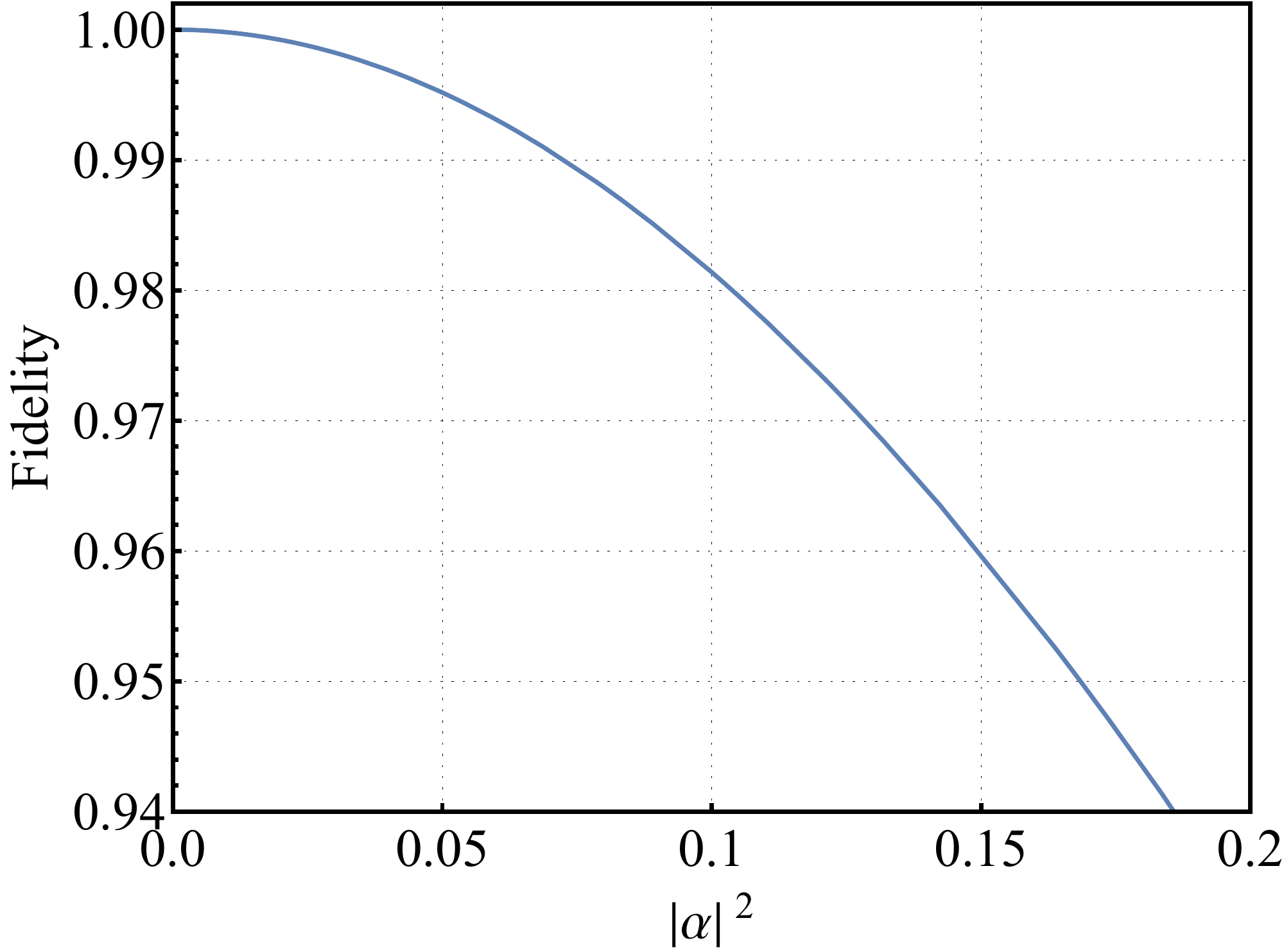}
  \label{Fidelity-vs-alpha}
  \caption{}
\end{subfigure}%
\begin{subfigure}{.5\textwidth}
  \centering
  \includegraphics[width=.8\linewidth]{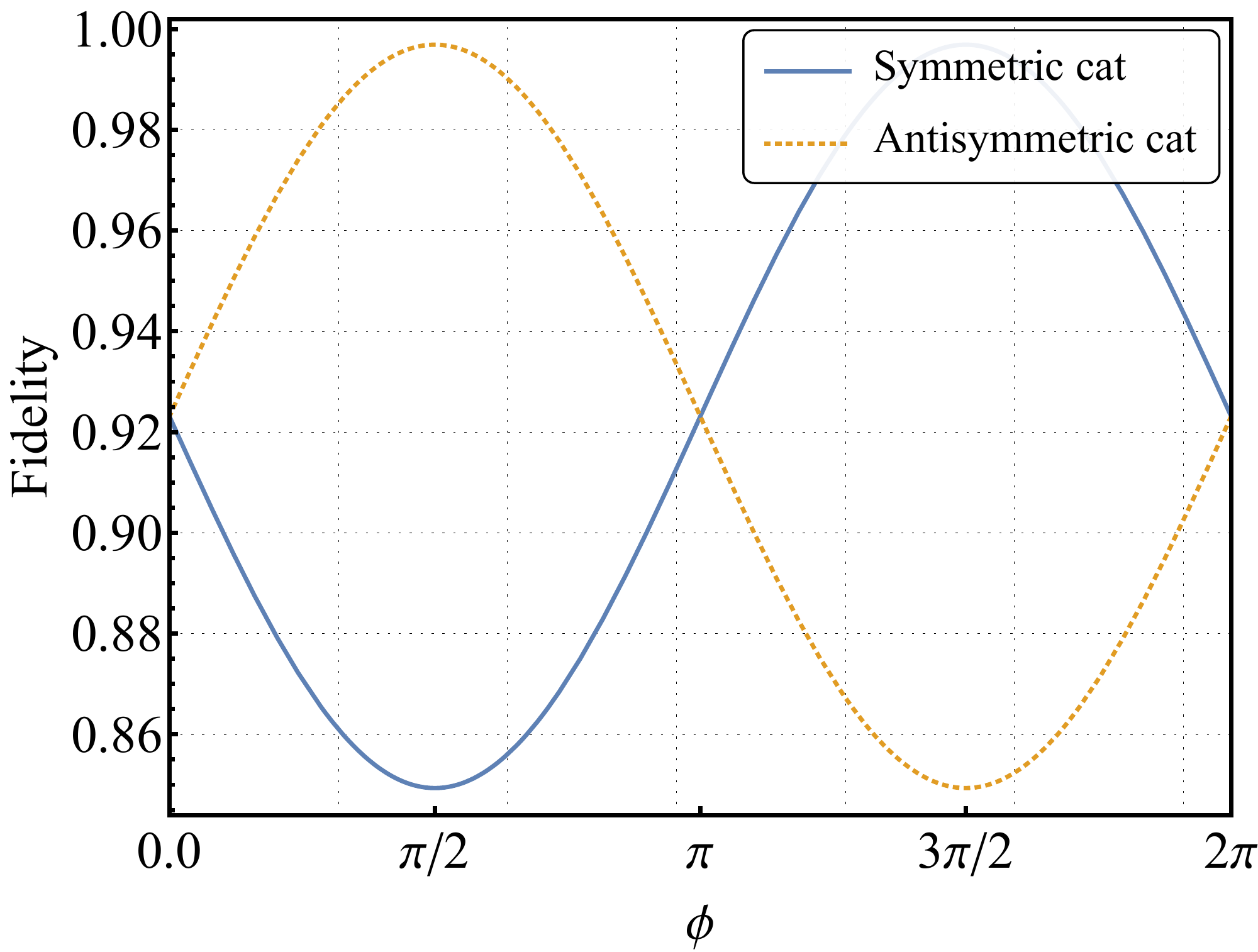}
  \label{Fidelity-vs-phi}
  \caption{}
\end{subfigure}
\caption{ \textbf{a)} The teleportation scheme to transfer the SCW phase
  from Charlie to Bob using entangled Schr\"odinger cat state. PM is
  the phase modulator; FF is the frequency filter; PS is the phase shifter;
  $D_1$ and $D_2$ are the photodetectors.  \textbf{b)}~Fidelity between
  $\ket{i\alpha}$ and $\frac{1}{\sqrt{2}}\left(\ket{\alpha}- i\ket{-\alpha} \right)$
  as a function of the amplitude $|\alpha|$ computed using Eq.~\eqref{eq:fid-tele-aa}.
  \textbf{c)} Dependence of the fidelity given by Eq.~\eqref{eq:fid-tele-aa}
  between $\ket{\alpha e^{i\phi}}$ and
  $\frac{1}{\sqrt{2}}(\ket{\alpha}_B \mp i \ket{-\alpha}_B)$  on the phase $\phi$
  at $|\alpha|=0.2$.
}
\label{fig:teleportation}
\end{figure*}

The scheme of the teleportation from Charlie to Bob is depicted in
Fig.~\ref{fig:teleportation}\textbf{a}.
Transfer of phase information from Charlie's SCW state to
Bob's state can be implemented in the following steps:
\begin{enumerate}
\item Alice uses the modulator with the modulation frequency and index
  identical to Charlie's ones to modulate her share of the state~\eqref{eq:Psi-AB-tele}
  at the phase $\phi_a$.
    \item Charlie's and Alice's states are interfered on $50:50$ beam splitter. 
    \item For the selected sideband, the presence of a single photon at the output channels of the beam
      splitter is monitored by two single photon detectors, $\text{D}_1$ and $\text{D}_2$.
    \item After a photon is registered by a photodetector, Alice communicates the result to Bob
      and he shifts the phase of his state depending on which detector is clicked.
      \textcolor{black}{Charlie}'s SCW phase thus appears to be transferred to the phase of the Bob's
      single-mode coherent state.
\end{enumerate}
Below we describe these steps in detail.
Phase modulation applied by Alice
transforms the
shared entangled state~\eqref{eq:entagled-cat}  as follows
\begin{align}
  &\ket{S_{CAB}^{(\nu)}}\mapsto
    \ket{\Psi_{CAB}^{(\nu)}}=\ket{\bs{\gamma}}_C\otimes\ket{\Psi_{\nu}^{(AB)}(\bs{\alpha},\beta)},\notag
    \\    \label{eq:CAB_modul-tele}
    &\alpha_{\mu}=\cnj{U_{\mu 0}^{(A)}}\alpha\approx J_\mu(m_C) e^{i\mu \phi_a} \alpha,
\end{align}
where the modulation indices of Charlie and Alice are assumed to be identical,
so that $|U_{\mu 0}^{(A)}|=|U_{\mu 0}^{(C)}|\equiv|U_{\mu 0}|$
and
$J_{\mu}(m_A)=J_{\mu}(m_C)$.

After Alice's and Charlie's modes
are brought into interference
onto the 50:50 beam splitter,
the output state after the beam splitter
%\begin{widetext}
\begin{align}
  &
  \label{eq:BS-tele-aa} 
    \hat{T}_{\bs{\gamma}\bs{\alpha}\to D_1D_2}\ket{\Psi_{CAB}^{(\nu)}}=
    \ket{\Psi_{D_1D_2B}^{(\nu)}}
    \notag
  \\
  &
    =
    \frac{1}{\sqrt{M_{\nu}(\alpha,\alpha)}}
    \bigr\{
    \ket{\bs{\gamma}_{+}}_{D_1}\otimes\ket{\bs{\gamma}_{-}}_{D_2}\otimes\ket{\alpha}_B
    \notag
  \\
  &
    +\nu
    \ket{\bs{\gamma}_{-}}_{D_1}\otimes\ket{\bs{\gamma}_{+}}_{D_2}\otimes\ket{-\alpha}_B
    \bigr\},
\end{align}
% \end{widetext}
where
$\bs{\gamma}_{\pm}=\frac{1}{\sqrt{2}}(\bs{\gamma}\pm \bs{\alpha})$,
has the two modes
$  \ket{\bs{\gamma}_{\pm}}_{D_1}$
and
$  \ket{\bs{\gamma}_{\pm}}_{D_2}$
monitored by the detectors $\text{D}_1$
and $\text{D}_2$, respectively.

For the sake of simplicity, in
the limit of low amplitudes $|\alpha|\ll 1$, we truncate the Hilbert space up to a single
photon.
For an ideal single photon detector that measures the sideband with index $\mu$ at
$D_1$ ($D_2$) channel,
detection of a single photon will project
the state~\eqref{eq:BS-tele-aa} onto
the one-photon state
$\ket{\Phi_{\mu}^{(1)}}=\ket{1_{\mu}}_{D_1}\otimes\ket{\vc{0}}_{D_2}$
(
$\ket{\Phi_{\mu}^{(2)}}=\ket{\vc{0}}_{D_1}\otimes\ket{1_{\mu}}_{D_2}$
),
where
$\ket{1_{\mu}}\equiv\ket{0,\ldots,0,1_\mu,0,\ldots,0}$
is the $\mu$th sideband one photon state, 
as follows
\begin{align}
  &
  \label{eq:pr-tele-aa}
  \ket{\Psi_{D_1D_2B}^{(\nu)}}\mapsto \ket{\Psi_{B}^{(\nu,\mu,i)}}=
    \avr{\Phi_{\mu}^{(i)}|\Psi_{D_1D_2B}^{(\nu)}}/\sqrt{P_{\mu}^{(\nu)}},
 \\
&
      \label{eq:prob-tele-aa}
      P_{\mu}^{(\nu)}=|\avr{\Phi_{\mu}^{(i)}|\Psi_{D_1D_2B}^{(\nu)}}|^2
=|U_{\mu0}|^2 P^{(\nu)},
  \\
  &
    \label{eq:prob-nu-tele-aa}
    P^{(\nu)}=
    \frac{|\alpha|^2}{1+\nu
    \ee^{-4|\alpha|^2}}\ee^{-2|\alpha|^2},
\end{align}
where $P_{\mu}^{(\nu)}$ is the probability to detect a single photon
in the $\mu$th sideband by a photodetector.
Since
$\gamma_{\mu}\pm\alpha_{\mu}=|U_{\mu}|\alpha (\exp(i\mu\phi_c)\pm\exp(i\mu\phi_a))$,
the post-measurement state of Bob (up to the global phase)
can be written in the following explicit form
\begin{align}
  &\ket{\Psi_{B}^{(\nu,\mu,1)}}\equiv\ket{\Psi_{B}^{(\nu\phi)}}=
    \cos(\nu\phi/2)\ket{\alpha}+i\sin(\nu\phi/2)\ket{-\alpha},\notag
    \\   \label{eq:Psi-numu1-1-tele-aa}
    &\phi=\mu(\phi_c-\phi_a),
  \\
  &
    \label{eq:Psi-numu2-1-tele-aa}
    \ket{\Psi_{B}^{(\nu,\mu,2)}}=\ket{\Psi_{B}^{(\nu,\mu,1)}}\Bigr|_{\alpha\to-\alpha}.
\end{align}
When the clicked detector is changed from $D_1$ to $D_2$,
relation~\eqref{eq:Psi-numu2-1-tele-aa}
shows that Bob needs to apply the $\pi$ shift changing $\alpha$ to $-\alpha$
so as to have the same teleported state.

\begin{table}[htp]
\centering
\begin{tabular}{|c|c|c|c|c|}
\hline
\multirow{2}{*}{$\phi_c$}   & \multicolumn{2}{|c|}{$D_{1}$ clicks}   &  \multicolumn{2}{|c|}{$D_{2}$ clicks} \\ 
  \cline{2-5} & +1 sideband  & -1 sideband & +1 sideband  & -1 sideband\\ \hline
$\phi_a$ &  $\ket{\alpha}$ & $\ket{\alpha}$ & $\ket{-\alpha}$ & $\ket{-\alpha}$ \\
$\phi_a+\pi$ &  $\ket{-\alpha}$ & $\ket{-\alpha}$ & $\ket{\alpha}$ & $\ket{\alpha}$ \\
$\phi_a+\pi/2$ &  $\ket{-i\alpha }$  &  $\ket{i\alpha}$ & $\ket{ i\alpha}$ & $\ket{-i\alpha}$ \\
$\phi_a+3\pi/2$ &  $\ket{i\alpha }$ & $\ket{-i\alpha}$ & $\ket{-i\alpha}$ & $\ket{i\alpha}$ \\  \hline
\end{tabular}
\caption[]{Truth table for the heralded state at Bob's site $\ket{\Psi_B}$ after performing
  teleportation with entangled coherent state $\ket{\Psi_{+}^{(AB)}}$ depending on
  the sideband registered, the photodetector that clicks, and the value of Charlie's SCW phase.}
\label{table:teleportation}
\end{table}

\begin{figure*}[!htb]
\centering
\begin{subfigure}{.5\textwidth}
  \centering
  \includegraphics[width=.8\linewidth]{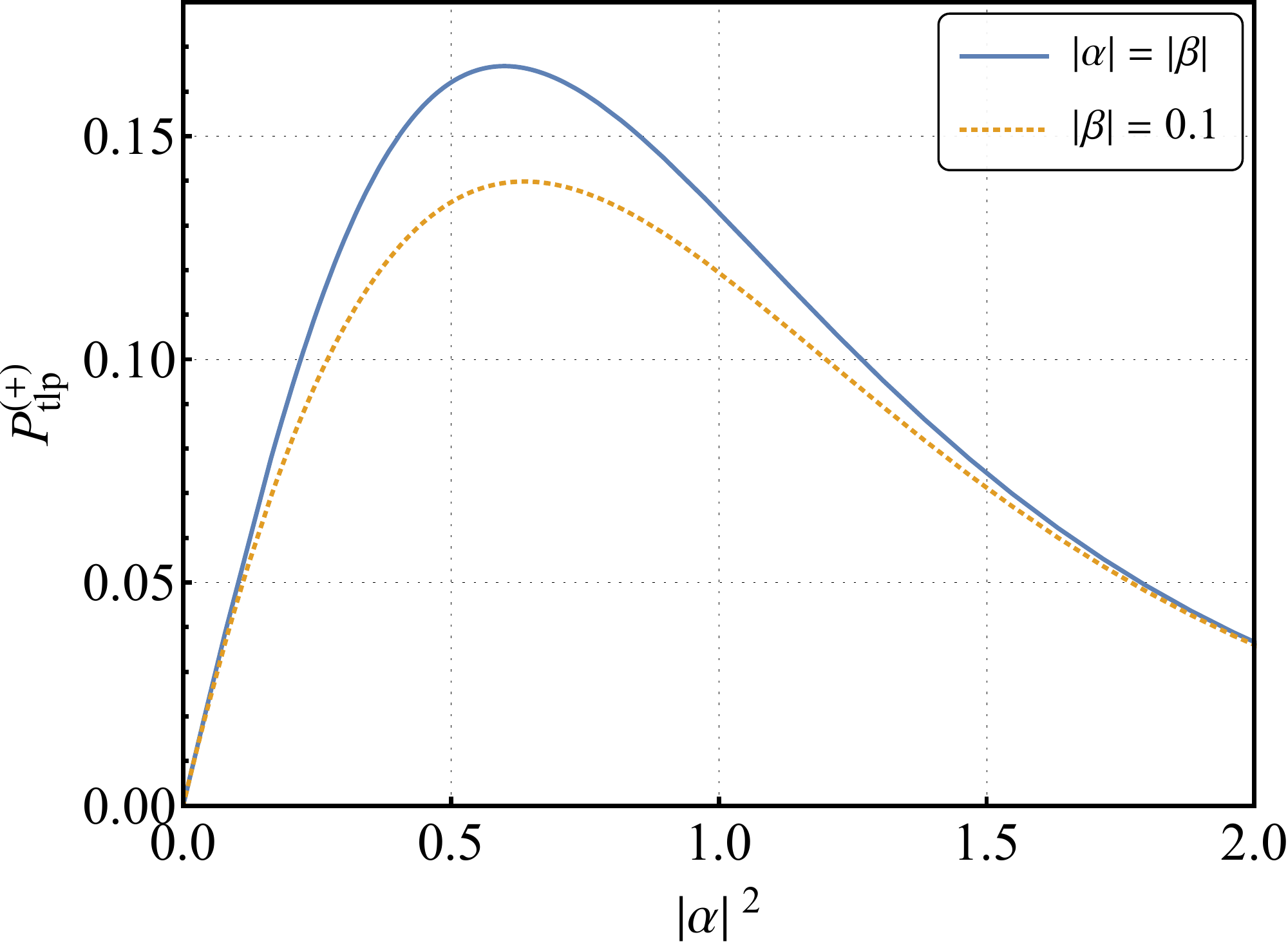}
  \label{p-tel-plus}
  \caption{}
\end{subfigure}%
\begin{subfigure}{.5\textwidth}
  \centering
  \includegraphics[width=.8\linewidth]{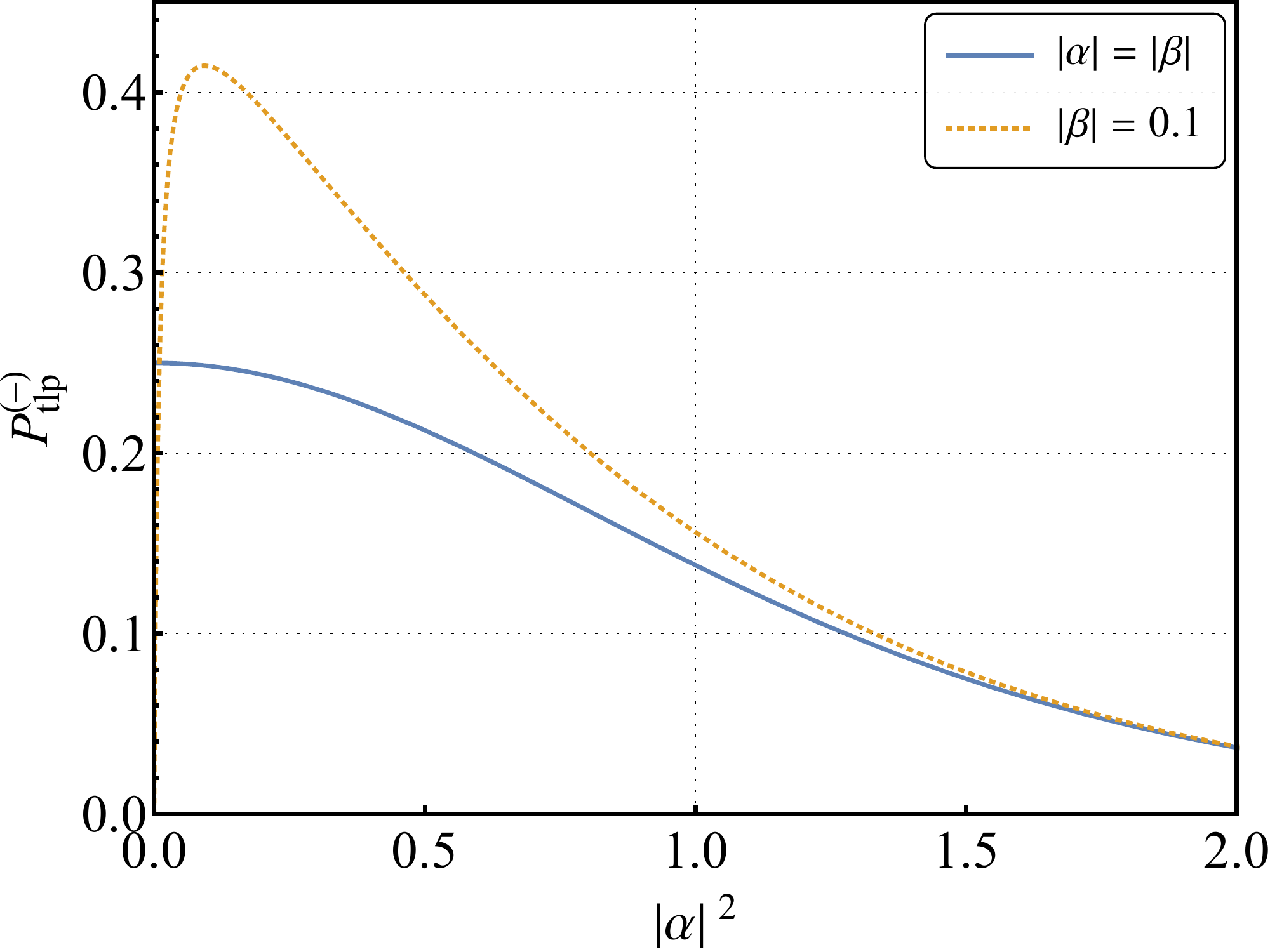}
  \label{p-tel-minus}
  \caption{}
\end{subfigure}
\caption{
  Dependence of the success probabilities of teleportation, $P^{(\pm)}_{\mathrm{tlp}}$, 
  on the squared amplitude $|\alpha|^2$ computed from Eq.~\eqref{eq:P_tlp-comp}.
} 
\label{p-telep}
\end{figure*}

For the special case, where
$\phi_c-\phi_a\in\{0,\pi/2,\pi,3\pi/2\}$
and $\mu=\pm 1$,
Table~\ref{table:teleportation}
summarizes how
Charlie's SCW phase is mapped into
the phase of Bob's coherent state
provided Alice's phase vanishes ($\phi_a = 0$)
and the cat state is symmetric ($\nu=1$).
Clearly, the teleported state exactly reproduces Charlie's phase
when it is either zero or $\pi$.
In this table, we have used the approximation 
$\frac{1}{\sqrt{2}}(\ket{\alpha}_B \pm i \ket{-\alpha}_B) \approx \ket{\mp i\alpha}_B$
with the fidelity plotted in
Figs.~\ref{fig:teleportation}\textbf{b}-\textbf{c}.

More generally,
in the low amplitude (mean photon number) region
where $|\alpha|$ is small,
the teleported state can be well
approximated by the coherent state 
\begin{align}
  &
  \label{eq:approx-tele-aa}
  \ket{\Psi_{B}^{(\phi)}}=
  \cos(\phi/2)\ket{\alpha}+i\sin(\phi/2)\ket{-\alpha}
  \approx\ket{\ee^{-i\phi}\alpha}.
\end{align}
The latter can be seen from the fidelity
between $\ket{\Psi_{B}^{(\phi)}}$
and $\ket{\ee^{-i\psi}\alpha}$
given by
%\begin{widetext}
\begin{align}
  &
    \label{eq:fid-tele-aa}
    |\avr{\ee^{-i\psi}\alpha|\Psi_{B}^{(\phi)}}|^2=
    \ee^{-2|\alpha|^2}
    \bigl|
    \cos(\phi/2)\exp(\ee^{i\psi}|\alpha|^2)
    \notag
  \\
  &
    +i\sin(\phi/2)\exp(-\ee^{i\psi}|\alpha|^2)
    \bigr|^2
    \notag
  \\
  &
    \approx
    1+2(\cos(\phi-\psi)-1)|\alpha|^2+\ldots,
\end{align}
%\end{widetext}
where the last equality is the fidelity expanded into a power series
over $|\alpha|$ up to second order.
Figure~\ref{fig:teleportation}\textbf{b} illustrates that,
at $\phi=-\pi/2$,
the fidelity can be higher than 99\% when the amplitude $|\alpha|$
is smaller than 0.25.

It is rather straightforward
to generalize our analysis to the non-symmetric case,
where the amplitudes,
$|\gamma|$, $|\alpha|$ and $|\beta|$,
that determine Charlie's, Alice's, and Bob's states,
respectively, differ from each other.
In this case, Alice and Bob share the state
\begin{align}
  &
  \ket{S_{AB}^{(\nu)}}=\ket{\Psi_{\nu}^{(AB)}(\alpha,\beta)}
    \notag \\ \label{eq:S-AB-ini-tele-gen} &=
    \frac{1}{\sqrt{M_{\nu}(\alpha,\beta)}}
    \{
    \ket{\alpha}_A\otimes\ket{\beta}_B+\nu
    \ket{-\alpha}_A\otimes\ket{-\beta}_B
    \},
\end{align}
where
$M_{\nu}(\alpha,\beta)=2(1+\nu\exp[-2(|\alpha|^2+|\beta|^2)])$,
while Charlie holds the state~\eqref{eq:Psi_SCW-tele}
with $\gamma_{\mu}=\cnj{U_{\mu 0}^{(C)}}\gamma$.

Once Alice has applied the phase modulation  to her mode and
interfered it with Charlie's state on 50:50 beam splitter,  
a single photon detection in the sideband $\mu$
heralds preparation of the following state at Bob's node
\begin{align}
  &
  \label{eq:Psi-numu1-tele-gen}
    \ket{\Psi_{B}^{(\nu,\mu,1)}}=
    \frac{(\gamma\ee^{i\phi_C}+\alpha\ee^{i\phi_A})\ket{\beta}+
    \nu(\gamma\ee^{i\phi_C}-\alpha\ee^{i\phi_A})\ket{-\beta}}{%
    \sqrt{2 (|\gamma|^2+|\alpha|^2+(|\gamma|^2-|\alpha|^2)\ee^{-2|\beta|^2})}},
  \\
  &
    \ket{\Psi_{B}^{(\nu,\mu,2)}}=\ket{\Psi_{B}^{(\nu,\mu,1)}}\Bigr|_{\beta\to-\beta}
\end{align}
with the probability of success given by
\begin{align}
      &P_{\mu,\;\mathrm{tlp}}^{(\nu)}=|U_{\mu0}|^2 P^{(\nu)}_{\mathrm{tlp}}, \notag
  \\
  \label{eq:prob-tele-gen}
    &P^{(\nu)}_{\mathrm{tlp}}=
    \frac{|\gamma|^2+|\alpha|^2+(|\gamma|^2-|\alpha|^2)\ee^{-2|\beta|^2}}{2(1+\nu
    \ee^{-2(|\alpha|^2+|\beta|^2)})}\ee^{-|\alpha|^2-|\gamma|^2},
  \end{align}

Note that, since $P_{\mathrm{tlp}}^{(-)}>P_{\mathrm{tlp}}^{(+)}$,
using the  antisymmetric entangled coherent state  $\ket{\Psi_{-}^{(AB)}}$
generally yields a higher teleportation rate as compared to the symmetric one
$\ket{\Psi_{+}^{(AB)}}$.
It can be seen from the curves
shown in Fig.~\ref{p-telep}.
These curves represent $|\alpha|^2$-dependencies
of $P_{\mathrm{tlp}}^{(\pm)}$ computed at
$|\alpha|=|\gamma|$ from the formula
\begin{align}
  &
    \label{eq:P_tlp-comp}
    P_{\ind{tlp}}^{(\nu)}(|\alpha|,|\beta|)=
  \frac{|\alpha|^2}{2(1+\nu
    \ee^{-2(|\alpha|^2+|\beta|^2)})}\ee^{-2 |\alpha|^2}.
\end{align}

Figure~\ref{p-telep} presents the results for two cases:
the curves computed at $|\alpha|=|\beta|$
($P_{\ind{tlp}}^{(\nu)}(|\alpha|,|\alpha|)$ is given
by Eq.~\eqref{eq:prob-nu-tele-aa})
and the ones where the amplitude of $\beta$
is fixed (see Eq.~\eqref{eq:P_tlp-comp}).
From Eq.~\eqref{eq:P_tlp-comp},
it is not difficult to see that 
$P_{\ind{tlp}}^{(+)}(|\alpha|,|\alpha|)>P_{\ind{tlp}}^{(+)}(|\alpha|,|\beta|)$
and $P_{\ind{tlp}}^{(-)}(|\alpha|,|\alpha|)<P_{\ind{tlp}}^{(-)}(|\alpha|,|\beta|)$
at $|\alpha|>|\beta|$.
As is shown in Fig.~\ref{p-telep}\textbf{a},
for symmetric cats,
all the curves are qualitatively similar
and the maximum value of $P_{\ind{tlp}}^{(+)}(|\alpha|,|\alpha|)$
reached at $|\alpha|=|\alpha|_{\ind{max}}$
will give the highest probability provided $|\beta|<|\alpha|_{\ind{max}}$.

For
teleportation using antisymmetric cats,
behavior of teleportation success probabilities
is strikingly different.
Referring to Fig.~\ref{p-telep}\textbf{b},
$P_{\ind{tlp}}^{(-)}(|\alpha|,|\alpha|)$
starts from its maximal value $1/4$
($P_{\ind{tlp}}^{(-)}(|\alpha|,|\alpha|)$ tends to $1/4$ at $|\alpha|\to 0$)
and monotonically decays zero as $|\alpha|$ increases.
By contrast,
the curve for $P_{\ind{tlp}}^{(-)}(|\alpha|,|\beta|)$ is similar to $P_{\ind{tlp}}^{(+)}$.
At sufficiently small $|\beta|$,
the maximum of $P_{\ind{tlp}}^{(-)}(|\alpha|,|\beta|)$ is above $1/4$
and can be close to the limiting value $1/2$.
This is the case giving the highest success probability of teleportation.

%\textit{
  Our concluding remark
  concerns
  the small mean photon number approximation~\eqref{eq:approx-tele-aa}
  applicable to Bob's state at $|\beta|< 0.2$.
  Such approximation may be of practical value
  as teleported Bob's state appears to be quasi-Gaussian
 with the corresponding benefits of Gaussian state control.
In particular, the phase of Bob's state can be measured with moderate resources  
and designing an apparatus for single-shot measurement of a phase for
Schr\"odinger cat and coherent state simultaneously is no longer required.  
%}

%%%%%%%%%%%%%%%%%
\section{Conclusions and discussion}
\label{sec:disc}
%%%%%%%%%%%%

In this work, we have proposed
to use phase-modulated multimode Schr\"odinger
cat states for producing entanglement between remote parties
(Alice and Bob) and for entanglement swapping.
This approach to quantum repeaters is based on multimode coherent states
generated by an electro-optic modulator
and may reveal a number of interesting potential extensions. 

\begin{figure*}[ht]
   \centering
     \includegraphics[width=0.7\textwidth]{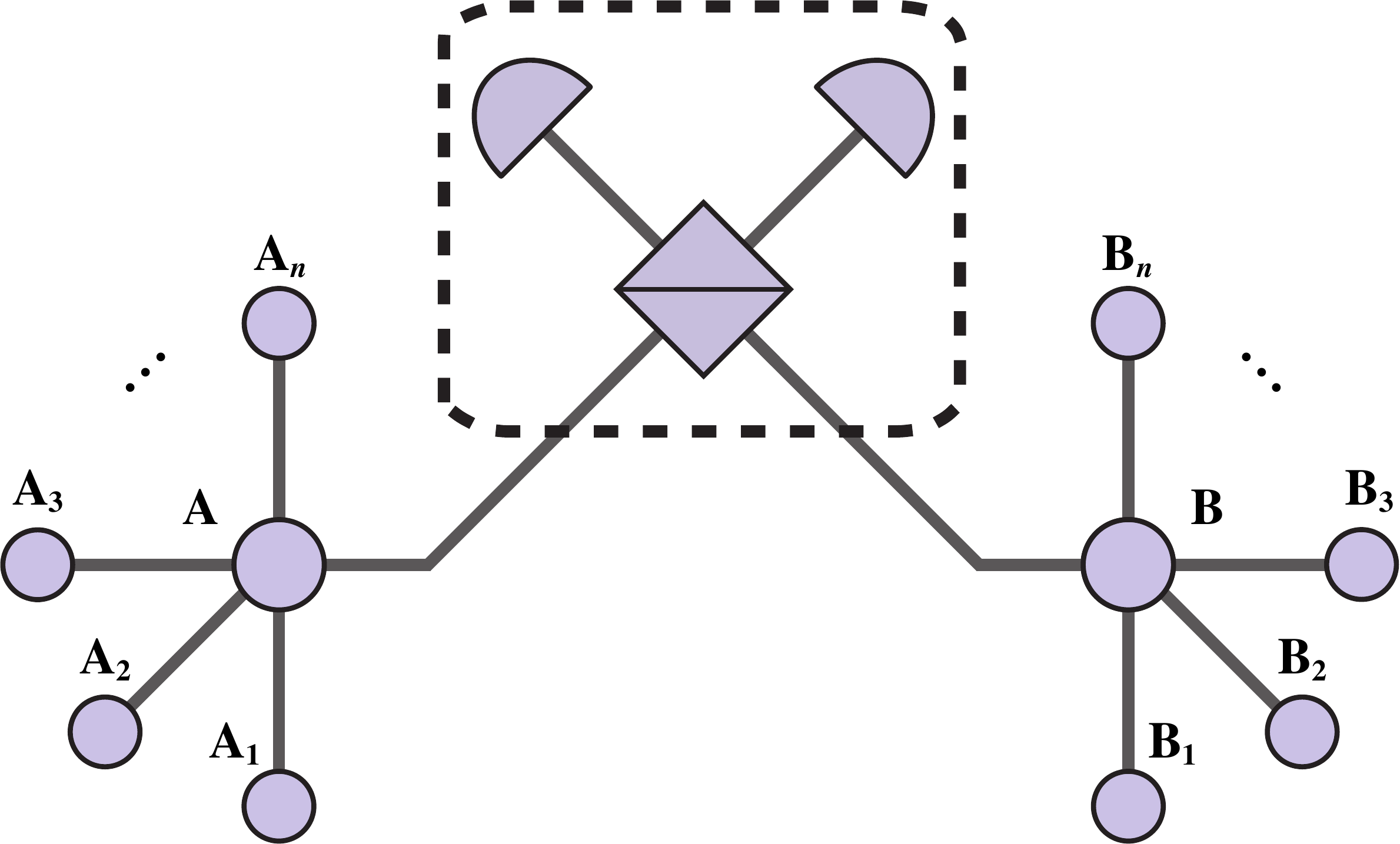}
\caption{Elementary link of an SCW quantum repeater in star topology.}
\label{scheme-2}
\end{figure*}

For instance, as is illustrated in Fig.~\ref{scheme-2},
one of the advantages of using  entangled multimode states 
is that the modes may, in principle,
be distributed and stored in quantum memories located at different
nodes of a star topology network.
Since these modes differ in frequency,
the DWDM demultiplexer can be employed to send them to
different locations. 

We have analyzed the heralded entanglement generation scheme
assuming that Alice and Bob send a part of modes
that enter their multimode cat states,
$\ket{\Psi_{\nu'}^{(A)}}$ and $\ket{\Psi_{\nu}^{(B)}}$,
to the symmetric ($50:50$) beam splitter 
(see Fig.~\ref{scheme-0} and Eq.~\eqref{eq:Psi_BS})
and have computed
the probabilities,
$P_{\mu}^{(\nu'\nu)}$,
for three orthogonal states
 (the vacuum state, $\ket{\vc{0}}$, the antisymmetric and modified symmetric cat states,
 $\ket{\Phi_{-}}$ and $\ket{\Phi_{+}}$) to occur at the output ports of
 the beam splitter for different couples
$(\nu',\nu)$ of input cats
 (see Eqs.~\eqref{eq:Pmum} and~\eqref{eq:Pmuc}).

 We have modeled
 the optical fiber quantum channel with the transmittance $\eta$
 using the Stinesping dilation based on a beam splitter transformation
 for an enlarged system supplemented with environmental (noisy) modes
 (see Eq.~\eqref{eq:channel}).
 For the heralding events determined by
 the parity of photons,
 $p_c\in\{\ind{odd},\ind{even}\}$,
 registered by a photodetector with the efficiency $\xi$,
 it was found that
 the conditional probabilities of
 detection~\eqref{eq:P-pc}
 give the probabilities of success~\eqref{eq:P-success}
 expressed in terms of $P_{\pm}^{(\nu'\nu)}$
 with the renormalized mean photon number ratio:
 $r_{\ind{bs}}\to\zeta=r_{\ind{bs}}\eta\xi$.

 It is shown that
 the fidelities, $F_{\mu}^{(\nu'\nu)}(p_c)$,
 between the density matrix of the heralded states~\eqref{eq:rho-pc}
 and the entangled cat states, $\Phi_{\mu}^{(AB)}$,
 can be described by two analytical expressions
 for the fidelities of
predominantly symmetric and antisymmetric cats:
 $F_{-}^{(\pm)}(\ind{odd})$ and $F_{+}^{(\pm)}(\ind{even})$ 
 given by Eqs.~\eqref{eq:F-m-odd} and~\eqref{eq:F-p-even},
 respectively.
 It turned out that these fidelities characterizing
 the quality of the generated entanglement
 are monotonically decreasing functions of the amplitude $|\alpha|$
 (see Fig.~\ref{fid-prob}).

 Our key analytical
 results are
 formulas~\eqref{eq:P-success}
 and~\eqref{eq:F-mp} for the success probabilities
 and the fidelities describing performance
 and quality of the entanglement generation procedure.
 These results imply that producing high-quality entangled states
 requires the amplitude $|\alpha|$ to be sufficiently small.
 In this region, the highest values of the probability of success
 correspond to odd number of clicks
 for either two antisymmetric input states (the heralded cat state is predominantly
 antisymmetric)
 or two states of different symmetry
 (the heralded cat state is predominantly symmetric).
 By contrast, in the low-amplitude region,
 the success probability for a couple of input symmetric cat states
 is negligibly small (see Eq.~\eqref{eq:Ps-p-vs-m}).

Note that, in reality,
 we have two detectors and each of the detectors
registers the odd (even) number of photons
with the same probability of success $P_{s}^{(\nu'\nu)}(\ind{odd})$
($P_{\pm}^{(\nu'\nu)}(\ind{even})$).
The difference between the corresponding heralded states,
$\ket{\Psi_{\mu}^{(AB)}(\bs{\alpha}_{\ind{qm}},\bs{\alpha}_{\ind{qm}})}$
and
$\ket{\Psi_{\mu}^{(AB)}(\bs{\alpha}_{\ind{qm}},-\bs{\alpha}_{\ind{qm}})}$,
can be corrected by applying $\pi$-shift to the phases
of the modes at either Alice's or Bob's site.
It means that total performance of entanglement generation can be
characterized by twice the success probability.

For generation of antisymmetric cats,
this probability cannot be higher than 50\%
because $P_s^{(-)}(\ind{odd})\le 1/4$
(see Fig.~\ref{ps-prob}).
Interestingly,
$P_s^{(-)}(\ind{odd})$ is identically equal to $1/4$
when the product $\zeta\equiv r_{\ind{bs}}\eta\xi$ is $1/2$
(or, equivalently, $r_{\ind{bs}}=r_{1/2}=1/(2\eta\xi)$).
From the other hand,
according to Fig.~\ref{fid-prob},
the photon number ratio $r_{\ind{bs}}$
should be lowered to enhance
the quality of entanglement with
the maximum value of fidelity $F_{-}^{(\pm)}(\ind{odd})$
exceeding 90\%.

By contrast,
when the input cat states differ in summery and
the heralded states are symmetric cats,
double the probability of success $2P_s^{(c)}(\ind{odd})$
is above $1/2$ at $r_{\ind{bs}}>r_{1/2}$
(see Fig.~\ref{ps-prob}).
In addition,
as is shown in Fig.~\ref{fid-prob},
the maximum value of $F_{+}^{(c)}(\ind{odd})$
reached at $|\alpha|=0$ is unity.
So, procedure leading to entangled symmetric cats is superior
to that for the case of antisymmetric ones in both performance and quality.

However,
given the symmetric cat states efficiently produced
to create entanglement between the nodes of elementary links,
the success probability for entanglement swapping
given by Eq.~\eqref{eq:Ps-12} will be close to zero.
Therefore, antisymmetric cat states play important part
in reaching reasonably high performance of the repeater.
An important point is that there is a tradeoff between the probability of success
and the fidelity~\eqref{eq:F_12} that needs to be optimized.

Part of our considerations assumes
that the modes can be stored in a multimode quantum memory~\cite{Moiseev2016,Moiseev_2021}.
For the repeater with two links,
we have estimated the so-called waiting time~\eqref{eq:T_prep_wait}
giving the lower bound for the storage time of the quantum memory.
The results presented in Fig.~\ref{t-wait}
show that
the waiting time falls within the range of hundreds of microseconds
for tens kilometer long distances.  
For instance, according to Ref.~\cite{Askarani:prl:2021},
QMs based on atomic frequency
comb protocol with rare-earth-ion doped crystals
are promising multimode QMs that
may tolerate such waiting times.
Efficiency of such QMs, however, still needs to be improved.

Alternatively, our system may operate without a multimode QM.
In this approach,
the carrier mode at the central frequency is the only mode to be stored in the QM,
whereas the sidebands are distributed over the repeater nodes.
After extracting this mode from the memory,
re-modulation can be carried out and then the carrier mode can be stored again.
High efficiency of the electro-optic modulator allows such process  to be recurring in cycles.

Thus,
our analysis suggests
feasibility of the proposed scheme
for antisymmetric cat states
and relatively short distances to the central node.
In general, approaches to QRs
using phase modulation to produce and control multimode states
may lead to promising methods for generation and distribution of
entanglement.

Quantum teleportation protocol that
uses the entangled cat states to
transfer the phase information between remote parties
discussed in Sec.~\ref{sec:teleporation}
exemplifies one of such methods.
Interestingly, one of our findings is that the antisymmetric cat states 
are superior to symmetric ones in teleportation performance.

% The produced entanglement may be also used for connecting
% remote SCW QKD networks by means of
% quantum teleportation.  

%\section*{Acknowledgements}
\begin{acknowledgements}
The work was done by Leading Research Center "National Center for Quantum Internet" of ITMO
University by order of JSCo Russian Railways.
The work of ADK was  also financially supported by
the Ministry of Education and Science of the Russian Federation
(Passport No. 2019-0903). 
ESM and SAM appreciate support within framework
project \# 00075-02-2020-051/1 from 02.03.2020.
\end{acknowledgements}

% \section*{Disclosures}
% The authors declare no conflicts of interest.

% \bibliography{article}

\begin{thebibliography}{73}%
\makeatletter
\providecommand \@ifxundefined [1]{%
 \@ifx{#1\undefined}
}%
\providecommand \@ifnum [1]{%
 \ifnum #1\expandafter \@firstoftwo
 \else \expandafter \@secondoftwo
 \fi
}%
\providecommand \@ifx [1]{%
 \ifx #1\expandafter \@firstoftwo
 \else \expandafter \@secondoftwo
 \fi
}%
\providecommand \natexlab [1]{#1}%
\providecommand \enquote  [1]{``#1''}%
\providecommand \bibnamefont  [1]{#1}%
\providecommand \bibfnamefont [1]{#1}%
\providecommand \citenamefont [1]{#1}%
\providecommand \href@noop [0]{\@secondoftwo}%
\providecommand \href [0]{\begingroup \@sanitize@url \@href}%
\providecommand \@href[1]{\@@startlink{#1}\@@href}%
\providecommand \@@href[1]{\endgroup#1\@@endlink}%
\providecommand \@sanitize@url [0]{\catcode `\\12\catcode `\$12\catcode
  `\&12\catcode `\#12\catcode `\^12\catcode `\_12\catcode `\%12\relax}%
\providecommand \@@startlink[1]{}%
\providecommand \@@endlink[0]{}%
\providecommand \url  [0]{\begingroup\@sanitize@url \@url }%
\providecommand \@url [1]{\endgroup\@href {#1}{\urlprefix }}%
\providecommand \urlprefix  [0]{URL }%
\providecommand \Eprint [0]{\href }%
\providecommand \doibase [0]{https://doi.org/}%
\providecommand \selectlanguage [0]{\@gobble}%
\providecommand \bibinfo  [0]{\@secondoftwo}%
\providecommand \bibfield  [0]{\@secondoftwo}%
\providecommand \translation [1]{[#1]}%
\providecommand \BibitemOpen [0]{}%
\providecommand \bibitemStop [0]{}%
\providecommand \bibitemNoStop [0]{.\EOS\space}%
\providecommand \EOS [0]{\spacefactor3000\relax}%
\providecommand \BibitemShut  [1]{\csname bibitem#1\endcsname}%
\let\auto@bib@innerbib\@empty
%</preamble>
\bibitem [{\citenamefont {Gisin}\ and\ \citenamefont
  {Thew}(2007)}]{Gisin:nphot:2007}%
  \BibitemOpen
  \bibfield  {author} {\bibinfo {author} {\bibfnamefont {N.}~\bibnamefont
  {Gisin}}\ and\ \bibinfo {author} {\bibfnamefont {R.}~\bibnamefont {Thew}},\
  }\bibfield  {title} {\bibinfo {title} {Quantum communication},\ }\href
  {https://doi.org/10.1038/nphoton.2007.22} {\bibfield  {journal} {\bibinfo
  {journal} {Nature Photonics}\ }\textbf {\bibinfo {volume} {1}},\ \bibinfo
  {pages} {165} (\bibinfo {year} {2007})}\BibitemShut {NoStop}%
\bibitem [{\citenamefont {Krenn}\ \emph {et~al.}(2016)\citenamefont {Krenn},
  \citenamefont {Malik}, \citenamefont {Scheidl}, \citenamefont {Ursin},\ and\
  \citenamefont {Zeilinger}}]{Krenn2016}%
  \BibitemOpen
  \bibfield  {author} {\bibinfo {author} {\bibfnamefont {M.}~\bibnamefont
  {Krenn}}, \bibinfo {author} {\bibfnamefont {M.}~\bibnamefont {Malik}},
  \bibinfo {author} {\bibfnamefont {T.}~\bibnamefont {Scheidl}}, \bibinfo
  {author} {\bibfnamefont {R.}~\bibnamefont {Ursin}},\ and\ \bibinfo {author}
  {\bibfnamefont {A.}~\bibnamefont {Zeilinger}},\ }\bibfield  {title} {\bibinfo
  {title} {{Quantum Communication with Photons}},\ }in\ \href
  {https://doi.org/10.1007/978-3-319-31903-2{\_}18} {\emph {\bibinfo
  {booktitle} {Optics in Our Time}}}\ (\bibinfo  {publisher} {Springer
  International Publishing},\ \bibinfo {address} {Cham},\ \bibinfo {year}
  {2016})\BibitemShut {NoStop}%
\bibitem [{\citenamefont {Gisin}\ \emph {et~al.}(2002)\citenamefont {Gisin},
  \citenamefont {Ribordy}, \citenamefont {Tittel},\ and\ \citenamefont
  {Zbinden}}]{Gisin:rmp:2002}%
  \BibitemOpen
  \bibfield  {author} {\bibinfo {author} {\bibfnamefont {N.}~\bibnamefont
  {Gisin}}, \bibinfo {author} {\bibfnamefont {G.}~\bibnamefont {Ribordy}},
  \bibinfo {author} {\bibfnamefont {W.}~\bibnamefont {Tittel}},\ and\ \bibinfo
  {author} {\bibfnamefont {H.}~\bibnamefont {Zbinden}},\ }\bibfield  {title}
  {\bibinfo {title} {Quantum cryptography},\ }\href
  {https://doi.org/10.1103/RevModPhys.74.145} {\bibfield  {journal} {\bibinfo
  {journal} {Rev. Mod. Phys.}\ }\textbf {\bibinfo {volume} {74}},\ \bibinfo
  {pages} {145} (\bibinfo {year} {2002})}\BibitemShut {NoStop}%
\bibitem [{\citenamefont {Xu}\ \emph {et~al.}(2020)\citenamefont {Xu},
  \citenamefont {Ma}, \citenamefont {Zhang}, \citenamefont {Lo},\ and\
  \citenamefont {Pan}}]{Xu:rmp:2020}%
  \BibitemOpen
  \bibfield  {author} {\bibinfo {author} {\bibfnamefont {F.}~\bibnamefont
  {Xu}}, \bibinfo {author} {\bibfnamefont {X.}~\bibnamefont {Ma}}, \bibinfo
  {author} {\bibfnamefont {Q.}~\bibnamefont {Zhang}}, \bibinfo {author}
  {\bibfnamefont {H.-K.}\ \bibnamefont {Lo}},\ and\ \bibinfo {author}
  {\bibfnamefont {J.-W.}\ \bibnamefont {Pan}},\ }\bibfield  {title} {\bibinfo
  {title} {Secure quantum key distribution with realistic devices},\ }\href
  {https://doi.org/10.1103/RevModPhys.92.025002} {\bibfield  {journal}
  {\bibinfo  {journal} {Rev. Mod. Phys.}\ }\textbf {\bibinfo {volume} {92}},\
  \bibinfo {pages} {025002} (\bibinfo {year} {2020})}\BibitemShut {NoStop}%
\bibitem [{\citenamefont {Giovannetti}\ \emph {et~al.}(201)\citenamefont
  {Giovannetti}, \citenamefont {Lloyd},\ and\ \citenamefont
  {Maccone}}]{Giovannetti:nphot:2011}%
  \BibitemOpen
  \bibfield  {author} {\bibinfo {author} {\bibfnamefont {V.}~\bibnamefont
  {Giovannetti}}, \bibinfo {author} {\bibfnamefont {S.}~\bibnamefont {Lloyd}},\
  and\ \bibinfo {author} {\bibfnamefont {L.}~\bibnamefont {Maccone}},\
  }\bibfield  {title} {\bibinfo {title} {Advances in quantum metrology},\
  }\href {https://doi.org/10.1038/nphoton.2011.35} {\bibfield  {journal}
  {\bibinfo  {journal} {Nature Photonics}\ }\textbf {\bibinfo {volume} {5}},\
  \bibinfo {pages} {222} (\bibinfo {year} {201})}\BibitemShut {NoStop}%
\bibitem [{\citenamefont {T\'oth}\ and\ \citenamefont
  {Apellaniz}(2014)}]{Toth:jpa:2014}%
  \BibitemOpen
  \bibfield  {author} {\bibinfo {author} {\bibfnamefont {G.}~\bibnamefont
  {T\'oth}}\ and\ \bibinfo {author} {\bibfnamefont {I.}~\bibnamefont
  {Apellaniz}},\ }\bibfield  {title} {\bibinfo {title} {Quantum metrology from
  a quantum information science perspective},\ }\href
  {https://doi.org/10.1088/1751-8113/47/42/424006} {\bibfield  {journal}
  {\bibinfo  {journal} {Journal of Physics A: Mathematical and Theoretical}\
  }\textbf {\bibinfo {volume} {47}},\ \bibinfo {pages} {424006} (\bibinfo
  {year} {2014})}\BibitemShut {NoStop}%
\bibitem [{\citenamefont {Khabiboulline}\ \emph {et~al.}(2019)\citenamefont
  {Khabiboulline}, \citenamefont {Borregaard}, \citenamefont {De~Greve},\ and\
  \citenamefont {Lukin}}]{Khabiboulline2019}%
  \BibitemOpen
  \bibfield  {author} {\bibinfo {author} {\bibfnamefont {E.~T.}\ \bibnamefont
  {Khabiboulline}}, \bibinfo {author} {\bibfnamefont {J.}~\bibnamefont
  {Borregaard}}, \bibinfo {author} {\bibfnamefont {K.}~\bibnamefont
  {De~Greve}},\ and\ \bibinfo {author} {\bibfnamefont {M.~D.}\ \bibnamefont
  {Lukin}},\ }\bibfield  {title} {\bibinfo {title} {{Optical Interferometry
  with Quantum Networks}},\ }\href
  {https://doi.org/10.1103/PhysRevLett.123.070504} {\bibfield  {journal}
  {\bibinfo  {journal} {Physical Review Letters}\ }\textbf {\bibinfo {volume}
  {123}},\ \bibinfo {pages} {070504} (\bibinfo {year} {2019})}\BibitemShut
  {NoStop}%
\bibitem [{\citenamefont {Van~Meter}\ and\ \citenamefont
  {Devitt}(2016)}]{Meter2016}%
  \BibitemOpen
  \bibfield  {author} {\bibinfo {author} {\bibfnamefont {R.}~\bibnamefont
  {Van~Meter}}\ and\ \bibinfo {author} {\bibfnamefont {S.~J.}\ \bibnamefont
  {Devitt}},\ }\bibfield  {title} {\bibinfo {title} {{The Path to Scalable
  Distributed Quantum Computing}},\ }\bibfield  {journal} {\bibinfo  {journal}
  {Computer}\ }\textbf {\bibinfo {volume} {49}},\ \href
  {https://doi.org/10.1109/MC.2016.291} {10.1109/MC.2016.291} (\bibinfo {year}
  {2016})\BibitemShut {NoStop}%
\bibitem [{\citenamefont {Yimsiriwattana}\ and\ \citenamefont
  {Lomonaco~Jr.}(2004)}]{Yimsiriwattana2004}%
  \BibitemOpen
  \bibfield  {author} {\bibinfo {author} {\bibfnamefont {A.}~\bibnamefont
  {Yimsiriwattana}}\ and\ \bibinfo {author} {\bibfnamefont {S.~J.}\
  \bibnamefont {Lomonaco~Jr.}},\ }\bibfield  {title} {\bibinfo {title}
  {{Distributed quantum computing: a distributed Shor algorithm}},\ }in\ \href
  {https://doi.org/10.1117/12.546504} {\emph {\bibinfo {booktitle} {Quantum
  Information and Computation II}}},\ \bibinfo {editor} {edited by\ \bibinfo
  {editor} {\bibfnamefont {E.}~\bibnamefont {Donkor}}, \bibinfo {editor}
  {\bibfnamefont {A.~R.}\ \bibnamefont {Pirich}},\ and\ \bibinfo {editor}
  {\bibfnamefont {H.~E.}\ \bibnamefont {Brandt}}}\ (\bibinfo {year} {2004})\
  pp.\ \bibinfo {pages} {360--372}\BibitemShut {NoStop}%
\bibitem [{\citenamefont {Briegel}\ \emph {et~al.}(1998)\citenamefont
  {Briegel}, \citenamefont {D\"ur}, \citenamefont {Cirac},\ and\ \citenamefont
  {Zoller}}]{Briegel:prl:1998}%
  \BibitemOpen
  \bibfield  {author} {\bibinfo {author} {\bibfnamefont {H.-J.}\ \bibnamefont
  {Briegel}}, \bibinfo {author} {\bibfnamefont {W.}~\bibnamefont {D\"ur}},
  \bibinfo {author} {\bibfnamefont {J.~I.}\ \bibnamefont {Cirac}},\ and\
  \bibinfo {author} {\bibfnamefont {P.}~\bibnamefont {Zoller}},\ }\bibfield
  {title} {\bibinfo {title} {Quantum repeaters: {T}he role of imperfect local
  operations in quantum communication},\ }\href
  {https://doi.org/10.1103/PhysRevLett.81.5932} {\bibfield  {journal} {\bibinfo
   {journal} {Phys. Rev. Lett.}\ }\textbf {\bibinfo {volume} {81}},\ \bibinfo
  {pages} {5932} (\bibinfo {year} {1998})}\BibitemShut {NoStop}%
\bibitem [{\citenamefont {Sangouard}\ \emph {et~al.}(2011)\citenamefont
  {Sangouard}, \citenamefont {Simon}, \citenamefont {de~Riedmatten},\ and\
  \citenamefont {Gisin}}]{Sangouard:rmp:2011}%
  \BibitemOpen
  \bibfield  {author} {\bibinfo {author} {\bibfnamefont {N.}~\bibnamefont
  {Sangouard}}, \bibinfo {author} {\bibfnamefont {C.}~\bibnamefont {Simon}},
  \bibinfo {author} {\bibfnamefont {H.}~\bibnamefont {de~Riedmatten}},\ and\
  \bibinfo {author} {\bibfnamefont {N.}~\bibnamefont {Gisin}},\ }\bibfield
  {title} {\bibinfo {title} {Quantum repeaters based on atomic ensembles and
  linear optics},\ }\href {https://doi.org/10.1103/RevModPhys.83.33} {\bibfield
   {journal} {\bibinfo  {journal} {Rev. Mod. Phys.}\ }\textbf {\bibinfo
  {volume} {83}},\ \bibinfo {pages} {33} (\bibinfo {year} {2011})}\BibitemShut
  {NoStop}%
\bibitem [{\citenamefont {Munro}\ \emph {et~al.}(2015)\citenamefont {Munro},
  \citenamefont {Azuma}, \citenamefont {Tamaki},\ and\ \citenamefont
  {Nemoto}}]{Munro2015}%
  \BibitemOpen
  \bibfield  {author} {\bibinfo {author} {\bibfnamefont {W.~J.}\ \bibnamefont
  {Munro}}, \bibinfo {author} {\bibfnamefont {K.}~\bibnamefont {Azuma}},
  \bibinfo {author} {\bibfnamefont {K.}~\bibnamefont {Tamaki}},\ and\ \bibinfo
  {author} {\bibfnamefont {K.}~\bibnamefont {Nemoto}},\ }\bibfield  {title}
  {\bibinfo {title} {{Inside Quantum Repeaters}},\ }\bibfield  {journal}
  {\bibinfo  {journal} {IEEE Journal of Selected Topics in Quantum
  Electronics}\ }\textbf {\bibinfo {volume} {21}},\ \href
  {https://doi.org/10.1109/JSTQE.2015.2392076} {10.1109/JSTQE.2015.2392076}
  (\bibinfo {year} {2015})\BibitemShut {NoStop}%
\bibitem [{\citenamefont {Muralidharan}\ \emph {et~al.}(2016)\citenamefont
  {Muralidharan}, \citenamefont {Li}, \citenamefont {Kim}, \citenamefont
  {L\"{u}tkenhaus}, \citenamefont {Lukin},\ and\ \citenamefont
  {Jiang}}]{Muralidharan:scirep:2016}%
  \BibitemOpen
  \bibfield  {author} {\bibinfo {author} {\bibfnamefont {S.}~\bibnamefont
  {Muralidharan}}, \bibinfo {author} {\bibfnamefont {L.}~\bibnamefont {Li}},
  \bibinfo {author} {\bibfnamefont {J.}~\bibnamefont {Kim}}, \bibinfo {author}
  {\bibfnamefont {N.}~\bibnamefont {L\"{u}tkenhaus}}, \bibinfo {author}
  {\bibfnamefont {M.~D.}\ \bibnamefont {Lukin}},\ and\ \bibinfo {author}
  {\bibfnamefont {L.}~\bibnamefont {Jiang}},\ }\bibfield  {title} {\bibinfo
  {title} {Optimal architectures for long distance quantum communication},\
  }\href {https://doi.org/10.1038/srep20463} {\bibfield  {journal} {\bibinfo
  {journal} {Scientific Reports}\ }\textbf {\bibinfo {volume} {6}},\ \bibinfo
  {pages} {20463} (\bibinfo {year} {2016})}\BibitemShut {NoStop}%
\bibitem [{\citenamefont {Azuma}\ \emph {et~al.}(2022)\citenamefont {Azuma},
  \citenamefont {Economou}, \citenamefont {Elkouss}, \citenamefont {Hilaire},
  \citenamefont {Jiang}, \citenamefont {Lo},\ and\ \citenamefont
  {Tzitrin}}]{Azuma:arxiv:2022}%
  \BibitemOpen
  \bibfield  {author} {\bibinfo {author} {\bibfnamefont {K.}~\bibnamefont
  {Azuma}}, \bibinfo {author} {\bibfnamefont {S.~E.}\ \bibnamefont {Economou}},
  \bibinfo {author} {\bibfnamefont {D.}~\bibnamefont {Elkouss}}, \bibinfo
  {author} {\bibfnamefont {P.}~\bibnamefont {Hilaire}}, \bibinfo {author}
  {\bibfnamefont {L.}~\bibnamefont {Jiang}}, \bibinfo {author} {\bibfnamefont
  {H.-K.}\ \bibnamefont {Lo}},\ and\ \bibinfo {author} {\bibfnamefont
  {I.}~\bibnamefont {Tzitrin}},\ }\href
  {https://doi.org/10.48550/ARXIV.2212.10820} {\bibinfo {title} {Quantum
  repeaters: {F}rom quantum networks to the quantum internet}},\ \bibinfo
  {howpublished} {arXiv:2212.10820} (\bibinfo {year} {2022})\BibitemShut
  {NoStop}%
\bibitem [{\citenamefont {Duan}\ \emph {et~al.}(2001)\citenamefont {Duan},
  \citenamefont {Lukin}, \citenamefont {Cirac},\ and\ \citenamefont
  {Zoller}}]{Duan:nature:2001}%
  \BibitemOpen
  \bibfield  {author} {\bibinfo {author} {\bibfnamefont {L.-M.}\ \bibnamefont
  {Duan}}, \bibinfo {author} {\bibfnamefont {M.~D.}\ \bibnamefont {Lukin}},
  \bibinfo {author} {\bibfnamefont {J.~I.}\ \bibnamefont {Cirac}},\ and\
  \bibinfo {author} {\bibfnamefont {P.}~\bibnamefont {Zoller}},\ }\bibfield
  {title} {\bibinfo {title} {Long-distance quantum communication with atomic
  ensembles and linear optics},\ }\href {https://doi.org/10.1038/35106500}
  {\bibfield  {journal} {\bibinfo  {journal} {Nature}\ }\textbf {\bibinfo
  {volume} {414}},\ \bibinfo {pages} {413} (\bibinfo {year}
  {2001})}\BibitemShut {NoStop}%
\bibitem [{\citenamefont {Lvovsky}\ \emph {et~al.}(2009)\citenamefont
  {Lvovsky}, \citenamefont {Sanders},\ and\ \citenamefont
  {Tittel}}]{Lvovsky:nature:2009}%
  \BibitemOpen
  \bibfield  {author} {\bibinfo {author} {\bibfnamefont {A.~I.}\ \bibnamefont
  {Lvovsky}}, \bibinfo {author} {\bibfnamefont {B.~C.}\ \bibnamefont
  {Sanders}},\ and\ \bibinfo {author} {\bibfnamefont {W.}~\bibnamefont
  {Tittel}},\ }\bibfield  {title} {\bibinfo {title} {Optical quantum memory},\
  }\href {https://doi.org/10.1038/nphoton.2009.231} {\bibfield  {journal}
  {\bibinfo  {journal} {Nature Photonics}\ }\textbf {\bibinfo {volume} {3}},\
  \bibinfo {pages} {706} (\bibinfo {year} {2009})}\BibitemShut {NoStop}%
\bibitem [{\citenamefont {Azuma}\ \emph {et~al.}(2015)\citenamefont {Azuma},
  \citenamefont {Tamaki},\ and\ \citenamefont {Lo}}]{Azuma:nature:2015}%
  \BibitemOpen
  \bibfield  {author} {\bibinfo {author} {\bibfnamefont {K.}~\bibnamefont
  {Azuma}}, \bibinfo {author} {\bibfnamefont {K.}~\bibnamefont {Tamaki}},\ and\
  \bibinfo {author} {\bibfnamefont {H.-K.}\ \bibnamefont {Lo}},\ }\bibfield
  {title} {\bibinfo {title} {All-photonic quantum repeaters},\ }\href
  {https://doi.org/10.1038/ncomms7787} {\bibfield  {journal} {\bibinfo
  {journal} {Nature Communications}\ }\textbf {\bibinfo {volume} {6}},\
  \bibinfo {pages} {6787} (\bibinfo {year} {2015})}\BibitemShut {NoStop}%
\bibitem [{\citenamefont {Zwerger}\ \emph {et~al.}(2018)\citenamefont
  {Zwerger}, \citenamefont {Pirker}, \citenamefont {Dunjko}, \citenamefont
  {Briegel},\ and\ \citenamefont {D\"ur}}]{Zwerger:prl:2018}%
  \BibitemOpen
  \bibfield  {author} {\bibinfo {author} {\bibfnamefont {M.}~\bibnamefont
  {Zwerger}}, \bibinfo {author} {\bibfnamefont {A.}~\bibnamefont {Pirker}},
  \bibinfo {author} {\bibfnamefont {V.}~\bibnamefont {Dunjko}}, \bibinfo
  {author} {\bibfnamefont {H.~J.}\ \bibnamefont {Briegel}},\ and\ \bibinfo
  {author} {\bibfnamefont {W.}~\bibnamefont {D\"ur}},\ }\bibfield  {title}
  {\bibinfo {title} {Long-range big quantum-data transmission},\ }\href
  {https://doi.org/10.1103/PhysRevLett.120.030503} {\bibfield  {journal}
  {\bibinfo  {journal} {Phys. Rev. Lett.}\ }\textbf {\bibinfo {volume} {120}},\
  \bibinfo {pages} {030503} (\bibinfo {year} {2018})}\BibitemShut {NoStop}%
\bibitem [{\citenamefont {Su}\ \emph {et~al.}(2018)\citenamefont {Su},
  \citenamefont {Guan},\ and\ \citenamefont {Li}}]{Su:pra:2018}%
  \BibitemOpen
  \bibfield  {author} {\bibinfo {author} {\bibfnamefont {Z.}~\bibnamefont
  {Su}}, \bibinfo {author} {\bibfnamefont {J.}~\bibnamefont {Guan}},\ and\
  \bibinfo {author} {\bibfnamefont {L.}~\bibnamefont {Li}},\ }\bibfield
  {title} {\bibinfo {title} {Efficient quantum repeater with respect to both
  entanglement-concentration rate and complexity of local operations and
  classical communication},\ }\href
  {https://doi.org/10.1103/PhysRevA.97.012325} {\bibfield  {journal} {\bibinfo
  {journal} {Phys. Rev. A}\ }\textbf {\bibinfo {volume} {97}},\ \bibinfo
  {pages} {012325} (\bibinfo {year} {2018})}\BibitemShut {NoStop}%
\bibitem [{\citenamefont {Wehner}\ \emph {et~al.}(2018)\citenamefont {Wehner},
  \citenamefont {Elkouss},\ and\ \citenamefont {Hanson}}]{Wehner2018}%
  \BibitemOpen
  \bibfield  {author} {\bibinfo {author} {\bibfnamefont {S.}~\bibnamefont
  {Wehner}}, \bibinfo {author} {\bibfnamefont {D.}~\bibnamefont {Elkouss}},\
  and\ \bibinfo {author} {\bibfnamefont {R.}~\bibnamefont {Hanson}},\
  }\bibfield  {title} {\bibinfo {title} {Quantum internet: A vision for the
  road ahead},\ }\href {https://doi.org/10.1126/science.aam9288} {\bibfield
  {journal} {\bibinfo  {journal} {Science}\ }\textbf {\bibinfo {volume}
  {362}},\ \bibinfo {pages} {eaam9288} (\bibinfo {year} {2018})}\BibitemShut
  {NoStop}%
\bibitem [{\citenamefont {Azuma}\ \emph {et~al.}(2021)\citenamefont {Azuma},
  \citenamefont {B\"auml}, \citenamefont {Coopmans}, \citenamefont {Elkouss},\
  and\ \citenamefont {Li}}]{Azuma:avs:2021}%
  \BibitemOpen
  \bibfield  {author} {\bibinfo {author} {\bibfnamefont {K.}~\bibnamefont
  {Azuma}}, \bibinfo {author} {\bibfnamefont {S.}~\bibnamefont {B\"auml}},
  \bibinfo {author} {\bibfnamefont {T.}~\bibnamefont {Coopmans}}, \bibinfo
  {author} {\bibfnamefont {D.}~\bibnamefont {Elkouss}},\ and\ \bibinfo {author}
  {\bibfnamefont {B.}~\bibnamefont {Li}},\ }\bibfield  {title} {\bibinfo
  {title} {Tools for quantum network design},\ }\href
  {https://doi.org/10.1116/5.0024062} {\bibfield  {journal} {\bibinfo
  {journal} {AVS Quantum Science}\ }\textbf {\bibinfo {volume} {3}},\ \bibinfo
  {pages} {014101} (\bibinfo {year} {2021})}\BibitemShut {NoStop}%
\bibitem [{\citenamefont {Wei}\ \emph {et~al.}(2022)\citenamefont {Wei},
  \citenamefont {Jing}, \citenamefont {Zhang}, \citenamefont {Liao},
  \citenamefont {Yuan}, \citenamefont {Fan}, \citenamefont {Lyu}, \citenamefont
  {Zhou}, \citenamefont {Wang}, \citenamefont {Deng}, \citenamefont {Song},
  \citenamefont {Oblak}, \citenamefont {Guo},\ and\ \citenamefont
  {Zhou}}]{Wei:lpor:2022}%
  \BibitemOpen
  \bibfield  {author} {\bibinfo {author} {\bibfnamefont {S.-H.}\ \bibnamefont
  {Wei}}, \bibinfo {author} {\bibfnamefont {B.}~\bibnamefont {Jing}}, \bibinfo
  {author} {\bibfnamefont {X.-Y.}\ \bibnamefont {Zhang}}, \bibinfo {author}
  {\bibfnamefont {J.-Y.}\ \bibnamefont {Liao}}, \bibinfo {author}
  {\bibfnamefont {C.-Z.}\ \bibnamefont {Yuan}}, \bibinfo {author}
  {\bibfnamefont {B.-Y.}\ \bibnamefont {Fan}}, \bibinfo {author} {\bibfnamefont
  {C.}~\bibnamefont {Lyu}}, \bibinfo {author} {\bibfnamefont {D.-L.}\
  \bibnamefont {Zhou}}, \bibinfo {author} {\bibfnamefont {Y.}~\bibnamefont
  {Wang}}, \bibinfo {author} {\bibfnamefont {G.-W.}\ \bibnamefont {Deng}},
  \bibinfo {author} {\bibfnamefont {H.-Z.}\ \bibnamefont {Song}}, \bibinfo
  {author} {\bibfnamefont {D.}~\bibnamefont {Oblak}}, \bibinfo {author}
  {\bibfnamefont {G.-C.}\ \bibnamefont {Guo}},\ and\ \bibinfo {author}
  {\bibfnamefont {Q.}~\bibnamefont {Zhou}},\ }\bibfield  {title} {\bibinfo
  {title} {Towards real-world quantum networks: {A} review},\ }\href
  {https://doi.org/10.1002/lpor.202100219} {\bibfield  {journal} {\bibinfo
  {journal} {Laser \& Photonics Reviews}\ }\textbf {\bibinfo {volume} {16}},\
  \bibinfo {pages} {2100219} (\bibinfo {year} {2022})}\BibitemShut {NoStop}%
\bibitem [{\citenamefont {van Loock}\ \emph {et~al.}(2008)\citenamefont {van
  Loock}, \citenamefont {L\"utkenhaus}, \citenamefont {Munro},\ and\
  \citenamefont {Nemoto}}]{Look:pra:2008}%
  \BibitemOpen
  \bibfield  {author} {\bibinfo {author} {\bibfnamefont {P.}~\bibnamefont {van
  Loock}}, \bibinfo {author} {\bibfnamefont {N.}~\bibnamefont {L\"utkenhaus}},
  \bibinfo {author} {\bibfnamefont {W.~J.}\ \bibnamefont {Munro}},\ and\
  \bibinfo {author} {\bibfnamefont {K.}~\bibnamefont {Nemoto}},\ }\bibfield
  {title} {\bibinfo {title} {Quantum repeaters using coherent-state
  communication},\ }\href {https://doi.org/10.1103/PhysRevA.78.062319}
  {\bibfield  {journal} {\bibinfo  {journal} {Phys. Rev. A}\ }\textbf {\bibinfo
  {volume} {78}},\ \bibinfo {pages} {062319} (\bibinfo {year}
  {2008})}\BibitemShut {NoStop}%
\bibitem [{\citenamefont {Sangouard}\ \emph {et~al.}(2010)\citenamefont
  {Sangouard}, \citenamefont {Simon}, \citenamefont {Gisin}, \citenamefont
  {Laurat}, \citenamefont {Tualle-Brouri},\ and\ \citenamefont
  {Grangier}}]{Sangouard:josab:2010}%
  \BibitemOpen
  \bibfield  {author} {\bibinfo {author} {\bibfnamefont {N.}~\bibnamefont
  {Sangouard}}, \bibinfo {author} {\bibfnamefont {C.}~\bibnamefont {Simon}},
  \bibinfo {author} {\bibfnamefont {N.}~\bibnamefont {Gisin}}, \bibinfo
  {author} {\bibfnamefont {J.}~\bibnamefont {Laurat}}, \bibinfo {author}
  {\bibfnamefont {R.}~\bibnamefont {Tualle-Brouri}},\ and\ \bibinfo {author}
  {\bibfnamefont {P.}~\bibnamefont {Grangier}},\ }\bibfield  {title} {\bibinfo
  {title} {Quantum repeaters with entangled coherent states},\ }\href
  {https://doi.org/10.1364/JOSAB.27.00A137} {\bibfield  {journal} {\bibinfo
  {journal} {J. Opt. Soc. Am. B}\ }\textbf {\bibinfo {volume} {27}},\ \bibinfo
  {pages} {A137} (\bibinfo {year} {2010})}\BibitemShut {NoStop}%
\bibitem [{\citenamefont {Ghasemi}\ and\ \citenamefont
  {Tavassoly}(2019)}]{Ghasemi:laserph:2019}%
  \BibitemOpen
  \bibfield  {author} {\bibinfo {author} {\bibfnamefont {M.}~\bibnamefont
  {Ghasemi}}\ and\ \bibinfo {author} {\bibfnamefont {M.~K.}\ \bibnamefont
  {Tavassoly}},\ }\bibfield  {title} {\bibinfo {title} {Toward a quantum
  repeater protocol based on the coherent state approach},\ }\bibfield
  {journal} {\bibinfo  {journal} {Laser Physics}\ }\textbf {\bibinfo {volume}
  {29}},\ \href {https://doi.org/10.1088/1555-6611/ab1cbc}
  {10.1088/1555-6611/ab1cbc} (\bibinfo {year} {2019})\BibitemShut {NoStop}%
\bibitem [{\citenamefont {Sanders}(1992)}]{Sanders:pra:1992}%
  \BibitemOpen
  \bibfield  {author} {\bibinfo {author} {\bibfnamefont {B.~C.}\ \bibnamefont
  {Sanders}},\ }\bibfield  {title} {\bibinfo {title} {Entangled coherent
  states},\ }\href {https://doi.org/10.1103/PhysRevA.45.6811} {\bibfield
  {journal} {\bibinfo  {journal} {Phys. Rev. A}\ }\textbf {\bibinfo {volume}
  {45}},\ \bibinfo {pages} {6811} (\bibinfo {year} {1992})}\BibitemShut
  {NoStop}%
\bibitem [{\citenamefont {Sanders}(2012)}]{Sanders:jpa:2012}%
  \BibitemOpen
  \bibfield  {author} {\bibinfo {author} {\bibfnamefont {B.~C.}\ \bibnamefont
  {Sanders}},\ }\bibfield  {title} {\bibinfo {title} {Review of entangled
  coherent states},\ }\href {https://doi.org/10.1088/1751-8113/45/24/244002}
  {\bibfield  {journal} {\bibinfo  {journal} {Journal of Physics A:
  Mathematical and Theoretical}\ }\textbf {\bibinfo {volume} {45}},\ \bibinfo
  {pages} {244002} (\bibinfo {year} {2012})}\BibitemShut {NoStop}%
\bibitem [{\citenamefont {Munro}\ \emph {et~al.}(2000)\citenamefont {Munro},
  \citenamefont {Milburn},\ and\ \citenamefont {Sanders}}]{Munro:pra:2000}%
  \BibitemOpen
  \bibfield  {author} {\bibinfo {author} {\bibfnamefont {W.~J.}\ \bibnamefont
  {Munro}}, \bibinfo {author} {\bibfnamefont {G.~J.}\ \bibnamefont {Milburn}},\
  and\ \bibinfo {author} {\bibfnamefont {B.~C.}\ \bibnamefont {Sanders}},\
  }\bibfield  {title} {\bibinfo {title} {Entangled coherent-state qubits in an
  ion trap},\ }\href {https://doi.org/10.1103/PhysRevA.62.052108} {\bibfield
  {journal} {\bibinfo  {journal} {Phys. Rev. A}\ }\textbf {\bibinfo {volume}
  {62}},\ \bibinfo {pages} {052108} (\bibinfo {year} {2000})}\BibitemShut
  {NoStop}%
\bibitem [{\citenamefont {Jeong}\ \emph {et~al.}(2001)\citenamefont {Jeong},
  \citenamefont {Kim},\ and\ \citenamefont {Lee}}]{Jeong:pra:2001}%
  \BibitemOpen
  \bibfield  {author} {\bibinfo {author} {\bibfnamefont {H.}~\bibnamefont
  {Jeong}}, \bibinfo {author} {\bibfnamefont {M.~S.}\ \bibnamefont {Kim}},\
  and\ \bibinfo {author} {\bibfnamefont {J.}~\bibnamefont {Lee}},\ }\bibfield
  {title} {\bibinfo {title} {Quantum-information processing for a coherent
  superposition state via a mixed entangled coherent channel},\ }\href
  {https://doi.org/10.1103/PhysRevA.64.052308} {\bibfield  {journal} {\bibinfo
  {journal} {Phys. Rev. A}\ }\textbf {\bibinfo {volume} {64}},\ \bibinfo
  {pages} {052308} (\bibinfo {year} {2001})}\BibitemShut {NoStop}%
\bibitem [{\citenamefont {van Enk}\ and\ \citenamefont
  {Hirota}(2001)}]{Enk:pra:2001}%
  \BibitemOpen
  \bibfield  {author} {\bibinfo {author} {\bibfnamefont {S.~J.}\ \bibnamefont
  {van Enk}}\ and\ \bibinfo {author} {\bibfnamefont {O.}~\bibnamefont
  {Hirota}},\ }\bibfield  {title} {\bibinfo {title} {Entangled coherent states:
  {T}eleportation and decoherence},\ }\href
  {https://doi.org/10.1103/PhysRevA.64.022313} {\bibfield  {journal} {\bibinfo
  {journal} {Phys. Rev. A}\ }\textbf {\bibinfo {volume} {64}},\ \bibinfo
  {pages} {022313} (\bibinfo {year} {2001})}\BibitemShut {NoStop}%
\bibitem [{\citenamefont {Liu}\ \emph {et~al.}(2016)\citenamefont {Liu},
  \citenamefont {Shi}, \citenamefont {Shi}, \citenamefont {Lv},\ and\
  \citenamefont {Guo}}]{Liu:chphys:2016}%
  \BibitemOpen
  \bibfield  {author} {\bibinfo {author} {\bibfnamefont {J.-L.}\ \bibnamefont
  {Liu}}, \bibinfo {author} {\bibfnamefont {R.-H.}\ \bibnamefont {Shi}},
  \bibinfo {author} {\bibfnamefont {J.-J.}\ \bibnamefont {Shi}}, \bibinfo
  {author} {\bibfnamefont {G.-L.}\ \bibnamefont {Lv}},\ and\ \bibinfo {author}
  {\bibfnamefont {Y.}~\bibnamefont {Guo}},\ }\bibfield  {title} {\bibinfo
  {title} {Quantum dual signature scheme based on coherent states with
  entanglement swapping},\ }\href
  {https://doi.org/10.1088/1674-1056/25/8/080306} {\bibfield  {journal}
  {\bibinfo  {journal} {Chinese Physics B}\ }\textbf {\bibinfo {volume} {25}},\
  \bibinfo {eid} {080306} (\bibinfo {year} {2016})}\BibitemShut {NoStop}%
\bibitem [{\citenamefont {Sisodia}\ \emph {et~al.}(2017)\citenamefont
  {Sisodia}, \citenamefont {Verma}, \citenamefont {Thapliyal},\ and\
  \citenamefont {Pathak}}]{Sisodia:qinf:2017}%
  \BibitemOpen
  \bibfield  {author} {\bibinfo {author} {\bibfnamefont {M.}~\bibnamefont
  {Sisodia}}, \bibinfo {author} {\bibfnamefont {V.}~\bibnamefont {Verma}},
  \bibinfo {author} {\bibfnamefont {K.}~\bibnamefont {Thapliyal}},\ and\
  \bibinfo {author} {\bibfnamefont {A.}~\bibnamefont {Pathak}},\ }\bibfield
  {title} {\bibinfo {title} {Teleportation of a qubit using entangled
  non-orthogonal states: a comparative study},\ }\href
  {https://doi.org/10.1007/s11128-017-1526-x} {\bibfield  {journal} {\bibinfo
  {journal} {Quantum Information Processing}\ }\textbf {\bibinfo {volume}
  {16}},\ \bibinfo {eid} {76} (\bibinfo {year} {2017})}\BibitemShut {NoStop}%
\bibitem [{\citenamefont {Miry}(2019)}]{Miry:tmph:2019}%
  \BibitemOpen
  \bibfield  {author} {\bibinfo {author} {\bibfnamefont {S.~R.}\ \bibnamefont
  {Miry}},\ }\bibfield  {title} {\bibinfo {title} {Superposition of entangled
  coherent states: Physical realization and properties},\ }\href
  {https://doi.org/10.1134/S0040577919070055} {\bibfield  {journal} {\bibinfo
  {journal} {Theoretical and Mathematical Physics}\ }\textbf {\bibinfo {volume}
  {100}},\ \bibinfo {pages} {1006} (\bibinfo {year} {2019})}\BibitemShut
  {NoStop}%
\bibitem [{\citenamefont {Ra}\ \emph {et~al.}(2020)\citenamefont {Ra},
  \citenamefont {Dufour}, \citenamefont {Walschaers}, \citenamefont {Jacquard},
  \citenamefont {Michel}, \citenamefont {Fabre},\ and\ \citenamefont
  {Treps}}]{Ra:nature:2020}%
  \BibitemOpen
  \bibfield  {author} {\bibinfo {author} {\bibfnamefont {Y.-S.}\ \bibnamefont
  {Ra}}, \bibinfo {author} {\bibfnamefont {A.}~\bibnamefont {Dufour}}, \bibinfo
  {author} {\bibfnamefont {M.}~\bibnamefont {Walschaers}}, \bibinfo {author}
  {\bibfnamefont {C.}~\bibnamefont {Jacquard}}, \bibinfo {author}
  {\bibfnamefont {T.}~\bibnamefont {Michel}}, \bibinfo {author} {\bibfnamefont
  {C.}~\bibnamefont {Fabre}},\ and\ \bibinfo {author} {\bibfnamefont
  {N.}~\bibnamefont {Treps}},\ }\bibfield  {title} {\bibinfo {title}
  {Non-{G}aussian quantum states of a multimode light field},\ }\href
  {https://doi.org/10.1038/s41567-019-0726-y} {\bibfield  {journal} {\bibinfo
  {journal} {Nature Physics}\ }\textbf {\bibinfo {volume} {16}},\ \bibinfo
  {pages} {144} (\bibinfo {year} {2020})}\BibitemShut {NoStop}%
\bibitem [{\citenamefont {M\'erolla}\ \emph {et~al.}(1999)\citenamefont
  {M\'erolla}, \citenamefont {Mazurenko}, \citenamefont {Goedgebuer},\ and\
  \citenamefont {Rhodes}}]{Merolla:prl:1999}%
  \BibitemOpen
  \bibfield  {author} {\bibinfo {author} {\bibfnamefont {J.-M.}\ \bibnamefont
  {M\'erolla}}, \bibinfo {author} {\bibfnamefont {Y.}~\bibnamefont
  {Mazurenko}}, \bibinfo {author} {\bibfnamefont {J.-P.}\ \bibnamefont
  {Goedgebuer}},\ and\ \bibinfo {author} {\bibfnamefont {W.~T.}\ \bibnamefont
  {Rhodes}},\ }\bibfield  {title} {\bibinfo {title} {Single-photon interference
  in sidebands of phase-modulated light for quantum cryptography},\ }\href
  {https://doi.org/10.1103/PhysRevLett.82.1656} {\bibfield  {journal} {\bibinfo
   {journal} {Phys. Rev. Lett.}\ }\textbf {\bibinfo {volume} {82}},\ \bibinfo
  {pages} {1656} (\bibinfo {year} {1999})}\BibitemShut {NoStop}%
\bibitem [{\citenamefont {Gleim}\ \emph {et~al.}(2016)\citenamefont {Gleim},
  \citenamefont {Egorov}, \citenamefont {Nazarov}, \citenamefont {Smirnov},
  \citenamefont {Chistyakov}, \citenamefont {Bannik}, \citenamefont {Anisimov},
  \citenamefont {Kynev}, \citenamefont {Ivanova}, \citenamefont {Collins},
  \citenamefont {Kozlov},\ and\ \citenamefont {Buller}}]{Gleim:16}%
  \BibitemOpen
  \bibfield  {author} {\bibinfo {author} {\bibfnamefont {A.~V.}\ \bibnamefont
  {Gleim}}, \bibinfo {author} {\bibfnamefont {V.~I.}\ \bibnamefont {Egorov}},
  \bibinfo {author} {\bibfnamefont {Y.~V.}\ \bibnamefont {Nazarov}}, \bibinfo
  {author} {\bibfnamefont {S.~V.}\ \bibnamefont {Smirnov}}, \bibinfo {author}
  {\bibfnamefont {V.~V.}\ \bibnamefont {Chistyakov}}, \bibinfo {author}
  {\bibfnamefont {O.~I.}\ \bibnamefont {Bannik}}, \bibinfo {author}
  {\bibfnamefont {A.~A.}\ \bibnamefont {Anisimov}}, \bibinfo {author}
  {\bibfnamefont {S.~M.}\ \bibnamefont {Kynev}}, \bibinfo {author}
  {\bibfnamefont {A.~E.}\ \bibnamefont {Ivanova}}, \bibinfo {author}
  {\bibfnamefont {R.~J.}\ \bibnamefont {Collins}}, \bibinfo {author}
  {\bibfnamefont {S.~A.}\ \bibnamefont {Kozlov}},\ and\ \bibinfo {author}
  {\bibfnamefont {G.~S.}\ \bibnamefont {Buller}},\ }\bibfield  {title}
  {\bibinfo {title} {Secure polarization-independent subcarrier quantum key
  distribution in optical fiber channel using bb84 protocol with a strong
  reference},\ }\href {https://doi.org/10.1364/OE.24.002619} {\bibfield
  {journal} {\bibinfo  {journal} {Opt. Express}\ }\textbf {\bibinfo {volume}
  {24}},\ \bibinfo {pages} {2619} (\bibinfo {year} {2016})}\BibitemShut
  {NoStop}%
\bibitem [{\citenamefont {Miroshnichenko}\ \emph {et~al.}(2018)\citenamefont
  {Miroshnichenko}, \citenamefont {Kozubov}, \citenamefont {Gaidash},
  \citenamefont {Gleim},\ and\ \citenamefont
  {Horoshko}}]{Miroshnichenko:optexp:2018}%
  \BibitemOpen
  \bibfield  {author} {\bibinfo {author} {\bibfnamefont {G.~P.}\ \bibnamefont
  {Miroshnichenko}}, \bibinfo {author} {\bibfnamefont {A.~V.}\ \bibnamefont
  {Kozubov}}, \bibinfo {author} {\bibfnamefont {A.~A.}\ \bibnamefont
  {Gaidash}}, \bibinfo {author} {\bibfnamefont {A.~V.}\ \bibnamefont {Gleim}},\
  and\ \bibinfo {author} {\bibfnamefont {D.~B.}\ \bibnamefont {Horoshko}},\
  }\bibfield  {title} {\bibinfo {title} {Security of subcarrier wave quantum
  key distribution against the collective beam-splitting attack},\ }\href
  {https://doi.org/10.1364/OE.26.011292} {\bibfield  {journal} {\bibinfo
  {journal} {Opt. Express}\ }\textbf {\bibinfo {volume} {26}},\ \bibinfo
  {pages} {11292} (\bibinfo {year} {2018})}\BibitemShut {NoStop}%
\bibitem [{\citenamefont {Gaidash}\ \emph {et~al.}(2022)\citenamefont
  {Gaidash}, \citenamefont {Miroshnichenko},\ and\ \citenamefont
  {Kozubov}}]{Gaidash:22}%
  \BibitemOpen
  \bibfield  {author} {\bibinfo {author} {\bibfnamefont {A.}~\bibnamefont
  {Gaidash}}, \bibinfo {author} {\bibfnamefont {G.}~\bibnamefont
  {Miroshnichenko}},\ and\ \bibinfo {author} {\bibfnamefont {A.}~\bibnamefont
  {Kozubov}},\ }\bibfield  {title} {\bibinfo {title} {Subcarrier wave quantum
  key distribution with leaky and flawed devices},\ }\href
  {https://doi.org/10.1364/JOSAB.439776} {\bibfield  {journal} {\bibinfo
  {journal} {J. Opt. Soc. Am. B}\ }\textbf {\bibinfo {volume} {39}},\ \bibinfo
  {pages} {577} (\bibinfo {year} {2022})}\BibitemShut {NoStop}%
\bibitem [{\citenamefont {Bannik}\ and\ \citenamefont
  {Moiseev}(2021)}]{Bannik:21}%
  \BibitemOpen
  \bibfield  {author} {\bibinfo {author} {\bibfnamefont {O.~I.}\ \bibnamefont
  {Bannik}}\ and\ \bibinfo {author} {\bibfnamefont {E.~S.}\ \bibnamefont
  {Moiseev}},\ }\bibfield  {title} {\bibinfo {title} {Plug\&play subcarrier
  wave quantum key distribution with deep modulation},\ }\href
  {https://doi.org/10.1364/OE.441619} {\bibfield  {journal} {\bibinfo
  {journal} {Opt. Express}\ }\textbf {\bibinfo {volume} {29}},\ \bibinfo
  {pages} {38858} (\bibinfo {year} {2021})}\BibitemShut {NoStop}%
\bibitem [{\citenamefont {Mel'nik}\ \emph {et~al.}(2018)\citenamefont
  {Mel'nik}, \citenamefont {Arslanov}, \citenamefont {Bannik}, \citenamefont
  {Gilyazov}, \citenamefont {Egorov}, \citenamefont {Gleim},\ and\
  \citenamefont {Moiseev}}]{Melnik2018}%
  \BibitemOpen
  \bibfield  {author} {\bibinfo {author} {\bibfnamefont {K.~S.}\ \bibnamefont
  {Mel'nik}}, \bibinfo {author} {\bibfnamefont {N.~M.}\ \bibnamefont
  {Arslanov}}, \bibinfo {author} {\bibfnamefont {O.~I.}\ \bibnamefont
  {Bannik}}, \bibinfo {author} {\bibfnamefont {L.~R.}\ \bibnamefont
  {Gilyazov}}, \bibinfo {author} {\bibfnamefont {V.~I.}\ \bibnamefont
  {Egorov}}, \bibinfo {author} {\bibfnamefont {A.~V.}\ \bibnamefont {Gleim}},\
  and\ \bibinfo {author} {\bibfnamefont {S.~A.}\ \bibnamefont {Moiseev}},\
  }\bibfield  {title} {\bibinfo {title} {Using a heterodyne detection scheme in
  a subcarrier wave quantum communication system},\ }\href
  {https://doi.org/10.3103/S1062873818080294} {\bibfield  {journal} {\bibinfo
  {journal} {Bulletin of the Russian Academy of Sciences: Physics}\ }\textbf
  {\bibinfo {volume} {82}},\ \bibinfo {pages} {1038} (\bibinfo {year}
  {2018})}\BibitemShut {NoStop}%
\bibitem [{\citenamefont {Samsonov}\ \emph {et~al.}(2020)\citenamefont
  {Samsonov}, \citenamefont {Goncharov}, \citenamefont {Gaidash}, \citenamefont
  {Kozubov}, \citenamefont {Egorov},\ and\ \citenamefont
  {Gleim}}]{Samsonov:scirep:2020}%
  \BibitemOpen
  \bibfield  {author} {\bibinfo {author} {\bibfnamefont {E.}~\bibnamefont
  {Samsonov}}, \bibinfo {author} {\bibfnamefont {R.}~\bibnamefont {Goncharov}},
  \bibinfo {author} {\bibfnamefont {A.}~\bibnamefont {Gaidash}}, \bibinfo
  {author} {\bibfnamefont {A.}~\bibnamefont {Kozubov}}, \bibinfo {author}
  {\bibfnamefont {V.}~\bibnamefont {Egorov}},\ and\ \bibinfo {author}
  {\bibfnamefont {A.}~\bibnamefont {Gleim}},\ }\bibfield  {title} {\bibinfo
  {title} {Subcarrier wave continuous variable quantum key distribution with
  discrete modulation: {M}athematical model and finite-key analysis},\ }\href
  {https://doi.org/10.1038/s41598-020-66948-0} {\bibfield  {journal} {\bibinfo
  {journal} {Scientific Reports}\ }\textbf {\bibinfo {volume} {10}},\ \bibinfo
  {pages} {10034} (\bibinfo {year} {2020})}\BibitemShut {NoStop}%
\bibitem [{\citenamefont {Samsonov}\ \emph {et~al.}(2021)\citenamefont
  {Samsonov}, \citenamefont {Goncharov}, \citenamefont {Fadeev}, \citenamefont
  {Zinoviev}, \citenamefont {Kirichenko}, \citenamefont {Nasedkin},
  \citenamefont {Kiselev},\ and\ \citenamefont {Egorov}}]{Samsonov:josab:2021}%
  \BibitemOpen
  \bibfield  {author} {\bibinfo {author} {\bibfnamefont {E.}~\bibnamefont
  {Samsonov}}, \bibinfo {author} {\bibfnamefont {R.}~\bibnamefont {Goncharov}},
  \bibinfo {author} {\bibfnamefont {M.}~\bibnamefont {Fadeev}}, \bibinfo
  {author} {\bibfnamefont {A.}~\bibnamefont {Zinoviev}}, \bibinfo {author}
  {\bibfnamefont {D.}~\bibnamefont {Kirichenko}}, \bibinfo {author}
  {\bibfnamefont {B.}~\bibnamefont {Nasedkin}}, \bibinfo {author}
  {\bibfnamefont {A.}~\bibnamefont {Kiselev}},\ and\ \bibinfo {author}
  {\bibfnamefont {V.}~\bibnamefont {Egorov}},\ }\bibfield  {title} {\bibinfo
  {title} {Coherent detection schemes for subcarrier wave continuous variable
  quantum key distribution},\ }\href@noop {} {\bibfield  {journal} {\bibinfo
  {journal} {Journal of the Optical Society of America B}\ }\textbf {\bibinfo
  {volume} {38}},\ \bibinfo {pages} {2215} (\bibinfo {year}
  {2021})}\BibitemShut {NoStop}%
\bibitem [{\citenamefont {Chistiakov}\ \emph {et~al.}(2019)\citenamefont
  {Chistiakov}, \citenamefont {Kozubov}, \citenamefont {Gaidash}, \citenamefont
  {Gleim},\ and\ \citenamefont {Miroshnichenko}}]{Chistiakov:19}%
  \BibitemOpen
  \bibfield  {author} {\bibinfo {author} {\bibfnamefont {V.}~\bibnamefont
  {Chistiakov}}, \bibinfo {author} {\bibfnamefont {A.}~\bibnamefont {Kozubov}},
  \bibinfo {author} {\bibfnamefont {A.}~\bibnamefont {Gaidash}}, \bibinfo
  {author} {\bibfnamefont {A.}~\bibnamefont {Gleim}},\ and\ \bibinfo {author}
  {\bibfnamefont {G.}~\bibnamefont {Miroshnichenko}},\ }\bibfield  {title}
  {\bibinfo {title} {Feasibility of twin-field quantum key distribution based
  on multi-mode coherent phase-coded states},\ }\href
  {https://doi.org/10.1364/OE.27.036551} {\bibfield  {journal} {\bibinfo
  {journal} {Opt. Express}\ }\textbf {\bibinfo {volume} {27}},\ \bibinfo
  {pages} {36551} (\bibinfo {year} {2019})}\BibitemShut {NoStop}%
\bibitem [{\citenamefont {Saglamyurek}\ \emph {et~al.}(2011)\citenamefont
  {Saglamyurek}, \citenamefont {Sinclair}, \citenamefont {Jin}, \citenamefont
  {Slater}, \citenamefont {Oblak}, \citenamefont {Bussi\'eres}, \citenamefont
  {George}, \citenamefont {Ricken}, \citenamefont {Sohler},\ and\ \citenamefont
  {Tittel}}]{Saglamyurek:nature:2011}%
  \BibitemOpen
  \bibfield  {author} {\bibinfo {author} {\bibfnamefont {E.}~\bibnamefont
  {Saglamyurek}}, \bibinfo {author} {\bibfnamefont {N.}~\bibnamefont
  {Sinclair}}, \bibinfo {author} {\bibfnamefont {J.}~\bibnamefont {Jin}},
  \bibinfo {author} {\bibfnamefont {J.~A.}\ \bibnamefont {Slater}}, \bibinfo
  {author} {\bibfnamefont {D.}~\bibnamefont {Oblak}}, \bibinfo {author}
  {\bibfnamefont {F.}~\bibnamefont {Bussi\'eres}}, \bibinfo {author}
  {\bibfnamefont {M.}~\bibnamefont {George}}, \bibinfo {author} {\bibfnamefont
  {R.}~\bibnamefont {Ricken}}, \bibinfo {author} {\bibfnamefont
  {W.}~\bibnamefont {Sohler}},\ and\ \bibinfo {author} {\bibfnamefont
  {W.}~\bibnamefont {Tittel}},\ }\bibfield  {title} {\bibinfo {title}
  {Broadband waveguide quantum memory for entangled photons},\ }\href
  {https://doi.org/10.1038/nature09719} {\bibfield  {journal} {\bibinfo
  {journal} {Nature}\ }\textbf {\bibinfo {volume} {469}},\ \bibinfo {pages}
  {512} (\bibinfo {year} {2011})}\BibitemShut {NoStop}%
\bibitem [{\citenamefont {Sinclair}\ \emph {et~al.}(2014)\citenamefont
  {Sinclair}, \citenamefont {Saglamyurek}, \citenamefont {Mallahzadeh},
  \citenamefont {Slater}, \citenamefont {George}, \citenamefont {Ricken},
  \citenamefont {Hedges}, \citenamefont {Oblak}, \citenamefont {Simon},
  \citenamefont {Sohler},\ and\ \citenamefont {Tittel}}]{Sinclair:prl:2014}%
  \BibitemOpen
  \bibfield  {author} {\bibinfo {author} {\bibfnamefont {N.}~\bibnamefont
  {Sinclair}}, \bibinfo {author} {\bibfnamefont {E.}~\bibnamefont
  {Saglamyurek}}, \bibinfo {author} {\bibfnamefont {H.}~\bibnamefont
  {Mallahzadeh}}, \bibinfo {author} {\bibfnamefont {J.~A.}\ \bibnamefont
  {Slater}}, \bibinfo {author} {\bibfnamefont {M.}~\bibnamefont {George}},
  \bibinfo {author} {\bibfnamefont {R.}~\bibnamefont {Ricken}}, \bibinfo
  {author} {\bibfnamefont {M.~P.}\ \bibnamefont {Hedges}}, \bibinfo {author}
  {\bibfnamefont {D.}~\bibnamefont {Oblak}}, \bibinfo {author} {\bibfnamefont
  {C.}~\bibnamefont {Simon}}, \bibinfo {author} {\bibfnamefont
  {W.}~\bibnamefont {Sohler}},\ and\ \bibinfo {author} {\bibfnamefont
  {W.}~\bibnamefont {Tittel}},\ }\bibfield  {title} {\bibinfo {title} {Spectral
  multiplexing for scalable quantum photonics using an atomic frequency comb
  quantum memory and feed-forward control},\ }\href
  {https://doi.org/10.1103/PhysRevLett.113.053603} {\bibfield  {journal}
  {\bibinfo  {journal} {Phys. Rev. Lett.}\ }\textbf {\bibinfo {volume} {113}},\
  \bibinfo {pages} {053603} (\bibinfo {year} {2014})}\BibitemShut {NoStop}%
\bibitem [{\citenamefont {Moiseev}\ and\ \citenamefont
  {Gleim}(2016)}]{Moiseev2016}%
  \BibitemOpen
  \bibfield  {author} {\bibinfo {author} {\bibfnamefont {S.~A.}\ \bibnamefont
  {Moiseev}}\ and\ \bibinfo {author} {\bibfnamefont {A.~V.}\ \bibnamefont
  {Gleim}},\ }\href@noop {} {\emph {\bibinfo {title} {25th Annual International
  Laser Physics Workshop (LPHYS'16)}}},\ \bibinfo {type} {Tech. Rep.}\
  (\bibinfo  {institution} {Yerevan Physics Institute},\ \bibinfo {address}
  {Yerevan},\ \bibinfo {year} {2016})\BibitemShut {NoStop}%
\bibitem [{\citenamefont {Kaczmarek}\ \emph {et~al.}(2018)\citenamefont
  {Kaczmarek}, \citenamefont {Ledingham}, \citenamefont {Brecht}, \citenamefont
  {Thomas}, \citenamefont {Thekkadath}, \citenamefont {Lazo-Arjona},
  \citenamefont {Munns}, \citenamefont {Poem}, \citenamefont {Feizpour},
  \citenamefont {Saunders}, \citenamefont {Nunn},\ and\ \citenamefont
  {Walmsley}}]{Kaczmarek:pra:2018}%
  \BibitemOpen
  \bibfield  {author} {\bibinfo {author} {\bibfnamefont {K.~T.}\ \bibnamefont
  {Kaczmarek}}, \bibinfo {author} {\bibfnamefont {P.~M.}\ \bibnamefont
  {Ledingham}}, \bibinfo {author} {\bibfnamefont {B.}~\bibnamefont {Brecht}},
  \bibinfo {author} {\bibfnamefont {S.~E.}\ \bibnamefont {Thomas}}, \bibinfo
  {author} {\bibfnamefont {G.~S.}\ \bibnamefont {Thekkadath}}, \bibinfo
  {author} {\bibfnamefont {O.}~\bibnamefont {Lazo-Arjona}}, \bibinfo {author}
  {\bibfnamefont {J.~H.~D.}\ \bibnamefont {Munns}}, \bibinfo {author}
  {\bibfnamefont {E.}~\bibnamefont {Poem}}, \bibinfo {author} {\bibfnamefont
  {A.}~\bibnamefont {Feizpour}}, \bibinfo {author} {\bibfnamefont {D.~J.}\
  \bibnamefont {Saunders}}, \bibinfo {author} {\bibfnamefont {J.}~\bibnamefont
  {Nunn}},\ and\ \bibinfo {author} {\bibfnamefont {I.~A.}\ \bibnamefont
  {Walmsley}},\ }\bibfield  {title} {\bibinfo {title} {High-speed noise-free
  optical quantum memory},\ }\href {https://doi.org/10.1103/PhysRevA.97.042316}
  {\bibfield  {journal} {\bibinfo  {journal} {Phys. Rev. A}\ }\textbf {\bibinfo
  {volume} {97}},\ \bibinfo {pages} {042316} (\bibinfo {year}
  {2018})}\BibitemShut {NoStop}%
\bibitem [{\citenamefont {Ikuta}\ \emph {et~al.}(2018)\citenamefont {Ikuta},
  \citenamefont {Kobayashi}, \citenamefont {Kawakami}, \citenamefont {Miki},
  \citenamefont {Yabuno}, \citenamefont {Yamashita}, \citenamefont {Terai},
  \citenamefont {Koashi}, \citenamefont {Mukai}, \citenamefont {Yamamoto},\
  and\ \citenamefont {Imoto}}]{Ikuta:nature:2018}%
  \BibitemOpen
  \bibfield  {author} {\bibinfo {author} {\bibfnamefont {R.}~\bibnamefont
  {Ikuta}}, \bibinfo {author} {\bibfnamefont {T.}~\bibnamefont {Kobayashi}},
  \bibinfo {author} {\bibfnamefont {T.}~\bibnamefont {Kawakami}}, \bibinfo
  {author} {\bibfnamefont {S.}~\bibnamefont {Miki}}, \bibinfo {author}
  {\bibfnamefont {M.}~\bibnamefont {Yabuno}}, \bibinfo {author} {\bibfnamefont
  {T.}~\bibnamefont {Yamashita}}, \bibinfo {author} {\bibfnamefont
  {H.}~\bibnamefont {Terai}}, \bibinfo {author} {\bibfnamefont
  {M.}~\bibnamefont {Koashi}}, \bibinfo {author} {\bibfnamefont
  {T.}~\bibnamefont {Mukai}}, \bibinfo {author} {\bibfnamefont
  {T.}~\bibnamefont {Yamamoto}},\ and\ \bibinfo {author} {\bibfnamefont
  {N.}~\bibnamefont {Imoto}},\ }\bibfield  {title} {\bibinfo {title}
  {Polarization insensitive frequency conversion for an atom-photon
  entanglement distribution via a telecom network},\ }\href
  {https://doi.org/10.1038/s41467-018-04338-x} {\bibfield  {journal} {\bibinfo
  {journal} {Nature Communications}\ }\textbf {\bibinfo {volume} {9}},\
  \bibinfo {pages} {1997} (\bibinfo {year} {2018})}\BibitemShut {NoStop}%
\bibitem [{\citenamefont {Davidson}\ \emph {et~al.}(2020)\citenamefont
  {Davidson}, \citenamefont {Lefebvre}, \citenamefont {Zhang}, \citenamefont
  {Oblak},\ and\ \citenamefont {Tittel}}]{Davidson:pra:2020}%
  \BibitemOpen
  \bibfield  {author} {\bibinfo {author} {\bibfnamefont {J.~H.}\ \bibnamefont
  {Davidson}}, \bibinfo {author} {\bibfnamefont {P.}~\bibnamefont {Lefebvre}},
  \bibinfo {author} {\bibfnamefont {J.}~\bibnamefont {Zhang}}, \bibinfo
  {author} {\bibfnamefont {D.}~\bibnamefont {Oblak}},\ and\ \bibinfo {author}
  {\bibfnamefont {W.}~\bibnamefont {Tittel}},\ }\bibfield  {title} {\bibinfo
  {title} {Improved light-matter interaction for storage of quantum states of
  light in a thulium-doped crystal cavity},\ }\href
  {https://doi.org/10.1103/PhysRevA.101.042333} {\bibfield  {journal} {\bibinfo
   {journal} {Phys. Rev. A}\ }\textbf {\bibinfo {volume} {101}},\ \bibinfo
  {pages} {042333} (\bibinfo {year} {2020})}\BibitemShut {NoStop}%
\bibitem [{\citenamefont {Moiseev}\ \emph {et~al.}(2021)\citenamefont
  {Moiseev}, \citenamefont {Tashchilina}, \citenamefont {Moiseev},\ and\
  \citenamefont {Sanders}}]{Moiseev_2021}%
  \BibitemOpen
  \bibfield  {author} {\bibinfo {author} {\bibfnamefont {E.~S.}\ \bibnamefont
  {Moiseev}}, \bibinfo {author} {\bibfnamefont {A.}~\bibnamefont
  {Tashchilina}}, \bibinfo {author} {\bibfnamefont {S.~A.}\ \bibnamefont
  {Moiseev}},\ and\ \bibinfo {author} {\bibfnamefont {B.~C.}\ \bibnamefont
  {Sanders}},\ }\bibfield  {title} {\bibinfo {title} {Broadband quantum memory
  in a cavity via zero spectral dispersion},\ }\href
  {https://doi.org/10.1088/1367-2630/ac0754} {\bibfield  {journal} {\bibinfo
  {journal} {New Journal of Physics}\ }\textbf {\bibinfo {volume} {23}},\
  \bibinfo {pages} {063071} (\bibinfo {year} {2021})}\BibitemShut {NoStop}%
\bibitem [{\citenamefont {Lago-Rivera}\ \emph {et~al.}(2021)\citenamefont
  {Lago-Rivera}, \citenamefont {Grandi}, \citenamefont {Rakonjac},
  \citenamefont {Seri},\ and\ \citenamefont
  {de~Riedmatten}}]{Lago:nature:2021}%
  \BibitemOpen
  \bibfield  {author} {\bibinfo {author} {\bibfnamefont {D.}~\bibnamefont
  {Lago-Rivera}}, \bibinfo {author} {\bibfnamefont {S.}~\bibnamefont {Grandi}},
  \bibinfo {author} {\bibfnamefont {J.~V.}\ \bibnamefont {Rakonjac}}, \bibinfo
  {author} {\bibfnamefont {A.}~\bibnamefont {Seri}},\ and\ \bibinfo {author}
  {\bibfnamefont {H.}~\bibnamefont {de~Riedmatten}},\ }\bibfield  {title}
  {\bibinfo {title} {Telecom-heralded entanglement between multimode
  solid-state quantum memories},\ }\href
  {https://doi.org/10.1038/s41586-021-03481-8} {\bibfield  {journal} {\bibinfo
  {journal} {Nature}\ }\textbf {\bibinfo {volume} {594}},\ \bibinfo {pages}
  {37} (\bibinfo {year} {2021})}\BibitemShut {NoStop}%
\bibitem [{\citenamefont {Askarani}\ \emph {et~al.}(2021)\citenamefont
  {Askarani}, \citenamefont {Das}, \citenamefont {Davidson}, \citenamefont
  {Amaral}, \citenamefont {Sinclair}, \citenamefont {Slater}, \citenamefont
  {Marzban}, \citenamefont {Thiel}, \citenamefont {Cone}, \citenamefont
  {Oblak},\ and\ \citenamefont {Tittel}}]{Askarani:prl:2021}%
  \BibitemOpen
  \bibfield  {author} {\bibinfo {author} {\bibfnamefont {M.~F.}\ \bibnamefont
  {Askarani}}, \bibinfo {author} {\bibfnamefont {A.}~\bibnamefont {Das}},
  \bibinfo {author} {\bibfnamefont {J.~H.}\ \bibnamefont {Davidson}}, \bibinfo
  {author} {\bibfnamefont {G.~C.}\ \bibnamefont {Amaral}}, \bibinfo {author}
  {\bibfnamefont {N.}~\bibnamefont {Sinclair}}, \bibinfo {author}
  {\bibfnamefont {J.~A.}\ \bibnamefont {Slater}}, \bibinfo {author}
  {\bibfnamefont {S.}~\bibnamefont {Marzban}}, \bibinfo {author} {\bibfnamefont
  {C.~W.}\ \bibnamefont {Thiel}}, \bibinfo {author} {\bibfnamefont {R.~L.}\
  \bibnamefont {Cone}}, \bibinfo {author} {\bibfnamefont {D.}~\bibnamefont
  {Oblak}},\ and\ \bibinfo {author} {\bibfnamefont {W.}~\bibnamefont
  {Tittel}},\ }\bibfield  {title} {\bibinfo {title} {Long-lived solid-state
  optical memory for high-rate quantum repeaters},\ }\href
  {https://doi.org/10.1103/PhysRevLett.127.220502} {\bibfield  {journal}
  {\bibinfo  {journal} {Phys. Rev. Lett.}\ }\textbf {\bibinfo {volume} {127}},\
  \bibinfo {pages} {220502} (\bibinfo {year} {2021})}\BibitemShut {NoStop}%
\bibitem [{\citenamefont {Wang}\ \emph {et~al.}(2021)\citenamefont {Wang},
  \citenamefont {Pietx-Casas}, \citenamefont {Askarani},\ and\ \citenamefont
  {do~Amaral}}]{Wang:josab:2021}%
  \BibitemOpen
  \bibfield  {author} {\bibinfo {author} {\bibfnamefont {P.-C.}\ \bibnamefont
  {Wang}}, \bibinfo {author} {\bibfnamefont {O.}~\bibnamefont {Pietx-Casas}},
  \bibinfo {author} {\bibfnamefont {M.~F.}\ \bibnamefont {Askarani}},\ and\
  \bibinfo {author} {\bibfnamefont {G.~C.}\ \bibnamefont {do~Amaral}},\
  }\bibfield  {title} {\bibinfo {title} {Proposal and proof-of-principle
  demonstration of fast-switching broadband frequency shifting for a
  frequency-multiplexed quantum repeater},\ }\href
  {https://doi.org/10.1364/JOSAB.412517} {\bibfield  {journal} {\bibinfo
  {journal} {J. Opt. Soc. Am. B}\ }\textbf {\bibinfo {volume} {38}},\ \bibinfo
  {pages} {1140} (\bibinfo {year} {2021})}\BibitemShut {NoStop}%
\bibitem [{\citenamefont {Bustard}\ \emph {et~al.}(2022)\citenamefont
  {Bustard}, \citenamefont {Bonsma-Fisher}, \citenamefont {Hnatovsky},
  \citenamefont {Grobnic}, \citenamefont {Mihailov}, \citenamefont {England},\
  and\ \citenamefont {Sussman}}]{Bustard:prl:2022}%
  \BibitemOpen
  \bibfield  {author} {\bibinfo {author} {\bibfnamefont {P.~J.}\ \bibnamefont
  {Bustard}}, \bibinfo {author} {\bibfnamefont {K.}~\bibnamefont
  {Bonsma-Fisher}}, \bibinfo {author} {\bibfnamefont {C.}~\bibnamefont
  {Hnatovsky}}, \bibinfo {author} {\bibfnamefont {D.}~\bibnamefont {Grobnic}},
  \bibinfo {author} {\bibfnamefont {S.~J.}\ \bibnamefont {Mihailov}}, \bibinfo
  {author} {\bibfnamefont {D.}~\bibnamefont {England}},\ and\ \bibinfo {author}
  {\bibfnamefont {B.~J.}\ \bibnamefont {Sussman}},\ }\bibfield  {title}
  {\bibinfo {title} {Toward a quantum memory in a fiber cavity controlled by
  intracavity frequency translation},\ }\href
  {https://doi.org/10.1103/PhysRevLett.128.120501} {\bibfield  {journal}
  {\bibinfo  {journal} {Phys. Rev. Lett.}\ }\textbf {\bibinfo {volume} {128}},\
  \bibinfo {pages} {120501} (\bibinfo {year} {2022})}\BibitemShut {NoStop}%
\bibitem [{\citenamefont {Businger}\ \emph {et~al.}(2022)\citenamefont
  {Businger}, \citenamefont {Nicolas}, \citenamefont {Mejia}, \citenamefont
  {Ferrier}, \citenamefont {Goldner},\ and\ \citenamefont
  {Afzelius}}]{Businger:nature:2022}%
  \BibitemOpen
  \bibfield  {author} {\bibinfo {author} {\bibfnamefont {M.}~\bibnamefont
  {Businger}}, \bibinfo {author} {\bibfnamefont {L.}~\bibnamefont {Nicolas}},
  \bibinfo {author} {\bibfnamefont {T.~S.}\ \bibnamefont {Mejia}}, \bibinfo
  {author} {\bibfnamefont {A.}~\bibnamefont {Ferrier}}, \bibinfo {author}
  {\bibfnamefont {P.}~\bibnamefont {Goldner}},\ and\ \bibinfo {author}
  {\bibfnamefont {M.}~\bibnamefont {Afzelius}},\ }\bibfield  {title} {\bibinfo
  {title} {Non-classical correlations over 1250 modes between telecom photons
  and 979-nm photons stored in {$^{171}$Yb$^{3+}$:Y$_2$SiO$_{5}$}},\ }\href
  {https://doi.org/10.1038/s41467-022-33929-y} {\bibfield  {journal} {\bibinfo
  {journal} {Nature Communications}\ }\textbf {\bibinfo {volume} {13}},\
  \bibinfo {pages} {6438} (\bibinfo {year} {2022})}\BibitemShut {NoStop}%
\bibitem [{\citenamefont {Dodonov}\ \emph {et~al.}(1974)\citenamefont
  {Dodonov}, \citenamefont {Malkin},\ and\ \citenamefont
  {Man'ko}}]{Dodonov:physica1974}%
  \BibitemOpen
  \bibfield  {author} {\bibinfo {author} {\bibfnamefont {V.~V.}\ \bibnamefont
  {Dodonov}}, \bibinfo {author} {\bibfnamefont {I.~A.}\ \bibnamefont
  {Malkin}},\ and\ \bibinfo {author} {\bibfnamefont {V.~I.}\ \bibnamefont
  {Man'ko}},\ }\bibfield  {title} {\bibinfo {title} {Even and odd coherent
  states and excitations of a singular oscillator},\ }\href
  {https://doi.org/10.1016/0031-8914(74)90215-8} {\bibfield  {journal}
  {\bibinfo  {journal} {Physica}\ }\textbf {\bibinfo {volume} {72}},\ \bibinfo
  {pages} {597} (\bibinfo {year} {1974})}\BibitemShut {NoStop}%
\bibitem [{\citenamefont {Neergaard-Nielsen}\ \emph {et~al.}(2006)\citenamefont
  {Neergaard-Nielsen}, \citenamefont {Nielsen}, \citenamefont {Hettich},
  \citenamefont {M\o{}lmer},\ and\ \citenamefont {Polzik}}]{Polzik:prl:2006}%
  \BibitemOpen
  \bibfield  {author} {\bibinfo {author} {\bibfnamefont {J.~S.}\ \bibnamefont
  {Neergaard-Nielsen}}, \bibinfo {author} {\bibfnamefont {B.~M.}\ \bibnamefont
  {Nielsen}}, \bibinfo {author} {\bibfnamefont {C.}~\bibnamefont {Hettich}},
  \bibinfo {author} {\bibfnamefont {K.}~\bibnamefont {M\o{}lmer}},\ and\
  \bibinfo {author} {\bibfnamefont {E.~S.}\ \bibnamefont {Polzik}},\ }\bibfield
   {title} {\bibinfo {title} {Generation of a superposition of odd photon
  number states for quantum information networks},\ }\href
  {https://doi.org/10.1103/PhysRevLett.97.083604} {\bibfield  {journal}
  {\bibinfo  {journal} {Phys. Rev. Lett.}\ }\textbf {\bibinfo {volume} {97}},\
  \bibinfo {pages} {083604} (\bibinfo {year} {2006})}\BibitemShut {NoStop}%
\bibitem [{\citenamefont {Ourjoumtsev}\ \emph {et~al.}(2009)\citenamefont
  {Ourjoumtsev}, \citenamefont {Ferreyrol}, \citenamefont {Tualle-Brouri},\
  and\ \citenamefont {Grangier}}]{Ourjoumtsev:nature:2009}%
  \BibitemOpen
  \bibfield  {author} {\bibinfo {author} {\bibfnamefont {A.}~\bibnamefont
  {Ourjoumtsev}}, \bibinfo {author} {\bibfnamefont {F.}~\bibnamefont
  {Ferreyrol}}, \bibinfo {author} {\bibfnamefont {R.}~\bibnamefont
  {Tualle-Brouri}},\ and\ \bibinfo {author} {\bibfnamefont {P.}~\bibnamefont
  {Grangier}},\ }\bibfield  {title} {\bibinfo {title} {Preparation of non-local
  superpositions of quasi-classical light states},\ }\href
  {https://doi.org/10.1038/nphys1199} {\bibfield  {journal} {\bibinfo
  {journal} {Nature Physics}\ }\textbf {\bibinfo {volume} {5}},\ \bibinfo
  {pages} {189} (\bibinfo {year} {2009})}\BibitemShut {NoStop}%
\bibitem [{\citenamefont {Lund}\ \emph {et~al.}(2013)\citenamefont {Lund},
  \citenamefont {Ralph},\ and\ \citenamefont {Jeong}}]{Lund:pra:2013}%
  \BibitemOpen
  \bibfield  {author} {\bibinfo {author} {\bibfnamefont {A.~P.}\ \bibnamefont
  {Lund}}, \bibinfo {author} {\bibfnamefont {T.~C.}\ \bibnamefont {Ralph}},\
  and\ \bibinfo {author} {\bibfnamefont {H.}~\bibnamefont {Jeong}},\ }\bibfield
   {title} {\bibinfo {title} {Generation of distributed entangled coherent
  states over a lossy environment with inefficient detectors},\ }\href
  {https://doi.org/10.1103/PhysRevA.88.052335} {\bibfield  {journal} {\bibinfo
  {journal} {Phys. Rev. A}\ }\textbf {\bibinfo {volume} {88}},\ \bibinfo
  {pages} {052335} (\bibinfo {year} {2013})}\BibitemShut {NoStop}%
\bibitem [{\citenamefont {Serikawa}\ \emph {et~al.}(2018)\citenamefont
  {Serikawa}, \citenamefont {Yoshikawa}, \citenamefont {Takeda}, \citenamefont
  {Yonezawa}, \citenamefont {Ralph}, \citenamefont {Huntington},\ and\
  \citenamefont {Furusawa}}]{Serikawa:prl:2018}%
  \BibitemOpen
  \bibfield  {author} {\bibinfo {author} {\bibfnamefont {T.}~\bibnamefont
  {Serikawa}}, \bibinfo {author} {\bibfnamefont {J.-i.}\ \bibnamefont
  {Yoshikawa}}, \bibinfo {author} {\bibfnamefont {S.}~\bibnamefont {Takeda}},
  \bibinfo {author} {\bibfnamefont {H.}~\bibnamefont {Yonezawa}}, \bibinfo
  {author} {\bibfnamefont {T.~C.}\ \bibnamefont {Ralph}}, \bibinfo {author}
  {\bibfnamefont {E.~H.}\ \bibnamefont {Huntington}},\ and\ \bibinfo {author}
  {\bibfnamefont {A.}~\bibnamefont {Furusawa}},\ }\bibfield  {title} {\bibinfo
  {title} {Generation of a cat state in an optical sideband},\ }\href
  {https://doi.org/10.1103/PhysRevLett.121.143602} {\bibfield  {journal}
  {\bibinfo  {journal} {Phys. Rev. Lett.}\ }\textbf {\bibinfo {volume} {121}},\
  \bibinfo {pages} {143602} (\bibinfo {year} {2018})}\BibitemShut {NoStop}%
\bibitem [{\citenamefont {Takase}\ \emph {et~al.}(2021)\citenamefont {Takase},
  \citenamefont {Yoshikawa}, \citenamefont {Asavanant}, \citenamefont {Endo},\
  and\ \citenamefont {Furusawa}}]{Takase:pra:2021}%
  \BibitemOpen
  \bibfield  {author} {\bibinfo {author} {\bibfnamefont {K.}~\bibnamefont
  {Takase}}, \bibinfo {author} {\bibfnamefont {J.-i.}\ \bibnamefont
  {Yoshikawa}}, \bibinfo {author} {\bibfnamefont {W.}~\bibnamefont
  {Asavanant}}, \bibinfo {author} {\bibfnamefont {M.}~\bibnamefont {Endo}},\
  and\ \bibinfo {author} {\bibfnamefont {A.}~\bibnamefont {Furusawa}},\
  }\bibfield  {title} {\bibinfo {title} {Generation of optical {S}chr\"odinger
  cat states by generalized photon subtraction},\ }\href
  {https://doi.org/10.1103/PhysRevA.103.013710} {\bibfield  {journal} {\bibinfo
   {journal} {Phys. Rev. A}\ }\textbf {\bibinfo {volume} {103}},\ \bibinfo
  {pages} {013710} (\bibinfo {year} {2021})}\BibitemShut {NoStop}%
\bibitem [{\citenamefont {Ourjoumtsev}\ \emph {et~al.}(2007)\citenamefont
  {Ourjoumtsev}, \citenamefont {Jeong}, \citenamefont {Tualle-Brouri},\ and\
  \citenamefont {Grangier}}]{Ourjoumtsev:nature:2007}%
  \BibitemOpen
  \bibfield  {author} {\bibinfo {author} {\bibfnamefont {A.}~\bibnamefont
  {Ourjoumtsev}}, \bibinfo {author} {\bibfnamefont {H.}~\bibnamefont {Jeong}},
  \bibinfo {author} {\bibfnamefont {R.}~\bibnamefont {Tualle-Brouri}},\ and\
  \bibinfo {author} {\bibfnamefont {P.}~\bibnamefont {Grangier}},\ }\bibfield
  {title} {\bibinfo {title} {Generation of optical {‘Schr\"odinger cats’}
  from photon number states},\ }\href {https://doi.org/10.1038/nature06054}
  {\bibfield  {journal} {\bibinfo  {journal} {Nature}\ }\textbf {\bibinfo
  {volume} {448}},\ \bibinfo {pages} {784} (\bibinfo {year}
  {2007})}\BibitemShut {NoStop}%
\bibitem [{\citenamefont {Puri}\ \emph {et~al.}(2017)\citenamefont {Puri},
  \citenamefont {Boutin},\ and\ \citenamefont {Blais}}]{Puri2017}%
  \BibitemOpen
  \bibfield  {author} {\bibinfo {author} {\bibfnamefont {S.}~\bibnamefont
  {Puri}}, \bibinfo {author} {\bibfnamefont {S.}~\bibnamefont {Boutin}},\ and\
  \bibinfo {author} {\bibfnamefont {A.}~\bibnamefont {Blais}},\ }\bibfield
  {title} {\bibinfo {title} {Engineering the quantum states of light in a
  {Kerr}-nonlinear resonator by two-photon driving},\ }\href
  {https://doi.org/10.1038/s41534-017-0019-1} {\bibfield  {journal} {\bibinfo
  {journal} {npj Quantum Information}\ }\textbf {\bibinfo {volume} {3}},\
  \bibinfo {pages} {18} (\bibinfo {year} {2017})}\BibitemShut {NoStop}%
\bibitem [{\citenamefont {Moiseev}\ \emph {et~al.}(2020)\citenamefont
  {Moiseev}, \citenamefont {Tashchilina}, \citenamefont {Moiseev},\ and\
  \citenamefont {Lvovsky}}]{Moiseev_2020}%
  \BibitemOpen
  \bibfield  {author} {\bibinfo {author} {\bibfnamefont {E.~S.}\ \bibnamefont
  {Moiseev}}, \bibinfo {author} {\bibfnamefont {A.}~\bibnamefont
  {Tashchilina}}, \bibinfo {author} {\bibfnamefont {S.~A.}\ \bibnamefont
  {Moiseev}},\ and\ \bibinfo {author} {\bibfnamefont {A.~I.}\ \bibnamefont
  {Lvovsky}},\ }\bibfield  {title} {\bibinfo {title} {Darkness of two-mode
  squeezed light in $\lambda$-type atomic system},\ }\href
  {https://doi.org/10.1088/1367-2630/ab5fac} {\bibfield  {journal} {\bibinfo
  {journal} {New Journal of Physics}\ }\textbf {\bibinfo {volume} {22}},\
  \bibinfo {pages} {013014} (\bibinfo {year} {2020})}\BibitemShut {NoStop}%
\bibitem [{\citenamefont {Grimm}\ \emph {et~al.}(2020)\citenamefont {Grimm},
  \citenamefont {Frattini}, \citenamefont {Puri}, \citenamefont {Mundhada},
  \citenamefont {Touzard}, \citenamefont {Mirrahimi}, \citenamefont {Girvin},
  \citenamefont {Shankar},\ and\ \citenamefont {Devoret}}]{Grimm2020}%
  \BibitemOpen
  \bibfield  {author} {\bibinfo {author} {\bibfnamefont {A.}~\bibnamefont
  {Grimm}}, \bibinfo {author} {\bibfnamefont {N.~E.}\ \bibnamefont {Frattini}},
  \bibinfo {author} {\bibfnamefont {S.}~\bibnamefont {Puri}}, \bibinfo {author}
  {\bibfnamefont {S.~O.}\ \bibnamefont {Mundhada}}, \bibinfo {author}
  {\bibfnamefont {S.}~\bibnamefont {Touzard}}, \bibinfo {author} {\bibfnamefont
  {M.}~\bibnamefont {Mirrahimi}}, \bibinfo {author} {\bibfnamefont {S.~M.}\
  \bibnamefont {Girvin}}, \bibinfo {author} {\bibfnamefont {S.}~\bibnamefont
  {Shankar}},\ and\ \bibinfo {author} {\bibfnamefont {M.~H.}\ \bibnamefont
  {Devoret}},\ }\bibfield  {title} {\bibinfo {title} {Stabilization and
  operation of a {Kerr}-cat qubit},\ }\href
  {https://doi.org/10.1038/s41586-020-2587-z} {\bibfield  {journal} {\bibinfo
  {journal} {Nature}\ }\textbf {\bibinfo {volume} {584}},\ \bibinfo {pages}
  {205} (\bibinfo {year} {2020})}\BibitemShut {NoStop}%
\bibitem [{\citenamefont {Sychev}\ \emph {et~al.}(2018)\citenamefont {Sychev},
  \citenamefont {Ulanov}, \citenamefont {Pushkina}, \citenamefont {Fedorov},
  \citenamefont {Richards}, \citenamefont {Grangier},\ and\ \citenamefont
  {Lvovsky}}]{Sychev:aipconf:2018}%
  \BibitemOpen
  \bibfield  {author} {\bibinfo {author} {\bibfnamefont {D.~V.}\ \bibnamefont
  {Sychev}}, \bibinfo {author} {\bibfnamefont {A.~E.}\ \bibnamefont {Ulanov}},
  \bibinfo {author} {\bibfnamefont {A.~A.}\ \bibnamefont {Pushkina}}, \bibinfo
  {author} {\bibfnamefont {I.~A.}\ \bibnamefont {Fedorov}}, \bibinfo {author}
  {\bibfnamefont {M.~W.}\ \bibnamefont {Richards}}, \bibinfo {author}
  {\bibfnamefont {P.}~\bibnamefont {Grangier}},\ and\ \bibinfo {author}
  {\bibfnamefont {A.~I.}\ \bibnamefont {Lvovsky}},\ }\bibfield  {title}
  {\bibinfo {title} {Generating and breeding optical {S}chr\"odinger’s cat
  states},\ }\href {https://doi.org/10.1063/1.5025456} {\bibfield  {journal}
  {\bibinfo  {journal} {AIP Conference Proceedings}\ }\textbf {\bibinfo
  {volume} {1936}},\ \bibinfo {pages} {020018} (\bibinfo {year}
  {2018})}\BibitemShut {NoStop}%
\bibitem [{\citenamefont {Wang}\ \emph {et~al.}(2022)\citenamefont {Wang},
  \citenamefont {Bao}, \citenamefont {Wu}, \citenamefont {Li}, \citenamefont
  {Cai}, \citenamefont {Wang}, \citenamefont {Ma}, \citenamefont {Cai},
  \citenamefont {Han}, \citenamefont {Wang}, \citenamefont {Song},
  \citenamefont {Sun}, \citenamefont {Zhang},\ and\ \citenamefont
  {Duan}}]{Zhiling:sciadv:2022}%
  \BibitemOpen
  \bibfield  {author} {\bibinfo {author} {\bibfnamefont {Z.}~\bibnamefont
  {Wang}}, \bibinfo {author} {\bibfnamefont {Z.}~\bibnamefont {Bao}}, \bibinfo
  {author} {\bibfnamefont {Y.}~\bibnamefont {Wu}}, \bibinfo {author}
  {\bibfnamefont {Y.}~\bibnamefont {Li}}, \bibinfo {author} {\bibfnamefont
  {W.}~\bibnamefont {Cai}}, \bibinfo {author} {\bibfnamefont {W.}~\bibnamefont
  {Wang}}, \bibinfo {author} {\bibfnamefont {Y.}~\bibnamefont {Ma}}, \bibinfo
  {author} {\bibfnamefont {T.}~\bibnamefont {Cai}}, \bibinfo {author}
  {\bibfnamefont {X.}~\bibnamefont {Han}}, \bibinfo {author} {\bibfnamefont
  {J.}~\bibnamefont {Wang}}, \bibinfo {author} {\bibfnamefont {Y.}~\bibnamefont
  {Song}}, \bibinfo {author} {\bibfnamefont {L.}~\bibnamefont {Sun}}, \bibinfo
  {author} {\bibfnamefont {H.}~\bibnamefont {Zhang}},\ and\ \bibinfo {author}
  {\bibfnamefont {L.}~\bibnamefont {Duan}},\ }\bibfield  {title} {\bibinfo
  {title} {A flying {S}chr\"odinger's cat in multipartite entangled states},\
  }\href {https://doi.org/10.1126/sciadv.abn1778} {\bibfield  {journal}
  {\bibinfo  {journal} {Science Advances}\ }\textbf {\bibinfo {volume} {8}},\
  \bibinfo {pages} {eabn1778} (\bibinfo {year} {2022})}\BibitemShut {NoStop}%
\bibitem [{\citenamefont {Miroshnichenko}\ \emph {et~al.}(2017)\citenamefont
  {Miroshnichenko}, \citenamefont {Kiselev}, \citenamefont {Trifanov},\ and\
  \citenamefont {Gleim}}]{Kiselev:josab:2017}%
  \BibitemOpen
  \bibfield  {author} {\bibinfo {author} {\bibfnamefont {G.~P.}\ \bibnamefont
  {Miroshnichenko}}, \bibinfo {author} {\bibfnamefont {A.~D.}\ \bibnamefont
  {Kiselev}}, \bibinfo {author} {\bibfnamefont {A.~I.}\ \bibnamefont
  {Trifanov}},\ and\ \bibinfo {author} {\bibfnamefont {A.~V.}\ \bibnamefont
  {Gleim}},\ }\bibfield  {title} {\bibinfo {title} {Algebraic approach to
  electro-optic modulation of light: exactly solvable multimode quantum
  model},\ }\href {https://doi.org/10.1364/JOSAB.34.001177} {\bibfield
  {journal} {\bibinfo  {journal} {J. Opt. Soc. Am. B}\ }\textbf {\bibinfo
  {volume} {34}},\ \bibinfo {pages} {1177} (\bibinfo {year}
  {2017})}\BibitemShut {NoStop}%
\bibitem [{\citenamefont {Kelley}\ and\ \citenamefont
  {Kleiner}(1964)}]{Kelley:pr:1964}%
  \BibitemOpen
  \bibfield  {author} {\bibinfo {author} {\bibfnamefont {P.~L.}\ \bibnamefont
  {Kelley}}\ and\ \bibinfo {author} {\bibfnamefont {W.~H.}\ \bibnamefont
  {Kleiner}},\ }\bibfield  {title} {\bibinfo {title} {Theory of electromagnetic
  field measurement and photoelectron counting},\ }\href
  {https://doi.org/10.1103/PhysRev.136.A316} {\bibfield  {journal} {\bibinfo
  {journal} {Phys. Rev.}\ }\textbf {\bibinfo {volume} {136}},\ \bibinfo {pages}
  {A316} (\bibinfo {year} {1964})}\BibitemShut {NoStop}%
\bibitem [{\citenamefont {Vogel}\ and\ \citenamefont
  {Welsch}(2006)}]{Vogel:bk:2006}%
  \BibitemOpen
  \bibfield  {author} {\bibinfo {author} {\bibfnamefont {W.}~\bibnamefont
  {Vogel}}\ and\ \bibinfo {author} {\bibfnamefont {D.-G.}\ \bibnamefont
  {Welsch}},\ }\href@noop {} {\emph {\bibinfo {title} {Quantum Optics}}},\
  \bibinfo {edition} {3rd}\ ed.\ (\bibinfo  {publisher} {Wiley-VCH},\ \bibinfo
  {address} {Berlin},\ \bibinfo {year} {2006})\ p.\ \bibinfo {pages}
  {508}\BibitemShut {NoStop}%
\bibitem [{\citenamefont {Wu}\ \emph {et~al.}(2020)\citenamefont {Wu},
  \citenamefont {Liu},\ and\ \citenamefont {Simon}}]{Wu:pra:2020}%
  \BibitemOpen
  \bibfield  {author} {\bibinfo {author} {\bibfnamefont {Y.}~\bibnamefont
  {Wu}}, \bibinfo {author} {\bibfnamefont {J.}~\bibnamefont {Liu}},\ and\
  \bibinfo {author} {\bibfnamefont {C.}~\bibnamefont {Simon}},\ }\bibfield
  {title} {\bibinfo {title} {Near-term performance of quantum repeaters with
  imperfect ensemble-based quantum memories},\ }\href
  {https://doi.org/10.1103/PhysRevA.101.042301} {\bibfield  {journal} {\bibinfo
   {journal} {Phys. Rev. A}\ }\textbf {\bibinfo {volume} {101}},\ \bibinfo
  {pages} {042301} (\bibinfo {year} {2020})}\BibitemShut {NoStop}%
\bibitem [{\citenamefont {Semenenko}\ \emph {et~al.}(2022)\citenamefont
  {Semenenko}, \citenamefont {Hu}, \citenamefont {Figueroa},\ and\
  \citenamefont {Perebeinos}}]{Semenenko:avs:2022}%
  \BibitemOpen
  \bibfield  {author} {\bibinfo {author} {\bibfnamefont {V.}~\bibnamefont
  {Semenenko}}, \bibinfo {author} {\bibfnamefont {X.}~\bibnamefont {Hu}},
  \bibinfo {author} {\bibfnamefont {E.}~\bibnamefont {Figueroa}},\ and\
  \bibinfo {author} {\bibfnamefont {V.}~\bibnamefont {Perebeinos}},\ }\bibfield
   {title} {{\selectlanguage {english}\bibinfo {title} {Entanglement generation
  in a quantum network with finite quantum memory lifetime}},\ }\href
  {https://doi.org/10.1116/5.0082239} {\bibfield  {journal} {\bibinfo
  {journal} {AVS Quantum Science}\ }\textbf {\bibinfo {volume} {4}},\ \bibinfo
  {pages} {012002} (\bibinfo {year} {2022})}\BibitemShut {NoStop}%
\bibitem [{\citenamefont {Mol}\ \emph {et~al.}(2023)\citenamefont {Mol},
  \citenamefont {Esguerra}, \citenamefont {Meister}, \citenamefont {Bruschi},
  \citenamefont {Schell}, \citenamefont {Wolters},\ and\ \citenamefont
  {Wörner}}]{Mol:qst:2023}%
  \BibitemOpen
  \bibfield  {author} {\bibinfo {author} {\bibfnamefont {J.-M.}\ \bibnamefont
  {Mol}}, \bibinfo {author} {\bibfnamefont {L.}~\bibnamefont {Esguerra}},
  \bibinfo {author} {\bibfnamefont {M.}~\bibnamefont {Meister}}, \bibinfo
  {author} {\bibfnamefont {D.~E.}\ \bibnamefont {Bruschi}}, \bibinfo {author}
  {\bibfnamefont {A.~W.}\ \bibnamefont {Schell}}, \bibinfo {author}
  {\bibfnamefont {J.}~\bibnamefont {Wolters}},\ and\ \bibinfo {author}
  {\bibfnamefont {L.}~\bibnamefont {Wörner}},\ }\bibfield  {title} {\bibinfo
  {title} {Quantum memories for fundamental science in space},\ }\href
  {https://doi.org/10.1088/2058-9565/acb2f1} {\bibfield  {journal} {\bibinfo
  {journal} {Quantum Science and Technology}\ }\textbf {\bibinfo {volume}
  {8}},\ \bibinfo {pages} {024006} (\bibinfo {year} {2023})}\BibitemShut
  {NoStop}%
\end{thebibliography}
\providecommand{\noopsort}[1]{}\providecommand{\singleletter}[1]{#1}%

\end{document}